\documentclass[
reprint,
aps,prc,
amssymb,amsmath,amsfonts
]{revtex4-2}

\usepackage{graphicx}
\usepackage{xcolor}
\usepackage[colorlinks=true,linkcolor=blue,urlcolor=blue,citecolor=blue,bookmarks=true,bookmarksopenlevel=2]{hyperref}
\usepackage{amsmath}

\usepackage[utf8]{inputenc}
\usepackage[T1]{fontenc}
\usepackage{microtype}

\usepackage{array,tabularx,booktabs,colortbl}
\usepackage{longtable}

\newcolumntype{L}{>{\raggedright\arraybackslash}X}
\newcolumntype{R}{>{\raggedleft\arraybackslash}X}
\newcolumntype{C}{>{\centering\arraybackslash}X}

\usepackage{rotating}

\newcommand{\ket}[1]{\left\vert#1\right\rangle}

\usepackage{xspace}
\usepackage{mhchem}
\newcommand{\grasp}{\textsc{Grasp2018}\xspace}
\newcommand{\fac}{\textsc{Fac}\xspace}

\newcommand{\secc}{s\textsuperscript{-1}\xspace}

\usepackage{siunitx}
\robustify\dots
\usepackage{tikz}
\usepackage[normalem]{ulem}
\usepackage{soul}
\soulregister\cite7
\soulregister\ref7
\soulregister\pageref7
\soulregister\ce7
\soulregister\num7
\definecolor{neongreen}{RGB}{195,255,0}
\definecolor{yellowgreen}{RGB}{154,255,135}
\definecolor{goldenrod}{RGB}{218,255,212}
\definecolor{burgundy}{RGB}{128, 0, 32}
\definecolor{purple}{RGB}{128, 0, 128}
\definecolor{OxfordBlue}{RGB}{0, 33, 71}
\definecolor{ElectricIndigo}{RGB}{111, 0, 255}
\definecolor{myred}{RGB}{200,0,0}

\usepackage[linecolor=red]{mdframed}

\begin{document}

\title{Nuclear Excitation by Near-Resonant Electron Transition in \ce{^{229}Th^{39+}} Ions}
\author{Karol Kozio{\l}}
\email{Karol.Koziol@ncbj.gov.pl}
\author{Jacek Rzadkiewicz} 
\email{Jacek.Rzadkiewicz@ncbj.gov.pl}
\affiliation{Narodowe Centrum Bada\'{n} J\k{a}drowych (NCBJ), Andrzeja So{\l}tana 7, 05-400 Otwock-\'{S}wierk, Poland}

\begin{abstract}
Theoretical considerations are made for the nuclear excitation from the ground state to the \SI{8}{eV} \ce{^{229m}Th} isomer via near-resonant electron transitions in Sb-like ($q=39+$) thorium ions. The energy of the first excited atomic state ($J=7/2$) in the \ce{^{229}Th^{39+}} ion is estimated to be \SI{8.308+-0.069}{eV}, which is very close to the new reference value for the \ce{^{229m}Th} nuclear isomer energy, \SI{8.356}{eV} [Zhang et al., Nature 633, 63 (2024)]. Therefore, the \ce{^{229}Th^{39+}} ion provides a unique opportunity to study the nuclear excitation by electron transition process under well-defined atomic conditions in the regime of extremely low nuclear excitation energy. Our results indicate that the upper theoretical limit for the \ce{^{229m}Th} isomer excitation rate reaches an enormous value of \SI{2.50e16}{\per\second} at resonance ($\Delta=0$~meV), but drops many orders of magnitude for the value of electronic transition energy within its uncertainty interval. Additionally, it was shown that using an electron beam ion trap, the production of the \ce{^{229}Th} isomer can reach rates ranging from a few to approximately \SI{5e20}{\per\second}.
\end{abstract}

\maketitle

%%%%%

\section{Introduction}

There is strong interdisciplinary interest in the \ce{^{229}Th} isotope due to its potential applications, such as serving as a nuclear clock frequency standard \cite{Campbell2012,Peik2021,Beeks2021} and allowing for precise determination of the temporal variation of fundamental constants \cite{Flambaum2006,Berengut2009,Thirolf2019}. This interest is particularly focused on its metastable first excited state, \ce{^{229m}Th} ($I^\pi=3/2^+$, $T_{1/2}$ = \SI{1740(50)}{s} \cite{Tiedau2024}), which is approximately \SI{8}{eV} above the nuclear ground state -- several orders of magnitude lower than typical nuclear excitation energies. 
Recent experimental values for the \ce{^{229m}Th} isomer excitation energy (since 2019) range from \SI{8.1}{eV} to \SI{8.3}{eV} \cite{Seiferle2019,Yamaguchi2019,Sikorsky2020,Kraemer2023}. 
The most recent experiments \cite{Tiedau2024,Zhang2024}, performed using a VUV frequency comb to directly excite the \ce{^{229m}Th} incorporated into \ce{CaF2} crystal, resulted in an excitation frequency of {\SI{2020407384335(2)}{kHz}} that corresponds to the energy of {\SI{8.355733552}{eV}}. That extraordinary accuracy in determining the nuclear excitation energy gives us a unique opportunity to study the interactions between the nucleus and electrons. 

Progress in experimental techniques, particularly in manipulating highly-charged ions (HCIs), has opened new avenues for driving the isomeric transition. 
Several methods have been proposed to excite the \ce{^{229}Th} nucleus from its ground state to various excited isomeric states \cite{Beeks2021}. Before performing successfully the excitation of the \ce{^{229}Th} ground state to the first excited state, a controlled excitation of the ground state to the second excited state ($I^\pi=5/2^+$, $E={\SI{29.19}{keV}}$, $T_{1/2}\sim{\SI{100}{ps}}$) has been achieved through x-ray pumping with synchrotron radiation \cite{Masuda2019}. Additionally, the production of \ce{^{229m}Th} by exciting the \ce{^{229}Th} ground state to the second excited state at \SI{29.19}{keV} via nuclear excitation by electron capture (NEEC) has been proposed \cite{Zhao2024}. 
Other proposed methods for exciting the \ce{^{229}Th} isomer include nuclear excitation by inelastic electron scattering (NEIES) \cite{Tkalya2020,Zhang2023}, via electronic resonance \cite{Porsev2010a,Porsev2010b,Muller2019,Bilous2020,Dzyublik2022}, and via defect states in doped crystals \cite{Nickerson2020}. The potential for nuclear excitation by electron transfer (NEET), two-photon electron transitions (NETP), or electronic bridge excitation (EBE) of the \ce{^{229}Th} isomer has also been theoretically investigated for \ce{Th^+} \cite{Porsev2010,Porsev2010b,Zhang2023}, \ce{Th^{2+}} \cite{Peik2015,Muller2019,Zhang2023}, \ce{Th^{3+}} \cite{Porsev2010a,Zhang2023}, and \ce{Th^{35+}} ions \cite{Bilous2020,Porsev2021}. 
Recently, the NEET, NEEC, and NEIES processes have been studied in the case of \ce{Th^+}, \ce{Th^{2+}}, and \ce{Th^{3+}} ions in plasma \cite{Zhang2023}. In that work for the NEET process, the radiative transitions from the high principal quantum number orbital have been used in order to match the expected nuclear isomer energy. However, as concluded in Ref. \cite{Zhang2023}, preparing a specific ionic excited state may be challenging, and an experimental setup with enhanced control over the ion's excited state, such as an electron beam ion trap (EBIT), could facilitate the NEET process. 
Until now, the NEET process has been studied for \ce{^235U} \cite{Morita1973,Izawa1979,Arutyunyan1991,Claverie2004,Chodash2016}, \ce{^237Np} \cite{Saito1980,Tkalya1992,Pisk1989,Ljubicic1991}, \ce{^189Os} \cite{Ahmad2000,Ljubicic1991,Tkalya1992}, \ce{^197Au} \cite{Kishimoto2000,Fujioka1984,Pisk1989,Ljubicic1991,Tkalya1992}, and \ce{^193Ir} \cite{Kishimoto2005,Tkalya1992} nuclei. 
The results of the NEET probability measurements vary by orders of magnitude and are sometimes even contradictory. For example, the probability results for the NEET process occurring at \ce{^{189}Os} are in the range of $10^{-6}$ to $10^{-10}$ as an upper limit (see Ref.~\cite{Sakabe2005} and references therein). Similarly, in the case of \ce{^{235}U} and \ce{^{237}Np}, the latest experimental studies \cite{Chodash2016,Shinohara1981} in which there was no evidence of NEET are in contradiction with earlier studies that reported the observation of the process \cite{Sakabe2005}. It seems that only for \ce{^{197}Au} and \ce{^{193}Ir} nuclei the observation of the NEET process has been successfully confirmed \cite{Kishimoto2000,Kishimoto2005}. However, both experimental results and theoretical predictions for \ce{^{197}Au} and \ce{^{193}Ir} differ by orders of magnitude (see Refs.~\cite{Kishimoto2000,Kishimoto2005,Tkalya1992} and references therein). The latest theoretical predictions are in reasonable agreement with the experimental results, although they still differ by a factor of up to several.
Despite the existence of abundant experimental data, they are highly inconsistent, which makes it practically impossible to verify the theory describing the NEET process thoroughly. Moreover, most of the NEET studies, except 77-eV \ce{^235U} case,  concern the excitation of the nuclear levels through high-energy x-ray core transitions. 
The \ce{^{229}Th^{39+}} isomer with its excitation energy known with exceptional precision provides a unique opportunity to study the NEET process under well-defined atomic conditions. Moreover, under these conditions, the NEET resonance structure has an extremely low width and reaches a very high rate at resonance.

\begin{figure}[!t]
\centering
\includegraphics[width=0.8\linewidth]{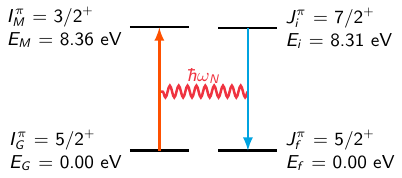}
\caption{NEET mechanism in \ce{^{229}Th^{39+}} ions. The symbols $G$ and $M$ denote the ground and isomeric states in the \ce{^{229}Th} nucleus, and $i$ and $f$ the ground and the first excited atomic states in the \ce{^{229}Th^{39+}} ion.}
\label{fig:neet_scheme}
\end{figure}

In this work, we investigate nuclear excitation by near-resonant electron transition in Sb-like (\ce{^{229}Th^{39+}}) thorium ions with the \ce{[Kr]{4d}^{10}{4f}^{5}} electronic configuration. In the NEET process, the energy of electron transition in an atom is transferred to the excitation of the nucleus. This transfer can be particularly effective if the nuclear excitation energy matches the electron transition energy. In the \ce{^{229}Th^{39+}}, the ground state transition from the first atomic excited state seems to meet this condition. Therefore we attempted to precisely determine the energy of this transition and its intensity (natural width of the state). Since experimental data are not available for the low-energy atomic states of \ce{Th^{39+}} ions, in our analyzes we employed the Multi-Configuration Dirac--Hartree--Fock (MCDHF) method with Configuration Interaction (CI). By using the atomic transition energy and the widths of the appropriate atomic states together with the most accurate estimates of the energy and width of the \ce{^{229m}Th} isomer \cite{Zhang2024}, we estimate the NEET rates for \ce{Th^{39+}} ions. In addition, the low-energy atomic structure of \ce{Th^{39+}} ions was analyzed. Using this structure, an experimental scenario for EBIT is considered. The scheme of the NEET mechanism for the \ce{^{229}Th^{39+}} ion is shown in Fig.~\ref{fig:neet_scheme}.

\section{MCDHF-CI calculations for \ce{Th^{39+}} ion}

The calculations of the energy levels for the \ce{Th^{39+}} thorium ion were performed using the \grasp code \cite{FroeseFischer2018}, which implements the MCDHF-CI method. The methodology employed in this study is similar to previous work (see, e.g., \cite{Grant2007,FroeseFischer2016,Kozio2018}) and is based on a fully relativistic Dirac Hamiltonian and a multi-configurational expansion of the Atomic State Function (ASF) of the total angular momentum $J$ and parity $p$: 
\begin{equation}
\Psi_{s} (J^{p} ) = \sum_{m} c_{sm} \psi ( \gamma_{m} J^{p} ) \;,
\end{equation}
where $\psi ( \gamma_{m} J^{p} )$ are the configuration state functions (CSFs), $c_{m} (s)$ are the configuration mixing coefficients for state $s$, and $\gamma_{m}$ represents all information required to define a certain CSF uniquely. 
In the present calculations, we used active spaces (AS) of virtual orbitals with principal quantum numbers $n$ up to 7 and angular momentum quantum numbers $l$ up to 5. The multireference (MR) set includes all CSFs corresponding to the \ce{[Kr]{4d}^{10}{4f}^{5}} electronic configuration. We explored the extension of the active space by increasing both the number of occupied orbitals, from which single (S) and double (D) substitutions are made, and the number of virtual orbitals to which these substitutions are applied. 
In this work, we utilized active spaces denoted as AS$x$(4f), AS$x$(4df), AS$x$(4pdf), and AS$x$(4spdf). The AS$x$(4f) active space includes SD substitutions (i.e., both S and D substitutions are allowed) from the 4f subshell, while AS$x$(4df) includes SD substitutions from both the 4d and 4f subshells, and so forth. The $n$ = 1, 2, 3 shells are treated as the inactive core. The notation $x$ refers to the highest principal quantum number of the virtual orbitals; for example, in the AS1 active space, the highest $n$ of the orbitals is 4, while in AS2 it is 5, and so on. 
The Extended Optimal Level (EOL) scheme \cite{Grant1984} was employed to calculate the radial wavefunctions of the considered levels. In this scheme, the radial wavefunctions are optimized for all states related to the given electron configuration, averaged with the weights of $(2J+1)$. We calculated the Breit term in the low-frequency limit, as a frequency-dependent term is not appropriate for virtual orbitals \cite{Si2018}. 

\begin{figure}[!t]
\centering
\includegraphics[width=\linewidth]{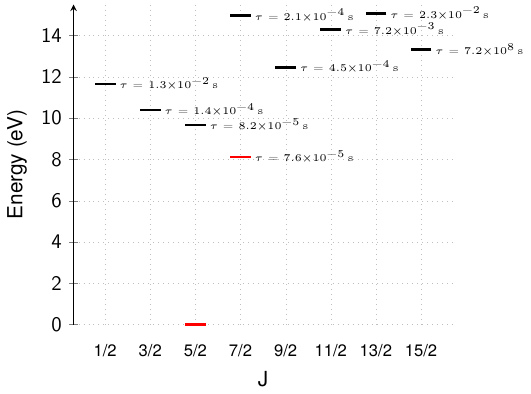}
\caption{Energy level diagram for the ten lowest-lying states of the \ce{[Kr]{4d}^{10}{4f}^{5}} configuration of \ce{Th^{39+}}. The radiative lifetime values are indicated next to the level markers.}
\label{fig:levels_low}
\end{figure}

In order to find the appropriate Th ion charge and transition of energy about \SI{8}{eV} we did a lot of trial and error calculations. We performed MCDHF calculations (without extensive CI) for all Th ions with charges from 31+ to 53+ and valence configurations from {\ce{[Kr]{4d}^{1}}} to {\ce{[Kr]{4d}^{10}{4f}^{13}}} and studied 1804 atomic levels and 61963 transitions within the electronic configurations presented above. As a result, the ground state transition of {Th$^{39+}$} has been chosen. 

\begin{table}[!t]
\caption{\label{tab:j7-j5-tran}Energy of the $J=7/2$ level relative to the $J=5/2$ ground state of \ce{Th^{39+}} for various active spaces.}
\begin{tabular*}{\linewidth}{@{} l @{\extracolsep{\fill}} llll r @{}}
\toprule
Active & \multicolumn{4}{c}{Excitations} & {$E$}\\
\cmidrule{2-5}
space & \multicolumn{2}{c}{from} & \multicolumn{2}{c}{to} & {(eV)} \\
\cmidrule{2-3}\cmidrule{4-5}
& SD & S only & SD & S only &\\
\midrule
MR\,=\,AS1(4f) & 4f && 4f & & 7.993 \\
AS2(4f) & 4f && $\le$5spdfg & & 8.140 \\
AS3(4f) & 4f && $\le$6spdfgh & & 8.150 \\
AS4(4f) & 4f && $\le$7spdfgh & & 8.151 \\
AS1(4df) & 4df && $\le$4df & & 8.177 \\
AS2(4df) & 4df && $\le$5spdfg & & 8.245 \\
AS3(4df) & 4df && $\le$6spdfgh & & 8.203 \\
AS4(4df) & 4df && $\le$7spdfgh & & 8.199 \\
AS1(4pdf) & 4pdf && $\le$4pdf & & 8.243 \\
AS2(4pdf) & 4pdf && $\le$5spdfg & & 8.386 \\
AS3(4pdf) & 4pdf && $\le$6spdfgh & & 8.290 \\
AS4(4pdf) & 4pdf && $\le$6spdfgh & 7spdfg & 8.288 \\
AS1(4spdf) & 4spdf && $\le$4spdf & & 8.261 \\
AS2(4spdf) & 4spdf && $\le$5spdfg & & 8.408 \\
AS1*(4spdf) & 4pdf & 4s & $\le$4spdf && 8.243 \\
AS2*(4spdf) & 4pdf & 4s & $\le$5spdf && 8.419 \\
AS3*(4spdf) & 4pdf & 4s & $\le$6spdfgh && 8.309 \\
AS4*(4spdf) & 4pdf & 4s & $\le$6spdfgh & 7spdfg & 8.308 \\
\multicolumn{5}{@{}l}{reference value} & 8.308(69) \\
\bottomrule
\end{tabular*}
\end{table}

The total number of calculated levels for the given $J^\pi$ symmetry of the \ce{Th^{39+}} {\ce{[Kr]{4d}^{10}{4f}^{5}}} configuration is as follows: 10 levels for $1/2^-$ symmetry, 21 for $3/2^-$, 28 for $5/2^-$, 30 for $7/2^-$, 29 for $9/2^-$, 26 for $11/2^-$, 20 for $13/2^-$, 16 for $15/2^-$, 9 for $17/2^-$, 5 for $19/2^-$, 3 for $21/2^-$, and 1 for $23/2^-$. Due to the rapid increase in the size of the expansions with the growth of the MR set, the AS2(4df) level of calculations was employed in this case. 
Calculated electronic energy levels for the ten lowest-lying states of the \ce{[Kr]{4d}^{10}{4f}^{5}} configuration of the \ce{Th^{39+}} ion, categorized by their angular momentum quantum number $J$ groups, are shown in Fig.~\ref{fig:levels_low}. 
Energies and $LS$-compositions of all \ce{[Kr]{4d}^{10}{4f}^{5}} levels are listed in the table in the Supplementary Material, and a graphical representation of the level arrangement is also presented in the Supplementary Material. 
As illustrated in Fig.~\ref{fig:levels_low}, the energy difference between the ground state of $J=5/2$ and the first excited state of $J=7/2$ is approximately \SI{8}{eV}, which is very close to the excitation energy of the isomeric state in the \ce{^{229}Th} nucleus. Notably, the $J=15/2$ atomic state at 13 eV has an extraordinarily long radiative lifetime, on the order of \SI{e8}{s}. The dominant radiative de-excitation channel for the $J=15/2$ state is the weak M3 transition to the $J=9/2$ state, with a rate of \SI{3.9e-20}{\per\s}. The sum of hyperfine-induced transition rates from the $J=15/2$ level to the possible lower levels (calculated using the \textsc{MitHit} code \cite{Li2020}, based on \grasp data) is \SI{1.4e-9}{\per\s}. 
The population of this $J=15/2$ state, along with all cascade feeding transitions, would limit the relative population of the first excited state ($J=7/2$), which is necessary to initiate the NEET process in the \ce{Th^{39+}} ion.

\begin{figure}[!t]
\centering
\includegraphics[width=\linewidth]{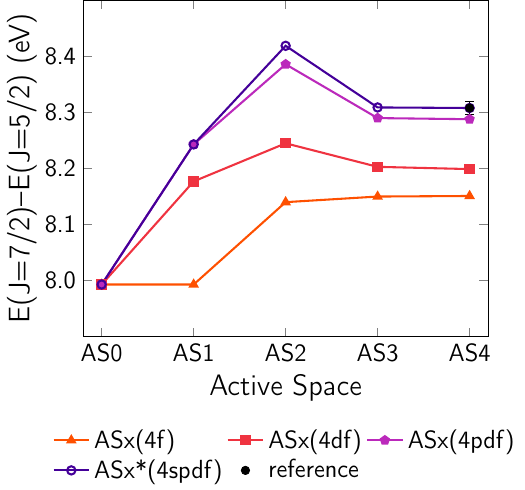}
\caption{Convergence of energies (in eV) for the first $J=7/2$ atomic level in \ce{Th^{39+}}, relative to the energy of the $J=5/2$ ground state, as the active spaces used in configuration interaction calculations are extended.}
\label{fig:lev-conv}
\end{figure}

As mentioned above, the recent analysis estimates the excitation energy for \ce{^{229}Th} to be \SI{8.356}{eV} \cite{Zhang2024}. Therefore, we focused on the atomic transitions between the first excited state ($J=7/2$) and the ground state ($J=5/2$) of the \ce{Th^{39+}} ion. Energies of the $J=7/2$ level related to the $J=5/2$ ground state for various active spaces are presented in Table~\ref{tab:j7-j5-tran} and Fig.~\ref{fig:lev-conv}. 
In the calculations, the radial wavefunctions were optimized separately for the $J=5/2$ ground and $J=7/2$ first excited states to account for the full orbital relaxation effect. The energy difference related to the frequency-dependent Breit term on energy levels has been examined at the MR level and found to be 4~meV. This value has been added to the numbers obtained for higher active spaces. 
Fig.~\ref{fig:lev-conv} presents the $J=7/2 \to J=5/2$ transition energies, which range from about \SI{8.0}{eV} to \SI{8.4}{eV}, depending on the size of the active space and the number of 4$l$ occupied active orbitals used in the calculations. As seen in the figure, the use of more 4$l$ occupied active orbitals in the MCDF-CI calculations results in a higher transition energy. A significant increase in the transition energy for AS1 and AS2 active spaces (except for AS1(4f)) is also clearly visible. 
For the AS$x$(4f), AS$x$(4df), and AS$x$(4pdf) estimates of the $J=7/2 \to J=5/2$ transition energy, it is evident that the AS4 stage is adequate to achieve level energy convergence. 

Due to computational limitations (specifically, the limitations of the parallel RCI procedure implemented in the \grasp code), we were not able to perform calculations at the AS3(4spdf) and AS4(4spdf) stages.
Instead of the full AS$x$(4spdf) active space, which accounts for double virtual excitations for all 4s, 4p, 4d, and 4f subshells, we used the slightly reduced AS$x$*(4spdf) active space. This space contains the double virtual excitations of the 4p, 4d, and 4f subshells, along with single excitations from the 4s subshell. The difference in transition energy calculated at the AS2(4spdf) and AS2*(4spdf) stages is 11~meV.
We assume that this value corresponds to the error in approximating the AS4(4spdf) value with the AS4*(4spdf) value, i.e., $\delta_\text{approx} = |E^\text{AS2(4spdf)}-E^\text{AS2*(4spdf)}|$. The uncertainty related to convergence with the size of the basis set is estimated as $\delta_\text{conv} = |E^\text{AS4*(4spdf)}-E^\text{AS3*(4spdf)}|$. This estimation yields 1~meV.  
Due to the computational limitations we were not able to perform calculations in which all shells are active. However, in order to estimate the error related to the missing correlation contributions resulting from excitations from core $K$, $L$, and $M$ shells we performed a series of limited calculations to probe core--valence and core--core correlations. In such calculations the active spaces of occupied orbitals are: \{1s, 4f\}, \{2s, 2p, 4f\}, \{3s, 3p, 4f\}, and \{3d, 4f\}, respectively, and SD substitutions are allowed. Then, the related uncertainties are calculated as $\delta_\text{corr.1s} = |E^\text{AS4(4f)}-E^\text{AS4(1s,4f)}|$ an so on. The obtained results are $\delta_\text{corr.1s}$ = 0.6~meV, $\delta_\text{corr.2sp}$ = 0.9~meV, $\delta_\text{corr.3sp}$ = 23~meV, and $\delta_\text{corr.3d}$ = 64~meV. 
The total theoretical uncertainty for the $J=7/2 \to J=5/2$ transition energy is then $\delta = \left[(\delta_\text{approx})^2+(\delta_\text{conv})^2\right.$ $\left.+(\delta_\text{corr.1s})^2+(\delta_\text{corr.2sp})^2+(\delta_\text{corr.3sp})^2+(\delta_\text{corr.3d})^2\right]^{1/2}$ = 69~meV. Thus the reference value of the $J=7/2 \to J=5/2$ transition energy resulting from our approach is \SI{8.308+-0.069}{eV}.

\section{NEET calculations}

Based on the calculated energy levels, we consider the NEET mechanism for the excitation of the nuclear ground state ($5/2^+$) to the \SI{8}{eV} isomeric state ($3/2^+$) in \ce{Th^{39+}} ions. The ground state electronic transition $\ket{i}\to\ket{f}$ is a part of a cascade decay of low-energy atomic states in \ce{Th^{39+}} ions. This transition can, therefore, spontaneously induce a NEET process from the nuclear ground state to the \ce{^{229m}Th} isomer (i.e., nuclear excitation $\ket{G}\to\ket{M}$). 
The general expression for the probability of the NEET process, $W_\text{NEET}$, can be found in Refs. \cite{Tkalya1992,Harston1999,Harston2001,Denis-Petit2017}. Because initial and final states of electronic transitions consist of more than one CSF, we use the multiconfigurational approach to the NEET rate calculations (see \cite{Harston1999,Denis-Petit2017} for details). Assuming that $\ket{i}$ and $\ket{f}$ are the initial (upper) and final (lower) electronic states of interest and $\ket{G}$ and $\ket{M}$ are the ground and metastable (isomeric) nuclear states, the rate of the spontaneous NEET process is expressed (in atomic units) as: 
\begin{equation}\label{eq:w_neet}
\begin{array}{>{\displaystyle}l>{\displaystyle }l}
W_\text{NEET} = & \frac{4\pi (\Gamma_i+\Gamma_f+\Gamma_M)}{(E_{if} - E_{GM})^2 + \tfrac{1}{4}(\Gamma_i+\Gamma_f+\Gamma_M)^2} \\[1.5em]
& \times \sum_{\lambda L} \left(\frac{E_{GM}}{c}\right)^{2L+2} \frac{R_{if}^2(\lambda L) B_{GM}(\lambda L)}{[(2L+1)!!]^2} \; ,
\end{array}
\end{equation}
where $\Gamma_i$/$\Gamma_f$/$\Gamma_M$ are the natural widths of the electronic and nuclear levels, $E_{if}$ is the electronic transition energy, $B_{GM}(\lambda L)$ is the reduced probability of the nuclear transition of the type $\lambda$ and multipolarity $L$ (M1 or E2 in our case, so $L = 1,2$), and $E_{GM}$ is the nuclear transition energy. Note that $B_{GM}(\lambda L) = \frac{(2I_N+1)}{(2I_M+1)}B_{MG}$. 
The squared matrix element $R_{if}^2(\lambda L)$ represents the electron part of electron-nucleus interaction and it is expressed as: 
\begin{equation}\label{eq:r_if}
\begin{array}{>{\displaystyle}l>{\displaystyle }l}
R_{if}^2(\lambda L) = & \sum_{t t'} \sum_{ab} c_{it} c_{ft'} d_{t t' ab}^L \\[1.5em]
& \times \left( j_a \tfrac{1}{2} L 0 | j_b \tfrac{1}{2} \right)^2 \left|M_{ab}(\lambda L)\right|^2 \;.
\end{array}
\end{equation}
The coefficients $c_{it}$ and $c_{ft'}$ are coefficients of expansion of ASFs $\ket{i}$ and $\ket{f}$ into CSFs $\ket{t}$ and $\ket{t'}$, respectively. The coefficients $d_{t t' ab}^L$ arising from expression of the reduced matrix elements involving CSFs $\ket{t}$ and $\ket{t'}$ in terms of the one electron reduced matrix elements involving orbitals $a$ and $b$ \cite{Grant2007}. In our case both $a$ and $b$ may be 4f$_{5/2}$ or 4f$_{7/2}$ orbitals. 
$M_{ab}(\lambda L)$ is the radial matrix element related to the $2^L$-pole atomic transition (E$L$ or M$L$) between orbitals $a$ and $b$ (see, e.g., \cite{Tkalya1992,Harston1999,Harston2001,Denis-Petit2017} for details) and $\left( j_a \tfrac{1}{2} L 0 | j_b \tfrac{1}{2} \right)$ is the Clebsch-Gordan coefficient. 
The probability of a nuclear transition into an excited state during the decay of an excited atomic level $\ket{i}$ is given by $P_\text{NEET} = W_\text{NEET}/{\Gamma_i}$. The enhancement factor $\beta$ \cite{Muller2019,Porsev2010a}, which is independent of the nuclear transition probability (a major source of uncertainty in calculating $W_\text{NEET}$), is defined as $\beta = W_\text{NEET}/\Gamma_M$, where $\Gamma_M$ is the width of the excited metastable nuclear level. The width of the nuclear level $\Gamma_{M}$, which is the sum of the radiative transitions (M1 and E2) between the isomeric state $\ket{M}$ and the ground state $\ket{G}$ in the \ce{^{229}Th} nucleus, is defined (in atomic units) as \cite{Ring1980}: 
\begin{equation}\label{eq:gamma_nucl}
\Gamma_{M} = \sum_{\lambda L} \frac{8\pi (L+1)}{L[(2L+1)!!]^2} \left(\frac{E_{GM}}{c} \right)^{2L+1} B_{MG}(\lambda L) \; .
\end{equation}
The NEET process has a resonant character, and its rate is strongly dependent on the value of $\Delta = (E_{if} - E_{GM})$, where $E_{if}$ is the energy of the electronic transition that facilitates the nuclear transition of energy $E_{GM}$. For this reason, an effort was made to identify an electronic transition in the Th ion with energy as close as possible to the nuclear excitation energy. 
The atomic state $\ket{i}$ can de-excite via an M1 or E2 transition to the ground atomic state, labeled $\ket{f}$. The $\ket{i}\to\ket{f}$ rate for the M1 transition is \SI{1.3e+4}{\per\s}, while for the E2 transition, it is \SI{1.1e-2}{\per\s}. The natural width of the $\ket{i}$ level is \SI{8.7e-12}{eV}, while the natural width of the $\ket{f}$ level is effectively zero.

\begin{table}[!htb]
\caption{\label{tab:neet}$W_\text{NEET}$ (in \si{\per\s}), $P_\text{NEET}$, $\beta_\text{NEET}$, and $\tilde{W}_\text{NEET}$ for various values of $|\Delta| = |E_{if} - E_{GM}|$.}
\sisetup{table-format=1.2e-1}
\begin{tabular*}{\linewidth}{@{} l @{\extracolsep{\fill}} *{3}{S[scientific-notation=true,round-mode=places,round-precision=2]} S[scientific-notation=true,round-mode=places,round-precision=2] @{}}
\toprule
$|\Delta|$ & {$W_\text{NEET}$} & {$P_\text{NEET}$} & {$\beta_\text{NEET}$} & {$\tilde{W}_\text{NEET}$} \\
(meV) & {(\secc)} &  &  & {(\secc)} \\
\midrule
0 & 2.4970E+16 & 1.8891E+12 & 6.1813E+19 & 5.2936E+20 \\
21 & 1.0447E-03 & 7.9041E-08 & 2.5862E+00 & 2.2148E+01 \\
48 & 2.0737E-04 & 1.5689E-08 & 5.1335E-01 & 4.3963E+00 \\
117 & 3.4674E-05 & 2.6233E-09 & 8.5836E-02 & 7.3509E-01 \\
\bottomrule
\end{tabular*}
\end{table}

\begin{figure}[!htb]
\centering
\includegraphics[width=\linewidth]{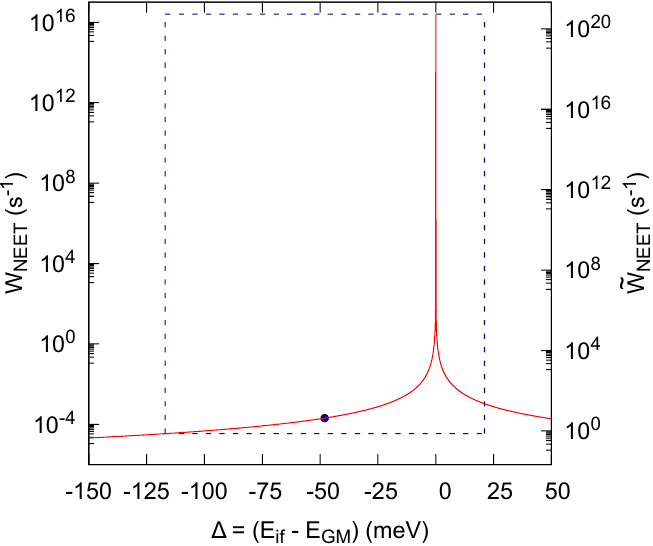}
\caption{
Possible values of the quantities $W_\text{NEET}$ and $\tilde{W}_\text{NEET}$ (in \si{\per\s}) for various values of $\Delta = (E_{if} - E_{GM})$. The dot marks the $\Delta$ value for which both $E_{if}$ and $E_{GM}$ are at the center of their uncertainty intervals. The dashed box delineates the possible values of $\Delta$ and $W_\text{NEET}$ based on the uncertainty intervals of the present calculated $E_{if}$ and $E_{GM}$ taken from Ref.~\cite{Zhang2024}. 
}
\label{fig:w_neet}
\end{figure}

The values of the quantities $W_\text{NEET}$, $P_\text{NEET}$, and $\beta_\text{NEET}$ for various values of $|\Delta| = |E_{if} - E_{GM}|$ are summarized in Table~\ref{tab:neet}. 
The calculated values of $\left|M_{4f_{7/2},4f_{5/2}}(M1)\right|^2$, $\left|M_{4f_{5/2},4f_{5/2}}(M1)\right|^2$, $\left|M_{4f_{7/2},4f_{7/2}}(M1)\right|^2$, $\left|M_{4f_{7/2},4f_{5/2}}(E2)\right|^2$, $\left|M_{4f_{5/2},4f_{5/2}}(E2)\right|^2$, and $\left|M_{4f_{7/2},4f_{7/2}}(E2)\right|^2$ are \SI{3.29E+09}{{a.u.}}, \SI{8.04E+12}{{a.u.}}, \SI{7.49E+12}{{a.u.}}, \SI{4.68E+20}{{a.u.}}, \SI{4.63E+20}{{a.u.}}, \SI{4.41E+20}{{a.u.}}, respectively. 
The probabilities of nuclear transitions $B_{M \to G}(M1)$ = \SI{0.022}{{W.u.}} and $B_{M \to G}(E2)$ = \SI{29}{{W.u.}} are taken from Ref.~\cite{Tiedau2024} and from Ref.~\cite{Minkov2019}, respectively. 
Notably, the M1 excitation accounts for 55\% of the NEET rate, while the E2 excitation contributes 45\%. 
In Table~\ref{tab:neet}, the $|\Delta|$ = 0 value corresponds to the NEET resonance, where the electronic transition energy $E_{if}$ exactly matches the nuclear transition energy $E_{GM}$. At resonance, the excitation rate of the \ce{^{229m}Th} isomer reaches a remarkable value of \SI{2.50e16}{\per\s}. For non-zero $\Delta$ values, the NEET probability in \ce{Th^{39+}} ions decreases by many orders of magnitude, as illustrated in Fig.~\ref{fig:w_neet}. 
The other $|\Delta|$ values are chosen based on $E_{GM}$ values from Ref.~\cite{Zhang2024} (the most recent experiment where \ce{^{229m}Th} ions were incorporated into \ce{CaF2} crystal). The $|\Delta|$ = \SI{48}{meV} case refers to when $E_{if}$ and $E_{GM}$ from Ref.~\cite{Zhang2024} are in the middle of their uncertainty intervals, with $E_{if}$ = \SI{8.308}{eV} and $E_{GM}$ = \SI{8.356}{eV}. For this $\Delta$ value, the NEET probability decreases to \SI{2.07e-4}{\per\s}. 
The $|\Delta|$ = \SI{21}{meV} and $|\Delta|$ = \SI{117}{meV} cases refer to the minimum and the maximum distances between $E_{if}$ and $E_{GM}$ according to the electronic transition uncertainty, with $E_{if}$ = \SI[parse-numbers = false]{(8.308+0.069)}{eV} and \SI[parse-numbers = false]{(8.308-0.069)}{eV}, respectively. 
For these $\Delta$ values, the NEET probabilities are \SI{1.04e-3}{\per\s} and \SI{3.47e-5}{\per\s}, respectively. 

\section{Proposed EBIT experiment}

In order to rigorously verify the theory of the NEET process, in the next step we examine the possibility of conducting an appropriate experiment in an EBIT device. To estimate the isomeric excitation rate in \ce{Th^{39+}} ions achievable in an EBIT, we assumed experimental conditions similar to those described in Ref.~\cite{Bilous2020}. Specifically, we considered an ion cloud with a diameter of \SI{100}{\micro\meter} containing approximately $N_\text{HCI} \simeq 10^6$ \ce{Th^{39+}} ions. Such a large number of exotic HCIs can be produced in an EBIT using a highly efficient technique based on in-trap laser-induced desorption of atoms from a sample positioned near the electron beam \cite{Schweiger2019}. 
Our collisional-radiative model simulations, conducted using the \fac code \cite{Gu2008}, indicate that the relative steady-state population of the first atomic excited $J=7/2$ level in \ce{Th^{39+}} ions is about $p_i \simeq 2.12\%$ at an electron energy of \SI{1.7}{keV} in the EBIT. Details of these simulations are provided in the Supplementary Material. The NEET rate achievable in the EBIT device is then calculated as $\tilde{W}_\text{NEET} = W_\text{NEET} \cdot p_i \cdot N_\text{HCI}$. 
Possible values of $\tilde{W}_\text{NEET}$ as a function of the energy difference $|\Delta| = |E_{if} - E_{GM}|$ are presented in Table~\ref{tab:neet} and Fig.~\ref{fig:w_neet} (refer to the scale on the right side of the figure). At resonance ($|\Delta|$ = 0), the upper limit of the NEET isomeric excitation rate in the \ce{^{229}Th} nucleus under EBIT conditions reaches an enormous value of \SI{5.29e20}{\per\s}, which is several orders of magnitude higher than the maximum rate (just over \SI{e5}{\per\s}) achievable using the EBE mechanism in \ce{Th^{35+}} ions, as described in Ref.~\cite{Bilous2020}. 
Within the energy uncertainty range of the nuclear and atomic transitions, the NEET rate under EBIT conditions decreases to \SI{4.40}{\per\s} for $|\Delta|$ = \SI{48}{meV}. 
Such huge differences in rates allow for an extremely rigorous test of the theory describing the NEET process under well-defined atomic conditions in the range of the lowest known nuclear excitation energies. 
Moreover, our proposed experimental scenario does not require the use of a tunable UV laser to initiate atomic and nuclear processes, which significantly simplifies the experimental setup. 

It should also be mentioned that in the proposed experiment it is not possible to completely exclude the contribution of other NEET channels in \ce{Th^{39+}} ion, which may be a source of additional systematic error. However, we assume that nearly the entire steady-state population of the \ce{Th^{39+}} ion is shared between the ground and few low-lying metastable states, according to the statement in Ref.~\cite{Bilous2020}. Therefore, we assume that NEET channels related to transitions from very highly excited states are negligible. Our analysis indicates that all states above the 20th excited state have populations below 1\% and state populations decrease drastically for highly excited states. We did not find any possible transition in \SIrange[range-units=single,range-phrase=--]{8.0}{8.6}{eV} energy range within the first 20 excited states, which can compete to the one considered in the manuscript. 

\section{Summary}

In summary, the NEET excitation of the \ce{^{229m}Th} isomer has been theoretically investigated for highly charged thorium ions. It was found that the atomic transition energy from the first excited state to the ground state in \ce{^{229}Th^{39+}} ions (\SI{8.308+-0.069}{eV}) closely matches the newly established experimental reference value for the nuclear isomer excitation energy (\SI{8.356}{eV}). The study demonstrated that NEET can produce the isomeric state in the \ce{^{229}Th} nucleus at rates reaching up to \SI{e16}{\per\s}, but dropping many orders of magnitude for the value of electronic transition energy within its uncertainty interval. 
General calculations for the experimental scenario in the EBIT facility were also performed. Within the range of energy uncertainties of nuclear and atomic transitions, NEET rates for EBIT conditions range from \SI{0.7}{\per\s} to \SI{5.3e20}{\per\s} at resonance. Thus, the NEET process in \ce{^{229}Th^{39+}} ions could be a unique opportunity to efficiently produce the \ce{^{229m}Th} isomer in the EBIT facility and thus to study the NEET process under well-defined conditions. 

The proposed experiment is expected to be an important test of the NEET theory under exceptional conditions associated with extremely low nuclear excitation energy. Such a test will be based on a comparison of experimental and theoretical rates, which strongly depend on the difference in the energies of nuclear and atomic transitions. Since the excitation energy of the \ce{^{229m}Th} isomer was determined with extremely high accuracy, the uncertainty associated with the difference in the energies of nuclear and atomic transitions is limited only to the uncertainty of the atomic transition ($J=7/2 \to J=5/2$), which according to our calculations does not exceed \SI{70}{meV}. Despite the relatively small uncertainty of the atomic transition, the estimated NEET rates range by orders of magnitude. In the extreme case, the resonance conditions ($\Delta=0$~meV) can be fully satisfied, which would lead to incredibly high NEET rates in \ce{^{229}Th} (\SI{5.3e20}{\per\s}). On the one hand, this makes testing the NEET theory somewhat difficult, but on the other hand, the potential possibility of achieving NEET resonance conditions makes the proposed experiment unprecedentedly attractive.

%\bibliographystyle{apsrev4-2}
%\bibliography{references}

\begin{thebibliography}{54}%
\makeatletter
\providecommand \@ifxundefined [1]{%
 \@ifx{#1\undefined}
}%
\providecommand \@ifnum [1]{%
 \ifnum #1\expandafter \@firstoftwo
 \else \expandafter \@secondoftwo
 \fi
}%
\providecommand \@ifx [1]{%
 \ifx #1\expandafter \@firstoftwo
 \else \expandafter \@secondoftwo
 \fi
}%
\providecommand \natexlab [1]{#1}%
\providecommand \enquote  [1]{``#1''}%
\providecommand \bibnamefont  [1]{#1}%
\providecommand \bibfnamefont [1]{#1}%
\providecommand \citenamefont [1]{#1}%
\providecommand \href@noop [0]{\@secondoftwo}%
\providecommand \href [0]{\begingroup \@sanitize@url \@href}%
\providecommand \@href[1]{\@@startlink{#1}\@@href}%
\providecommand \@@href[1]{\endgroup#1\@@endlink}%
\providecommand \@sanitize@url [0]{\catcode `\\12\catcode `\$12\catcode
  `\&12\catcode `\#12\catcode `\^12\catcode `\_12\catcode `\%12\relax}%
\providecommand \@@startlink[1]{}%
\providecommand \@@endlink[0]{}%
\providecommand \url  [0]{\begingroup\@sanitize@url \@url }%
\providecommand \@url [1]{\endgroup\@href {#1}{\urlprefix }}%
\providecommand \urlprefix  [0]{URL }%
\providecommand \Eprint [0]{\href }%
\providecommand \doibase [0]{https://doi.org/}%
\providecommand \selectlanguage [0]{\@gobble}%
\providecommand \bibinfo  [0]{\@secondoftwo}%
\providecommand \bibfield  [0]{\@secondoftwo}%
\providecommand \translation [1]{[#1]}%
\providecommand \BibitemOpen [0]{}%
\providecommand \bibitemStop [0]{}%
\providecommand \bibitemNoStop [0]{.\EOS\space}%
\providecommand \EOS [0]{\spacefactor3000\relax}%
\providecommand \BibitemShut  [1]{\csname bibitem#1\endcsname}%
\let\auto@bib@innerbib\@empty
%</preamble>
\bibitem [{\citenamefont {Campbell}\ \emph {et~al.}(2012)\citenamefont
  {Campbell}, \citenamefont {Radnaev}, \citenamefont {Kuzmich}, \citenamefont
  {Dzuba}, \citenamefont {Flambaum},\ and\ \citenamefont
  {Derevianko}}]{Campbell2012}%
  \BibitemOpen
  \bibfield  {author} {\bibinfo {author} {\bibfnamefont {C.~J.}\ \bibnamefont
  {Campbell}}, \bibinfo {author} {\bibfnamefont {A.~G.}\ \bibnamefont
  {Radnaev}}, \bibinfo {author} {\bibfnamefont {A.}~\bibnamefont {Kuzmich}},
  \bibinfo {author} {\bibfnamefont {V.~A.}\ \bibnamefont {Dzuba}}, \bibinfo
  {author} {\bibfnamefont {V.~V.}\ \bibnamefont {Flambaum}},\ and\ \bibinfo
  {author} {\bibfnamefont {A.}~\bibnamefont {Derevianko}},\ }\href
  {https://doi.org/10.1103/PhysRevLett.108.120802} {\bibfield  {journal}
  {\bibinfo  {journal} {Physical Review Letters}\ }\textbf {\bibinfo {volume}
  {108}},\ \bibinfo {pages} {120802} (\bibinfo {year} {2012})}\BibitemShut
  {NoStop}%
\bibitem [{\citenamefont {Peik}\ \emph {et~al.}(2021)\citenamefont {Peik},
  \citenamefont {Schumm}, \citenamefont {Safronova}, \citenamefont
  {P{\'{a}}lffy}, \citenamefont {Weitenberg},\ and\ \citenamefont
  {Thirolf}}]{Peik2021}%
  \BibitemOpen
  \bibfield  {author} {\bibinfo {author} {\bibfnamefont {E.}~\bibnamefont
  {Peik}}, \bibinfo {author} {\bibfnamefont {T.}~\bibnamefont {Schumm}},
  \bibinfo {author} {\bibfnamefont {M.~S.}\ \bibnamefont {Safronova}}, \bibinfo
  {author} {\bibfnamefont {A.}~\bibnamefont {P{\'{a}}lffy}}, \bibinfo {author}
  {\bibfnamefont {J.}~\bibnamefont {Weitenberg}},\ and\ \bibinfo {author}
  {\bibfnamefont {P.~G.}\ \bibnamefont {Thirolf}},\ }\href
  {https://doi.org/10.1088/2058-9565/abe9c2} {\bibfield  {journal} {\bibinfo
  {journal} {Quantum Science and Technology}\ }\textbf {\bibinfo {volume}
  {6}},\ \bibinfo {pages} {034002} (\bibinfo {year} {2021})},\ \Eprint
  {https://arxiv.org/abs/2012.09304} {arXiv:2012.09304} \BibitemShut {NoStop}%
\bibitem [{\citenamefont {Beeks}\ \emph {et~al.}(2021)\citenamefont {Beeks},
  \citenamefont {Sikorsky}, \citenamefont {Schumm}, \citenamefont {Thielking},
  \citenamefont {Okhapkin},\ and\ \citenamefont {Peik}}]{Beeks2021}%
  \BibitemOpen
  \bibfield  {author} {\bibinfo {author} {\bibfnamefont {K.}~\bibnamefont
  {Beeks}}, \bibinfo {author} {\bibfnamefont {T.}~\bibnamefont {Sikorsky}},
  \bibinfo {author} {\bibfnamefont {T.}~\bibnamefont {Schumm}}, \bibinfo
  {author} {\bibfnamefont {J.}~\bibnamefont {Thielking}}, \bibinfo {author}
  {\bibfnamefont {M.~V.}\ \bibnamefont {Okhapkin}},\ and\ \bibinfo {author}
  {\bibfnamefont {E.}~\bibnamefont {Peik}},\ }\href
  {https://doi.org/10.1038/s42254-021-00286-6} {\bibfield  {journal} {\bibinfo
  {journal} {Nature Reviews Physics}\ }\textbf {\bibinfo {volume} {3}},\
  \bibinfo {pages} {238} (\bibinfo {year} {2021})}\BibitemShut {NoStop}%
\bibitem [{\citenamefont {Flambaum}(2006)}]{Flambaum2006}%
  \BibitemOpen
  \bibfield  {author} {\bibinfo {author} {\bibfnamefont {V.~V.}\ \bibnamefont
  {Flambaum}},\ }\href {https://doi.org/10.1103/PhysRevLett.97.092502}
  {\bibfield  {journal} {\bibinfo  {journal} {Physical Review Letters}\
  }\textbf {\bibinfo {volume} {97}},\ \bibinfo {pages} {092502} (\bibinfo
  {year} {2006})}\BibitemShut {NoStop}%
\bibitem [{\citenamefont {Berengut}\ \emph {et~al.}(2009)\citenamefont
  {Berengut}, \citenamefont {Dzuba}, \citenamefont {Flambaum},\ and\
  \citenamefont {Porsev}}]{Berengut2009}%
  \BibitemOpen
  \bibfield  {author} {\bibinfo {author} {\bibfnamefont {J.~C.}\ \bibnamefont
  {Berengut}}, \bibinfo {author} {\bibfnamefont {V.~A.}\ \bibnamefont {Dzuba}},
  \bibinfo {author} {\bibfnamefont {V.~V.}\ \bibnamefont {Flambaum}},\ and\
  \bibinfo {author} {\bibfnamefont {S.~G.}\ \bibnamefont {Porsev}},\ }\href
  {https://doi.org/10.1103/PhysRevLett.102.210801} {\bibfield  {journal}
  {\bibinfo  {journal} {Physical Review Letters}\ }\textbf {\bibinfo {volume}
  {102}},\ \bibinfo {pages} {210801} (\bibinfo {year} {2009})}\BibitemShut
  {NoStop}%
\bibitem [{\citenamefont {Thirolf}\ \emph {et~al.}(2019)\citenamefont
  {Thirolf}, \citenamefont {Seiferle},\ and\ \citenamefont {von~der
  Wense}}]{Thirolf2019}%
  \BibitemOpen
  \bibfield  {author} {\bibinfo {author} {\bibfnamefont {P.~G.}\ \bibnamefont
  {Thirolf}}, \bibinfo {author} {\bibfnamefont {B.}~\bibnamefont {Seiferle}},\
  and\ \bibinfo {author} {\bibfnamefont {L.}~\bibnamefont {von~der Wense}},\
  }\href {https://doi.org/10.1002/andp.201800381} {\bibfield  {journal}
  {\bibinfo  {journal} {Annalen der Physik}\ }\textbf {\bibinfo {volume}
  {531}},\ \bibinfo {pages} {1800381} (\bibinfo {year} {2019})}\BibitemShut
  {NoStop}%
\bibitem [{\citenamefont {Tiedau}\ \emph {et~al.}(2024)\citenamefont {Tiedau},
  \citenamefont {Okhapkin}, \citenamefont {Zhang}, \citenamefont {Thielking},
  \citenamefont {Zitzer}, \citenamefont {Peik}, \citenamefont {Schaden},
  \citenamefont {Pronebner}, \citenamefont {Morawetz}, \citenamefont {{De
  Col}}, \citenamefont {Schneider}, \citenamefont {Leitner}, \citenamefont
  {Pressler}, \citenamefont {Kazakov}, \citenamefont {Beeks}, \citenamefont
  {Sikorsky},\ and\ \citenamefont {Schumm}}]{Tiedau2024}%
  \BibitemOpen
  \bibfield  {author} {\bibinfo {author} {\bibfnamefont {J.}~\bibnamefont
  {Tiedau}}, \bibinfo {author} {\bibfnamefont {M.~V.}\ \bibnamefont
  {Okhapkin}}, \bibinfo {author} {\bibfnamefont {K.}~\bibnamefont {Zhang}},
  \bibinfo {author} {\bibfnamefont {J.}~\bibnamefont {Thielking}}, \bibinfo
  {author} {\bibfnamefont {G.}~\bibnamefont {Zitzer}}, \bibinfo {author}
  {\bibfnamefont {E.}~\bibnamefont {Peik}}, \bibinfo {author} {\bibfnamefont
  {F.}~\bibnamefont {Schaden}}, \bibinfo {author} {\bibfnamefont
  {T.}~\bibnamefont {Pronebner}}, \bibinfo {author} {\bibfnamefont
  {I.}~\bibnamefont {Morawetz}}, \bibinfo {author} {\bibfnamefont {L.~T.}\
  \bibnamefont {{De Col}}}, \bibinfo {author} {\bibfnamefont {F.}~\bibnamefont
  {Schneider}}, \bibinfo {author} {\bibfnamefont {A.}~\bibnamefont {Leitner}},
  \bibinfo {author} {\bibfnamefont {M.}~\bibnamefont {Pressler}}, \bibinfo
  {author} {\bibfnamefont {G.~A.}\ \bibnamefont {Kazakov}}, \bibinfo {author}
  {\bibfnamefont {K.}~\bibnamefont {Beeks}}, \bibinfo {author} {\bibfnamefont
  {T.}~\bibnamefont {Sikorsky}},\ and\ \bibinfo {author} {\bibfnamefont
  {T.}~\bibnamefont {Schumm}},\ }\href
  {https://doi.org/10.1103/PhysRevLett.132.182501} {\bibfield  {journal}
  {\bibinfo  {journal} {Physical Review Letters}\ }\textbf {\bibinfo {volume}
  {132}},\ \bibinfo {pages} {182501} (\bibinfo {year} {2024})}\BibitemShut
  {NoStop}%
\bibitem [{\citenamefont {Seiferle}\ \emph {et~al.}(2019)\citenamefont
  {Seiferle}, \citenamefont {von~der Wense}, \citenamefont {Bilous},
  \citenamefont {Amersdorffer}, \citenamefont {Lemell}, \citenamefont
  {Libisch}, \citenamefont {Stellmer}, \citenamefont {Schumm}, \citenamefont
  {D{\"{u}}llmann}, \citenamefont {P{\'{a}}lffy},\ and\ \citenamefont
  {Thirolf}}]{Seiferle2019}%
  \BibitemOpen
  \bibfield  {author} {\bibinfo {author} {\bibfnamefont {B.}~\bibnamefont
  {Seiferle}}, \bibinfo {author} {\bibfnamefont {L.}~\bibnamefont {von~der
  Wense}}, \bibinfo {author} {\bibfnamefont {P.~V.}\ \bibnamefont {Bilous}},
  \bibinfo {author} {\bibfnamefont {I.}~\bibnamefont {Amersdorffer}}, \bibinfo
  {author} {\bibfnamefont {C.}~\bibnamefont {Lemell}}, \bibinfo {author}
  {\bibfnamefont {F.}~\bibnamefont {Libisch}}, \bibinfo {author} {\bibfnamefont
  {S.}~\bibnamefont {Stellmer}}, \bibinfo {author} {\bibfnamefont
  {T.}~\bibnamefont {Schumm}}, \bibinfo {author} {\bibfnamefont {C.~E.}\
  \bibnamefont {D{\"{u}}llmann}}, \bibinfo {author} {\bibfnamefont
  {A.}~\bibnamefont {P{\'{a}}lffy}},\ and\ \bibinfo {author} {\bibfnamefont
  {P.~G.}\ \bibnamefont {Thirolf}},\ }\href
  {https://doi.org/10.1038/s41586-019-1533-4} {\bibfield  {journal} {\bibinfo
  {journal} {Nature}\ }\textbf {\bibinfo {volume} {573}},\ \bibinfo {pages}
  {243} (\bibinfo {year} {2019})},\ \Eprint {https://arxiv.org/abs/1905.06308}
  {arXiv:1905.06308} \BibitemShut {NoStop}%
\bibitem [{\citenamefont {Yamaguchi}\ \emph {et~al.}(2019)\citenamefont
  {Yamaguchi}, \citenamefont {Muramatsu}, \citenamefont {Hayashi},
  \citenamefont {Yuasa}, \citenamefont {Nakamura}, \citenamefont {Takimoto},
  \citenamefont {Haba}, \citenamefont {Konashi}, \citenamefont {Watanabe},
  \citenamefont {Kikunaga}, \citenamefont {Maehata}, \citenamefont {Yamasaki},\
  and\ \citenamefont {Mitsuda}}]{Yamaguchi2019}%
  \BibitemOpen
  \bibfield  {author} {\bibinfo {author} {\bibfnamefont {A.}~\bibnamefont
  {Yamaguchi}}, \bibinfo {author} {\bibfnamefont {H.}~\bibnamefont
  {Muramatsu}}, \bibinfo {author} {\bibfnamefont {T.}~\bibnamefont {Hayashi}},
  \bibinfo {author} {\bibfnamefont {N.}~\bibnamefont {Yuasa}}, \bibinfo
  {author} {\bibfnamefont {K.}~\bibnamefont {Nakamura}}, \bibinfo {author}
  {\bibfnamefont {M.}~\bibnamefont {Takimoto}}, \bibinfo {author}
  {\bibfnamefont {H.}~\bibnamefont {Haba}}, \bibinfo {author} {\bibfnamefont
  {K.}~\bibnamefont {Konashi}}, \bibinfo {author} {\bibfnamefont
  {M.}~\bibnamefont {Watanabe}}, \bibinfo {author} {\bibfnamefont
  {H.}~\bibnamefont {Kikunaga}}, \bibinfo {author} {\bibfnamefont
  {K.}~\bibnamefont {Maehata}}, \bibinfo {author} {\bibfnamefont {N.~Y.}\
  \bibnamefont {Yamasaki}},\ and\ \bibinfo {author} {\bibfnamefont
  {K.}~\bibnamefont {Mitsuda}},\ }\href
  {https://doi.org/10.1103/PhysRevLett.123.222501} {\bibfield  {journal}
  {\bibinfo  {journal} {Physical Review Letters}\ }\textbf {\bibinfo {volume}
  {123}},\ \bibinfo {pages} {222501} (\bibinfo {year} {2019})}\BibitemShut
  {NoStop}%
\bibitem [{\citenamefont {Sikorsky}\ \emph {et~al.}(2020)\citenamefont
  {Sikorsky}, \citenamefont {Geist}, \citenamefont {Hengstler}, \citenamefont
  {Kempf}, \citenamefont {Gastaldo}, \citenamefont {Enss}, \citenamefont
  {Mokry}, \citenamefont {Runke}, \citenamefont {D{\"{u}}llmann}, \citenamefont
  {Wobrauschek}, \citenamefont {Beeks}, \citenamefont {Rosecker}, \citenamefont
  {Sterba}, \citenamefont {Kazakov}, \citenamefont {Schumm},\ and\
  \citenamefont {Fleischmann}}]{Sikorsky2020}%
  \BibitemOpen
  \bibfield  {author} {\bibinfo {author} {\bibfnamefont {T.}~\bibnamefont
  {Sikorsky}}, \bibinfo {author} {\bibfnamefont {J.}~\bibnamefont {Geist}},
  \bibinfo {author} {\bibfnamefont {D.}~\bibnamefont {Hengstler}}, \bibinfo
  {author} {\bibfnamefont {S.}~\bibnamefont {Kempf}}, \bibinfo {author}
  {\bibfnamefont {L.}~\bibnamefont {Gastaldo}}, \bibinfo {author}
  {\bibfnamefont {C.}~\bibnamefont {Enss}}, \bibinfo {author} {\bibfnamefont
  {C.}~\bibnamefont {Mokry}}, \bibinfo {author} {\bibfnamefont
  {J.}~\bibnamefont {Runke}}, \bibinfo {author} {\bibfnamefont {C.~E.}\
  \bibnamefont {D{\"{u}}llmann}}, \bibinfo {author} {\bibfnamefont
  {P.}~\bibnamefont {Wobrauschek}}, \bibinfo {author} {\bibfnamefont
  {K.}~\bibnamefont {Beeks}}, \bibinfo {author} {\bibfnamefont
  {V.}~\bibnamefont {Rosecker}}, \bibinfo {author} {\bibfnamefont {J.~H.}\
  \bibnamefont {Sterba}}, \bibinfo {author} {\bibfnamefont {G.}~\bibnamefont
  {Kazakov}}, \bibinfo {author} {\bibfnamefont {T.}~\bibnamefont {Schumm}},\
  and\ \bibinfo {author} {\bibfnamefont {A.}~\bibnamefont {Fleischmann}},\
  }\href {https://doi.org/10.1103/PhysRevLett.125.142503} {\bibfield  {journal}
  {\bibinfo  {journal} {Physical Review Letters}\ }\textbf {\bibinfo {volume}
  {125}},\ \bibinfo {pages} {142503} (\bibinfo {year} {2020})}\BibitemShut
  {NoStop}%
\bibitem [{\citenamefont {Kraemer}\ \emph {et~al.}(2023)\citenamefont
  {Kraemer}, \citenamefont {Moens}, \citenamefont {Athanasakis-Kaklamanakis},
  \citenamefont {Bara}, \citenamefont {Beeks}, \citenamefont {Chhetri},
  \citenamefont {Chrysalidis}, \citenamefont {Claessens}, \citenamefont
  {Cocolios}, \citenamefont {Correia}, \citenamefont {Witte}, \citenamefont
  {Ferrer}, \citenamefont {Geldhof}, \citenamefont {Heinke}, \citenamefont
  {Hosseini}, \citenamefont {Huyse}, \citenamefont {K{\"{o}}ster},
  \citenamefont {Kudryavtsev}, \citenamefont {Laatiaoui}, \citenamefont {Lica},
  \citenamefont {Magchiels}, \citenamefont {Manea}, \citenamefont {Merckling},
  \citenamefont {Pereira}, \citenamefont {Raeder}, \citenamefont {Schumm},
  \citenamefont {Sels}, \citenamefont {Thirolf}, \citenamefont {Tunhuma},
  \citenamefont {{Van Den Bergh}}, \citenamefont {{Van Duppen}}, \citenamefont
  {Vantomme}, \citenamefont {Verlinde}, \citenamefont {Villarreal},\ and\
  \citenamefont {Wahl}}]{Kraemer2023}%
  \BibitemOpen
  \bibfield  {author} {\bibinfo {author} {\bibfnamefont {S.}~\bibnamefont
  {Kraemer}}, \bibinfo {author} {\bibfnamefont {J.}~\bibnamefont {Moens}},
  \bibinfo {author} {\bibfnamefont {M.}~\bibnamefont
  {Athanasakis-Kaklamanakis}}, \bibinfo {author} {\bibfnamefont
  {S.}~\bibnamefont {Bara}}, \bibinfo {author} {\bibfnamefont {K.}~\bibnamefont
  {Beeks}}, \bibinfo {author} {\bibfnamefont {P.}~\bibnamefont {Chhetri}},
  \bibinfo {author} {\bibfnamefont {K.}~\bibnamefont {Chrysalidis}}, \bibinfo
  {author} {\bibfnamefont {A.}~\bibnamefont {Claessens}}, \bibinfo {author}
  {\bibfnamefont {T.~E.}\ \bibnamefont {Cocolios}}, \bibinfo {author}
  {\bibfnamefont {J.~G.~M.}\ \bibnamefont {Correia}}, \bibinfo {author}
  {\bibfnamefont {H.~D.}\ \bibnamefont {Witte}}, \bibinfo {author}
  {\bibfnamefont {R.}~\bibnamefont {Ferrer}}, \bibinfo {author} {\bibfnamefont
  {S.}~\bibnamefont {Geldhof}}, \bibinfo {author} {\bibfnamefont
  {R.}~\bibnamefont {Heinke}}, \bibinfo {author} {\bibfnamefont
  {N.}~\bibnamefont {Hosseini}}, \bibinfo {author} {\bibfnamefont
  {M.}~\bibnamefont {Huyse}}, \bibinfo {author} {\bibfnamefont
  {U.}~\bibnamefont {K{\"{o}}ster}}, \bibinfo {author} {\bibfnamefont
  {Y.}~\bibnamefont {Kudryavtsev}}, \bibinfo {author} {\bibfnamefont
  {M.}~\bibnamefont {Laatiaoui}}, \bibinfo {author} {\bibfnamefont
  {R.}~\bibnamefont {Lica}}, \bibinfo {author} {\bibfnamefont {G.}~\bibnamefont
  {Magchiels}}, \bibinfo {author} {\bibfnamefont {V.}~\bibnamefont {Manea}},
  \bibinfo {author} {\bibfnamefont {C.}~\bibnamefont {Merckling}}, \bibinfo
  {author} {\bibfnamefont {L.~M.~C.}\ \bibnamefont {Pereira}}, \bibinfo
  {author} {\bibfnamefont {S.}~\bibnamefont {Raeder}}, \bibinfo {author}
  {\bibfnamefont {T.}~\bibnamefont {Schumm}}, \bibinfo {author} {\bibfnamefont
  {S.}~\bibnamefont {Sels}}, \bibinfo {author} {\bibfnamefont {P.~G.}\
  \bibnamefont {Thirolf}}, \bibinfo {author} {\bibfnamefont {S.~M.}\
  \bibnamefont {Tunhuma}}, \bibinfo {author} {\bibfnamefont {P.}~\bibnamefont
  {{Van Den Bergh}}}, \bibinfo {author} {\bibfnamefont {P.}~\bibnamefont {{Van
  Duppen}}}, \bibinfo {author} {\bibfnamefont {A.}~\bibnamefont {Vantomme}},
  \bibinfo {author} {\bibfnamefont {M.}~\bibnamefont {Verlinde}}, \bibinfo
  {author} {\bibfnamefont {R.}~\bibnamefont {Villarreal}},\ and\ \bibinfo
  {author} {\bibfnamefont {U.}~\bibnamefont {Wahl}},\ }\href
  {https://doi.org/10.1038/s41586-023-05894-z} {\bibfield  {journal} {\bibinfo
  {journal} {Nature}\ }\textbf {\bibinfo {volume} {617}},\ \bibinfo {pages}
  {706} (\bibinfo {year} {2023})},\ \Eprint {https://arxiv.org/abs/2209.10276}
  {arXiv:2209.10276} \BibitemShut {NoStop}%
\bibitem [{\citenamefont {Zhang}\ \emph {et~al.}(2024)\citenamefont {Zhang},
  \citenamefont {Ooi}, \citenamefont {Higgins}, \citenamefont {Doyle},
  \citenamefont {von~der Wense}, \citenamefont {Beeks}, \citenamefont
  {Leitner}, \citenamefont {Kazakov}, \citenamefont {Li}, \citenamefont
  {Thirolf}, \citenamefont {Schumm},\ and\ \citenamefont {Ye}}]{Zhang2024}%
  \BibitemOpen
  \bibfield  {author} {\bibinfo {author} {\bibfnamefont {C.}~\bibnamefont
  {Zhang}}, \bibinfo {author} {\bibfnamefont {T.}~\bibnamefont {Ooi}}, \bibinfo
  {author} {\bibfnamefont {J.~S.}\ \bibnamefont {Higgins}}, \bibinfo {author}
  {\bibfnamefont {J.~F.}\ \bibnamefont {Doyle}}, \bibinfo {author}
  {\bibfnamefont {L.}~\bibnamefont {von~der Wense}}, \bibinfo {author}
  {\bibfnamefont {K.}~\bibnamefont {Beeks}}, \bibinfo {author} {\bibfnamefont
  {A.}~\bibnamefont {Leitner}}, \bibinfo {author} {\bibfnamefont
  {G.}~\bibnamefont {Kazakov}}, \bibinfo {author} {\bibfnamefont
  {P.}~\bibnamefont {Li}}, \bibinfo {author} {\bibfnamefont {P.~G.}\
  \bibnamefont {Thirolf}}, \bibinfo {author} {\bibfnamefont {T.}~\bibnamefont
  {Schumm}},\ and\ \bibinfo {author} {\bibfnamefont {J.}~\bibnamefont {Ye}},\
  }\href {https://doi.org/10.1038/s41586-024-07839-6} {\bibfield  {journal}
  {\bibinfo  {journal} {Nature}\ }\textbf {\bibinfo {volume} {633}},\ \bibinfo
  {pages} {63} (\bibinfo {year} {2024})},\ \Eprint
  {https://arxiv.org/abs/2406.18719} {arXiv:2406.18719} \BibitemShut {NoStop}%
\bibitem [{\citenamefont {Masuda}\ \emph {et~al.}(2019)\citenamefont {Masuda},
  \citenamefont {Yoshimi}, \citenamefont {Fujieda}, \citenamefont {Fujimoto},
  \citenamefont {Haba}, \citenamefont {Hara}, \citenamefont {Hiraki},
  \citenamefont {Kaino}, \citenamefont {Kasamatsu}, \citenamefont {Kitao},
  \citenamefont {Konashi}, \citenamefont {Miyamoto}, \citenamefont {Okai},
  \citenamefont {Okubo}, \citenamefont {Sasao}, \citenamefont {Seto},
  \citenamefont {Schumm}, \citenamefont {Shigekawa}, \citenamefont {Suzuki},
  \citenamefont {Stellmer}, \citenamefont {Tamasaku}, \citenamefont {Uetake},
  \citenamefont {Watanabe}, \citenamefont {Watanabe}, \citenamefont {Yasuda},
  \citenamefont {Yamaguchi}, \citenamefont {Yoda}, \citenamefont {Yokokita},
  \citenamefont {Yoshimura},\ and\ \citenamefont {Yoshimura}}]{Masuda2019}%
  \BibitemOpen
  \bibfield  {author} {\bibinfo {author} {\bibfnamefont {T.}~\bibnamefont
  {Masuda}}, \bibinfo {author} {\bibfnamefont {A.}~\bibnamefont {Yoshimi}},
  \bibinfo {author} {\bibfnamefont {A.}~\bibnamefont {Fujieda}}, \bibinfo
  {author} {\bibfnamefont {H.}~\bibnamefont {Fujimoto}}, \bibinfo {author}
  {\bibfnamefont {H.}~\bibnamefont {Haba}}, \bibinfo {author} {\bibfnamefont
  {H.}~\bibnamefont {Hara}}, \bibinfo {author} {\bibfnamefont {T.}~\bibnamefont
  {Hiraki}}, \bibinfo {author} {\bibfnamefont {H.}~\bibnamefont {Kaino}},
  \bibinfo {author} {\bibfnamefont {Y.}~\bibnamefont {Kasamatsu}}, \bibinfo
  {author} {\bibfnamefont {S.}~\bibnamefont {Kitao}}, \bibinfo {author}
  {\bibfnamefont {K.}~\bibnamefont {Konashi}}, \bibinfo {author} {\bibfnamefont
  {Y.}~\bibnamefont {Miyamoto}}, \bibinfo {author} {\bibfnamefont
  {K.}~\bibnamefont {Okai}}, \bibinfo {author} {\bibfnamefont {S.}~\bibnamefont
  {Okubo}}, \bibinfo {author} {\bibfnamefont {N.}~\bibnamefont {Sasao}},
  \bibinfo {author} {\bibfnamefont {M.}~\bibnamefont {Seto}}, \bibinfo {author}
  {\bibfnamefont {T.}~\bibnamefont {Schumm}}, \bibinfo {author} {\bibfnamefont
  {Y.}~\bibnamefont {Shigekawa}}, \bibinfo {author} {\bibfnamefont
  {K.}~\bibnamefont {Suzuki}}, \bibinfo {author} {\bibfnamefont
  {S.}~\bibnamefont {Stellmer}}, \bibinfo {author} {\bibfnamefont
  {K.}~\bibnamefont {Tamasaku}}, \bibinfo {author} {\bibfnamefont
  {S.}~\bibnamefont {Uetake}}, \bibinfo {author} {\bibfnamefont
  {M.}~\bibnamefont {Watanabe}}, \bibinfo {author} {\bibfnamefont
  {T.}~\bibnamefont {Watanabe}}, \bibinfo {author} {\bibfnamefont
  {Y.}~\bibnamefont {Yasuda}}, \bibinfo {author} {\bibfnamefont
  {A.}~\bibnamefont {Yamaguchi}}, \bibinfo {author} {\bibfnamefont
  {Y.}~\bibnamefont {Yoda}}, \bibinfo {author} {\bibfnamefont {T.}~\bibnamefont
  {Yokokita}}, \bibinfo {author} {\bibfnamefont {M.}~\bibnamefont
  {Yoshimura}},\ and\ \bibinfo {author} {\bibfnamefont {K.}~\bibnamefont
  {Yoshimura}},\ }\href {https://doi.org/10.1038/s41586-019-1542-3} {\bibfield
  {journal} {\bibinfo  {journal} {Nature}\ }\textbf {\bibinfo {volume} {573}},\
  \bibinfo {pages} {238} (\bibinfo {year} {2019})}\BibitemShut {NoStop}%
\bibitem [{\citenamefont {Zhao}\ \emph {et~al.}(2024)\citenamefont {Zhao},
  \citenamefont {P{\'{a}}lffy}, \citenamefont {Keitel},\ and\ \citenamefont
  {Wu}}]{Zhao2024}%
  \BibitemOpen
  \bibfield  {author} {\bibinfo {author} {\bibfnamefont {J.}~\bibnamefont
  {Zhao}}, \bibinfo {author} {\bibfnamefont {A.}~\bibnamefont {P{\'{a}}lffy}},
  \bibinfo {author} {\bibfnamefont {C.~H.}\ \bibnamefont {Keitel}},\ and\
  \bibinfo {author} {\bibfnamefont {Y.}~\bibnamefont {Wu}},\ }\href
  {https://doi.org/10.1103/PhysRevC.110.014330} {\bibfield  {journal} {\bibinfo
   {journal} {Physical Review C}\ }\textbf {\bibinfo {volume} {110}},\ \bibinfo
  {pages} {014330} (\bibinfo {year} {2024})}\BibitemShut {NoStop}%
\bibitem [{\citenamefont {Tkalya}(2020)}]{Tkalya2020}%
  \BibitemOpen
  \bibfield  {author} {\bibinfo {author} {\bibfnamefont {E.~V.}\ \bibnamefont
  {Tkalya}},\ }\href {https://doi.org/10.1103/PhysRevLett.124.242501}
  {\bibfield  {journal} {\bibinfo  {journal} {Physical Review Letters}\
  }\textbf {\bibinfo {volume} {124}},\ \bibinfo {pages} {242501} (\bibinfo
  {year} {2020})}\BibitemShut {NoStop}%
\bibitem [{\citenamefont {Zhang}\ and\ \citenamefont {Wang}(2023)}]{Zhang2023}%
  \BibitemOpen
  \bibfield  {author} {\bibinfo {author} {\bibfnamefont {H.}~\bibnamefont
  {Zhang}}\ and\ \bibinfo {author} {\bibfnamefont {X.}~\bibnamefont {Wang}},\
  }\href {https://doi.org/10.3389/fphy.2023.1166566} {\bibfield  {journal}
  {\bibinfo  {journal} {Frontiers in Physics}\ }\textbf {\bibinfo {volume}
  {11}},\ \bibinfo {pages} {1} (\bibinfo {year} {2023})}\BibitemShut {NoStop}%
\bibitem [{\citenamefont {Porsev}\ and\ \citenamefont
  {Flambaum}(2010{\natexlab{a}})}]{Porsev2010a}%
  \BibitemOpen
  \bibfield  {author} {\bibinfo {author} {\bibfnamefont {S.~G.}\ \bibnamefont
  {Porsev}}\ and\ \bibinfo {author} {\bibfnamefont {V.~V.}\ \bibnamefont
  {Flambaum}},\ }\href {https://doi.org/10.1103/PhysRevA.81.032504} {\bibfield
  {journal} {\bibinfo  {journal} {Physical Review A}\ }\textbf {\bibinfo
  {volume} {81}},\ \bibinfo {pages} {032504} (\bibinfo {year}
  {2010}{\natexlab{a}})}\BibitemShut {NoStop}%
\bibitem [{\citenamefont {Porsev}\ \emph {et~al.}(2010)\citenamefont {Porsev},
  \citenamefont {Flambaum}, \citenamefont {Peik},\ and\ \citenamefont
  {Tamm}}]{Porsev2010b}%
  \BibitemOpen
  \bibfield  {author} {\bibinfo {author} {\bibfnamefont {S.~G.}\ \bibnamefont
  {Porsev}}, \bibinfo {author} {\bibfnamefont {V.~V.}\ \bibnamefont
  {Flambaum}}, \bibinfo {author} {\bibfnamefont {E.}~\bibnamefont {Peik}},\
  and\ \bibinfo {author} {\bibfnamefont {C.}~\bibnamefont {Tamm}},\ }\href
  {https://doi.org/10.1103/PhysRevLett.105.182501} {\bibfield  {journal}
  {\bibinfo  {journal} {Physical Review Letters}\ }\textbf {\bibinfo {volume}
  {105}},\ \bibinfo {pages} {182501} (\bibinfo {year} {2010})}\BibitemShut
  {NoStop}%
\bibitem [{\citenamefont {M{\"{u}}ller}\ \emph {et~al.}(2019)\citenamefont
  {M{\"{u}}ller}, \citenamefont {Volotka},\ and\ \citenamefont
  {Surzhykov}}]{Muller2019}%
  \BibitemOpen
  \bibfield  {author} {\bibinfo {author} {\bibfnamefont {R.~A.}\ \bibnamefont
  {M{\"{u}}ller}}, \bibinfo {author} {\bibfnamefont {A.~V.}\ \bibnamefont
  {Volotka}},\ and\ \bibinfo {author} {\bibfnamefont {A.}~\bibnamefont
  {Surzhykov}},\ }\href {https://doi.org/10.1103/PhysRevA.99.042517} {\bibfield
   {journal} {\bibinfo  {journal} {Physical Review A}\ }\textbf {\bibinfo
  {volume} {99}},\ \bibinfo {pages} {042517} (\bibinfo {year} {2019})},\
  \Eprint {https://arxiv.org/abs/1902.05459} {arXiv:1902.05459} \BibitemShut
  {NoStop}%
\bibitem [{\citenamefont {Bilous}\ \emph {et~al.}(2020)\citenamefont {Bilous},
  \citenamefont {Bekker}, \citenamefont {Berengut}, \citenamefont {Seiferle},
  \citenamefont {von~der Wense}, \citenamefont {Thirolf}, \citenamefont
  {Pfeifer}, \citenamefont {L{\'{o}}pez-Urrutia},\ and\ \citenamefont
  {P{\'{a}}lffy}}]{Bilous2020}%
  \BibitemOpen
  \bibfield  {author} {\bibinfo {author} {\bibfnamefont {P.~V.}\ \bibnamefont
  {Bilous}}, \bibinfo {author} {\bibfnamefont {H.}~\bibnamefont {Bekker}},
  \bibinfo {author} {\bibfnamefont {J.~C.}\ \bibnamefont {Berengut}}, \bibinfo
  {author} {\bibfnamefont {B.}~\bibnamefont {Seiferle}}, \bibinfo {author}
  {\bibfnamefont {L.}~\bibnamefont {von~der Wense}}, \bibinfo {author}
  {\bibfnamefont {P.~G.}\ \bibnamefont {Thirolf}}, \bibinfo {author}
  {\bibfnamefont {T.}~\bibnamefont {Pfeifer}}, \bibinfo {author} {\bibfnamefont
  {J.~R.~C.}\ \bibnamefont {L{\'{o}}pez-Urrutia}},\ and\ \bibinfo {author}
  {\bibfnamefont {A.}~\bibnamefont {P{\'{a}}lffy}},\ }\href
  {https://doi.org/10.1103/PhysRevLett.124.192502} {\bibfield  {journal}
  {\bibinfo  {journal} {Physical Review Letters}\ }\textbf {\bibinfo {volume}
  {124}},\ \bibinfo {pages} {192502} (\bibinfo {year} {2020})},\ \Eprint
  {https://arxiv.org/abs/2001.06421} {arXiv:2001.06421} \BibitemShut {NoStop}%
\bibitem [{\citenamefont {Dzyublik}(2022)}]{Dzyublik2022}%
  \BibitemOpen
  \bibfield  {author} {\bibinfo {author} {\bibfnamefont {A.~Y.}\ \bibnamefont
  {Dzyublik}},\ }\href {https://doi.org/10.1103/PhysRevC.106.064608} {\bibfield
   {journal} {\bibinfo  {journal} {Physical Review C}\ }\textbf {\bibinfo
  {volume} {106}},\ \bibinfo {pages} {064608} (\bibinfo {year}
  {2022})}\BibitemShut {NoStop}%
\bibitem [{\citenamefont {Nickerson}\ \emph {et~al.}(2020)\citenamefont
  {Nickerson}, \citenamefont {Pimon}, \citenamefont {Bilous}, \citenamefont
  {Gugler}, \citenamefont {Beeks}, \citenamefont {Sikorsky}, \citenamefont
  {Mohn}, \citenamefont {Schumm},\ and\ \citenamefont
  {P{\'{a}}lffy}}]{Nickerson2020}%
  \BibitemOpen
  \bibfield  {author} {\bibinfo {author} {\bibfnamefont {B.~S.}\ \bibnamefont
  {Nickerson}}, \bibinfo {author} {\bibfnamefont {M.}~\bibnamefont {Pimon}},
  \bibinfo {author} {\bibfnamefont {P.~V.}\ \bibnamefont {Bilous}}, \bibinfo
  {author} {\bibfnamefont {J.}~\bibnamefont {Gugler}}, \bibinfo {author}
  {\bibfnamefont {K.}~\bibnamefont {Beeks}}, \bibinfo {author} {\bibfnamefont
  {T.}~\bibnamefont {Sikorsky}}, \bibinfo {author} {\bibfnamefont
  {P.}~\bibnamefont {Mohn}}, \bibinfo {author} {\bibfnamefont {T.}~\bibnamefont
  {Schumm}},\ and\ \bibinfo {author} {\bibfnamefont {A.}~\bibnamefont
  {P{\'{a}}lffy}},\ }\href {https://doi.org/10.1103/PhysRevLett.125.032501}
  {\bibfield  {journal} {\bibinfo  {journal} {Physical Review Letters}\
  }\textbf {\bibinfo {volume} {125}},\ \bibinfo {pages} {032501} (\bibinfo
  {year} {2020})}\BibitemShut {NoStop}%
\bibitem [{\citenamefont {Porsev}\ and\ \citenamefont
  {Flambaum}(2010{\natexlab{b}})}]{Porsev2010}%
  \BibitemOpen
  \bibfield  {author} {\bibinfo {author} {\bibfnamefont {S.~G.}\ \bibnamefont
  {Porsev}}\ and\ \bibinfo {author} {\bibfnamefont {V.~V.}\ \bibnamefont
  {Flambaum}},\ }\href {https://doi.org/10.1103/PhysRevA.81.042516} {\bibfield
  {journal} {\bibinfo  {journal} {Physical Review A}\ }\textbf {\bibinfo
  {volume} {81}},\ \bibinfo {pages} {042516} (\bibinfo {year}
  {2010}{\natexlab{b}})}\BibitemShut {NoStop}%
\bibitem [{\citenamefont {Peik}\ and\ \citenamefont
  {Okhapkin}(2015)}]{Peik2015}%
  \BibitemOpen
  \bibfield  {author} {\bibinfo {author} {\bibfnamefont {E.}~\bibnamefont
  {Peik}}\ and\ \bibinfo {author} {\bibfnamefont {M.}~\bibnamefont
  {Okhapkin}},\ }\href {https://doi.org/10.1016/j.crhy.2015.02.007} {\bibfield
  {journal} {\bibinfo  {journal} {Comptes Rendus Physique}\ }\textbf {\bibinfo
  {volume} {16}},\ \bibinfo {pages} {516} (\bibinfo {year} {2015})}\BibitemShut
  {NoStop}%
\bibitem [{\citenamefont {Porsev}\ \emph {et~al.}(2021)\citenamefont {Porsev},
  \citenamefont {Cheung},\ and\ \citenamefont {Safronova}}]{Porsev2021}%
  \BibitemOpen
  \bibfield  {author} {\bibinfo {author} {\bibfnamefont {S.~G.}\ \bibnamefont
  {Porsev}}, \bibinfo {author} {\bibfnamefont {C.}~\bibnamefont {Cheung}},\
  and\ \bibinfo {author} {\bibfnamefont {M.~S.}\ \bibnamefont {Safronova}},\
  }\href {https://doi.org/10.1088/2058-9565/ac08f1} {\bibfield  {journal}
  {\bibinfo  {journal} {Quantum Science and Technology}\ }\textbf {\bibinfo
  {volume} {6}},\ \bibinfo {pages} {034014} (\bibinfo {year}
  {2021})}\BibitemShut {NoStop}%
\bibitem [{\citenamefont {Morita}(1973)}]{Morita1973}%
  \BibitemOpen
  \bibfield  {author} {\bibinfo {author} {\bibfnamefont {M.}~\bibnamefont
  {Morita}},\ }\href {https://doi.org/10.1143/PTP.49.1574} {\bibfield
  {journal} {\bibinfo  {journal} {Progress of Theoretical Physics}\ }\textbf
  {\bibinfo {volume} {49}},\ \bibinfo {pages} {1574} (\bibinfo {year}
  {1973})}\BibitemShut {NoStop}%
\bibitem [{\citenamefont {Izawa}\ and\ \citenamefont
  {Yamanaka}(1979)}]{Izawa1979}%
  \BibitemOpen
  \bibfield  {author} {\bibinfo {author} {\bibfnamefont {Y.}~\bibnamefont
  {Izawa}}\ and\ \bibinfo {author} {\bibfnamefont {C.}~\bibnamefont
  {Yamanaka}},\ }\href {https://doi.org/10.1016/0370-2693(79)90113-8}
  {\bibfield  {journal} {\bibinfo  {journal} {Physics Letters B}\ }\textbf
  {\bibinfo {volume} {88}},\ \bibinfo {pages} {59} (\bibinfo {year}
  {1979})}\BibitemShut {NoStop}%
\bibitem [{\citenamefont {Arutyunyan}\ \emph {et~al.}(1991)\citenamefont
  {Arutyunyan}, \citenamefont {Bol'shov}, \citenamefont {Vikharev},
  \citenamefont {Dorshakov}, \citenamefont {Kornilo}, \citenamefont
  {Krivolapov}, \citenamefont {Smirnov},\ and\ \citenamefont
  {Tkalya}}]{Arutyunyan1991}%
  \BibitemOpen
  \bibfield  {author} {\bibinfo {author} {\bibfnamefont {R.}~\bibnamefont
  {Arutyunyan}}, \bibinfo {author} {\bibfnamefont {L.}~\bibnamefont
  {Bol'shov}}, \bibinfo {author} {\bibfnamefont {V.}~\bibnamefont {Vikharev}},
  \bibinfo {author} {\bibfnamefont {S.}~\bibnamefont {Dorshakov}}, \bibinfo
  {author} {\bibfnamefont {V.}~\bibnamefont {Kornilo}}, \bibinfo {author}
  {\bibfnamefont {A.}~\bibnamefont {Krivolapov}}, \bibinfo {author}
  {\bibfnamefont {V.}~\bibnamefont {Smirnov}},\ and\ \bibinfo {author}
  {\bibfnamefont {E.}~\bibnamefont {Tkalya}},\ }\href
  {https://inis.iaea.org/search/search.aspx?orig{\_}q=RN:23033001} {\bibfield
  {journal} {\bibinfo  {journal} {Soviet Journal of Nuclear Physics}\ }\textbf
  {\bibinfo {volume} {53}},\ \bibinfo {pages} {23} (\bibinfo {year}
  {1991})}\BibitemShut {NoStop}%
\bibitem [{\citenamefont {Claverie}\ \emph {et~al.}(2004)\citenamefont
  {Claverie}, \citenamefont {Al{\'{e}}onard}, \citenamefont {Chemin},
  \citenamefont {Gobet}, \citenamefont {Hannachi}, \citenamefont {Harston},
  \citenamefont {Malka}, \citenamefont {Scheurer}, \citenamefont {Morel},\ and\
  \citenamefont {M{\'{e}}ot}}]{Claverie2004}%
  \BibitemOpen
  \bibfield  {author} {\bibinfo {author} {\bibfnamefont {G.}~\bibnamefont
  {Claverie}}, \bibinfo {author} {\bibfnamefont {M.~M.}\ \bibnamefont
  {Al{\'{e}}onard}}, \bibinfo {author} {\bibfnamefont {J.~F.}\ \bibnamefont
  {Chemin}}, \bibinfo {author} {\bibfnamefont {F.}~\bibnamefont {Gobet}},
  \bibinfo {author} {\bibfnamefont {F.}~\bibnamefont {Hannachi}}, \bibinfo
  {author} {\bibfnamefont {M.~R.}\ \bibnamefont {Harston}}, \bibinfo {author}
  {\bibfnamefont {G.}~\bibnamefont {Malka}}, \bibinfo {author} {\bibfnamefont
  {J.~N.}\ \bibnamefont {Scheurer}}, \bibinfo {author} {\bibfnamefont
  {P.}~\bibnamefont {Morel}},\ and\ \bibinfo {author} {\bibfnamefont
  {V.}~\bibnamefont {M{\'{e}}ot}},\ }\href
  {https://doi.org/10.1103/PhysRevC.70.044303} {\bibfield  {journal} {\bibinfo
  {journal} {Physical Review C}\ }\textbf {\bibinfo {volume} {70}},\ \bibinfo
  {pages} {044303} (\bibinfo {year} {2004})}\BibitemShut {NoStop}%
\bibitem [{\citenamefont {Chodash}\ \emph {et~al.}(2016)\citenamefont
  {Chodash}, \citenamefont {Burke}, \citenamefont {Norman}, \citenamefont
  {Wilks}, \citenamefont {Casperson}, \citenamefont {Fisher}, \citenamefont
  {Holliday}, \citenamefont {Jeffries},\ and\ \citenamefont
  {Wakeling}}]{Chodash2016}%
  \BibitemOpen
  \bibfield  {author} {\bibinfo {author} {\bibfnamefont {P.~A.}\ \bibnamefont
  {Chodash}}, \bibinfo {author} {\bibfnamefont {J.~T.}\ \bibnamefont {Burke}},
  \bibinfo {author} {\bibfnamefont {E.~B.}\ \bibnamefont {Norman}}, \bibinfo
  {author} {\bibfnamefont {S.~C.}\ \bibnamefont {Wilks}}, \bibinfo {author}
  {\bibfnamefont {R.~J.}\ \bibnamefont {Casperson}}, \bibinfo {author}
  {\bibfnamefont {S.~E.}\ \bibnamefont {Fisher}}, \bibinfo {author}
  {\bibfnamefont {K.~S.}\ \bibnamefont {Holliday}}, \bibinfo {author}
  {\bibfnamefont {J.~R.}\ \bibnamefont {Jeffries}},\ and\ \bibinfo {author}
  {\bibfnamefont {M.~A.}\ \bibnamefont {Wakeling}},\ }\href
  {https://doi.org/10.1103/PhysRevC.93.034610} {\bibfield  {journal} {\bibinfo
  {journal} {Physical Review C}\ }\textbf {\bibinfo {volume} {93}},\ \bibinfo
  {pages} {034610} (\bibinfo {year} {2016})}\BibitemShut {NoStop}%
\bibitem [{\citenamefont {Saito}\ \emph {et~al.}(1980)\citenamefont {Saito},
  \citenamefont {Shinohara},\ and\ \citenamefont {Otozai}}]{Saito1980}%
  \BibitemOpen
  \bibfield  {author} {\bibinfo {author} {\bibfnamefont {T.}~\bibnamefont
  {Saito}}, \bibinfo {author} {\bibfnamefont {A.}~\bibnamefont {Shinohara}},\
  and\ \bibinfo {author} {\bibfnamefont {K.}~\bibnamefont {Otozai}},\ }\href
  {https://doi.org/10.1016/0370-2693(80)90267-1} {\bibfield  {journal}
  {\bibinfo  {journal} {Physics Letters B}\ }\textbf {\bibinfo {volume} {92}},\
  \bibinfo {pages} {293} (\bibinfo {year} {1980})}\BibitemShut {NoStop}%
\bibitem [{\citenamefont {Tkalya}(1992)}]{Tkalya1992}%
  \BibitemOpen
  \bibfield  {author} {\bibinfo {author} {\bibfnamefont {E.~V.}\ \bibnamefont
  {Tkalya}},\ }\href {https://doi.org/10.1016/0375-9474(92)90267-N} {\bibfield
  {journal} {\bibinfo  {journal} {Nuclear Physics A}\ }\textbf {\bibinfo
  {volume} {539}},\ \bibinfo {pages} {209} (\bibinfo {year}
  {1992})}\BibitemShut {NoStop}%
\bibitem [{\citenamefont {Pisk}\ \emph {et~al.}(1989)\citenamefont {Pisk},
  \citenamefont {Kaliman},\ and\ \citenamefont {Logan}}]{Pisk1989}%
  \BibitemOpen
  \bibfield  {author} {\bibinfo {author} {\bibfnamefont {K.}~\bibnamefont
  {Pisk}}, \bibinfo {author} {\bibfnamefont {Z.}~\bibnamefont {Kaliman}},\ and\
  \bibinfo {author} {\bibfnamefont {B.}~\bibnamefont {Logan}},\ }\href
  {https://doi.org/10.1016/0375-9474(89)90284-4} {\bibfield  {journal}
  {\bibinfo  {journal} {Nuclear Physics A}\ }\textbf {\bibinfo {volume}
  {504}},\ \bibinfo {pages} {103} (\bibinfo {year} {1989})}\BibitemShut
  {NoStop}%
\bibitem [{\citenamefont {Ljubi{\v{c}}i{\'{c}}}\ \emph
  {et~al.}(1991)\citenamefont {Ljubi{\v{c}}i{\'{c}}}, \citenamefont {Kekez},\
  and\ \citenamefont {Logan}}]{Ljubicic1991}%
  \BibitemOpen
  \bibfield  {author} {\bibinfo {author} {\bibfnamefont {A.}~\bibnamefont
  {Ljubi{\v{c}}i{\'{c}}}}, \bibinfo {author} {\bibfnamefont {D.}~\bibnamefont
  {Kekez}},\ and\ \bibinfo {author} {\bibfnamefont {B.}~\bibnamefont {Logan}},\
  }\href {https://doi.org/10.1016/0370-2693(91)91002-D} {\bibfield  {journal}
  {\bibinfo  {journal} {Physics Letters B}\ }\textbf {\bibinfo {volume}
  {272}},\ \bibinfo {pages} {1} (\bibinfo {year} {1991})}\BibitemShut {NoStop}%
\bibitem [{\citenamefont {Ahmad}\ \emph {et~al.}(2000)\citenamefont {Ahmad},
  \citenamefont {Dunford}, \citenamefont {Esbensen}, \citenamefont {Gemmell},
  \citenamefont {Kanter}, \citenamefont {R{\"{u}}tt},\ and\ \citenamefont
  {Southworth}}]{Ahmad2000}%
  \BibitemOpen
  \bibfield  {author} {\bibinfo {author} {\bibfnamefont {I.}~\bibnamefont
  {Ahmad}}, \bibinfo {author} {\bibfnamefont {R.~W.}\ \bibnamefont {Dunford}},
  \bibinfo {author} {\bibfnamefont {H.}~\bibnamefont {Esbensen}}, \bibinfo
  {author} {\bibfnamefont {D.~S.}\ \bibnamefont {Gemmell}}, \bibinfo {author}
  {\bibfnamefont {E.~P.}\ \bibnamefont {Kanter}}, \bibinfo {author}
  {\bibfnamefont {U.}~\bibnamefont {R{\"{u}}tt}},\ and\ \bibinfo {author}
  {\bibfnamefont {S.~H.}\ \bibnamefont {Southworth}},\ }\href
  {https://doi.org/10.1103/PhysRevC.61.051304} {\bibfield  {journal} {\bibinfo
  {journal} {Physical Review C}\ }\textbf {\bibinfo {volume} {61}},\ \bibinfo
  {pages} {051304} (\bibinfo {year} {2000})}\BibitemShut {NoStop}%
\bibitem [{\citenamefont {Kishimoto}\ \emph {et~al.}(2000)\citenamefont
  {Kishimoto}, \citenamefont {Yoda}, \citenamefont {Seto}, \citenamefont
  {Kobayashi}, \citenamefont {Kitao}, \citenamefont {Haruki}, \citenamefont
  {Kawauchi}, \citenamefont {Fukutani},\ and\ \citenamefont
  {Okano}}]{Kishimoto2000}%
  \BibitemOpen
  \bibfield  {author} {\bibinfo {author} {\bibfnamefont {S.}~\bibnamefont
  {Kishimoto}}, \bibinfo {author} {\bibfnamefont {Y.}~\bibnamefont {Yoda}},
  \bibinfo {author} {\bibfnamefont {M.}~\bibnamefont {Seto}}, \bibinfo {author}
  {\bibfnamefont {Y.}~\bibnamefont {Kobayashi}}, \bibinfo {author}
  {\bibfnamefont {S.}~\bibnamefont {Kitao}}, \bibinfo {author} {\bibfnamefont
  {R.}~\bibnamefont {Haruki}}, \bibinfo {author} {\bibfnamefont
  {T.}~\bibnamefont {Kawauchi}}, \bibinfo {author} {\bibfnamefont
  {K.}~\bibnamefont {Fukutani}},\ and\ \bibinfo {author} {\bibfnamefont
  {T.}~\bibnamefont {Okano}},\ }\href
  {https://doi.org/10.1103/PhysRevLett.85.1831} {\bibfield  {journal} {\bibinfo
   {journal} {Physical Review Letters}\ }\textbf {\bibinfo {volume} {85}},\
  \bibinfo {pages} {1831} (\bibinfo {year} {2000})}\BibitemShut {NoStop}%
\bibitem [{\citenamefont {Fujioka}\ \emph {et~al.}(1984)\citenamefont
  {Fujioka}, \citenamefont {Ura}, \citenamefont {Shinohara}, \citenamefont
  {Saito},\ and\ \citenamefont {Otozai}}]{Fujioka1984}%
  \BibitemOpen
  \bibfield  {author} {\bibinfo {author} {\bibfnamefont {H.}~\bibnamefont
  {Fujioka}}, \bibinfo {author} {\bibfnamefont {K.}~\bibnamefont {Ura}},
  \bibinfo {author} {\bibfnamefont {A.}~\bibnamefont {Shinohara}}, \bibinfo
  {author} {\bibfnamefont {T.}~\bibnamefont {Saito}},\ and\ \bibinfo {author}
  {\bibfnamefont {K.}~\bibnamefont {Otozai}},\ }\href
  {https://doi.org/10.1007/BF01436218} {\bibfield  {journal} {\bibinfo
  {journal} {Zeitschrift f{\"{u}}r Physik A Atoms and Nuclei}\ }\textbf
  {\bibinfo {volume} {315}},\ \bibinfo {pages} {121} (\bibinfo {year}
  {1984})}\BibitemShut {NoStop}%
\bibitem [{\citenamefont {Kishimoto}\ \emph {et~al.}(2005)\citenamefont
  {Kishimoto}, \citenamefont {Yoda}, \citenamefont {Kobayashi}, \citenamefont
  {Kitao}, \citenamefont {Haruki},\ and\ \citenamefont {Seto}}]{Kishimoto2005}%
  \BibitemOpen
  \bibfield  {author} {\bibinfo {author} {\bibfnamefont {S.}~\bibnamefont
  {Kishimoto}}, \bibinfo {author} {\bibfnamefont {Y.}~\bibnamefont {Yoda}},
  \bibinfo {author} {\bibfnamefont {Y.}~\bibnamefont {Kobayashi}}, \bibinfo
  {author} {\bibfnamefont {S.}~\bibnamefont {Kitao}}, \bibinfo {author}
  {\bibfnamefont {R.}~\bibnamefont {Haruki}},\ and\ \bibinfo {author}
  {\bibfnamefont {M.}~\bibnamefont {Seto}},\ }\href
  {https://doi.org/10.1016/j.nuclphysa.2004.10.016} {\bibfield  {journal}
  {\bibinfo  {journal} {Nuclear Physics A}\ }\textbf {\bibinfo {volume}
  {748}},\ \bibinfo {pages} {3} (\bibinfo {year} {2005})}\BibitemShut {NoStop}%
\bibitem [{\citenamefont {Sakabe}\ \emph {et~al.}(2005)\citenamefont {Sakabe},
  \citenamefont {Takahashi}, \citenamefont {Hashida}, \citenamefont {Shimizu},\
  and\ \citenamefont {Iida}}]{Sakabe2005}%
  \BibitemOpen
  \bibfield  {author} {\bibinfo {author} {\bibfnamefont {S.}~\bibnamefont
  {Sakabe}}, \bibinfo {author} {\bibfnamefont {K.}~\bibnamefont {Takahashi}},
  \bibinfo {author} {\bibfnamefont {M.}~\bibnamefont {Hashida}}, \bibinfo
  {author} {\bibfnamefont {S.}~\bibnamefont {Shimizu}},\ and\ \bibinfo {author}
  {\bibfnamefont {T.}~\bibnamefont {Iida}},\ }\href
  {https://doi.org/10.1016/j.adt.2005.07.002} {\bibfield  {journal} {\bibinfo
  {journal} {Atomic Data and Nuclear Data Tables}\ }\textbf {\bibinfo {volume}
  {91}},\ \bibinfo {pages} {1} (\bibinfo {year} {2005})}\BibitemShut {NoStop}%
\bibitem [{\citenamefont {Shinohara}\ \emph {et~al.}(1981)\citenamefont
  {Shinohara}, \citenamefont {Saito}, \citenamefont {Arakawa}, \citenamefont
  {Otozai}, \citenamefont {Baba}, \citenamefont {Hata},\ and\ \citenamefont
  {Suzuki}}]{Shinohara1981}%
  \BibitemOpen
  \bibfield  {author} {\bibinfo {author} {\bibfnamefont {A.}~\bibnamefont
  {Shinohara}}, \bibinfo {author} {\bibfnamefont {T.}~\bibnamefont {Saito}},
  \bibinfo {author} {\bibfnamefont {R.}~\bibnamefont {Arakawa}}, \bibinfo
  {author} {\bibfnamefont {K.}~\bibnamefont {Otozai}}, \bibinfo {author}
  {\bibfnamefont {H.}~\bibnamefont {Baba}}, \bibinfo {author} {\bibfnamefont
  {K.}~\bibnamefont {Hata}},\ and\ \bibinfo {author} {\bibfnamefont
  {T.}~\bibnamefont {Suzuki}},\ }\href@noop {} {\emph {\bibinfo {title} {{Japan
  Atomic Energy Research Institute report, JAERI-M No. 9362}}}},\ \bibinfo
  {type} {Tech. Rep.}\ (\bibinfo {year} {1981})\BibitemShut {NoStop}%
\bibitem [{\citenamefont {{Froese Fischer}}\ \emph {et~al.}(2019)\citenamefont
  {{Froese Fischer}}, \citenamefont {Gaigalas}, \citenamefont {J{\"{o}}nsson},\
  and\ \citenamefont {Biero{\'{n}}}}]{FroeseFischer2018}%
  \BibitemOpen
  \bibfield  {author} {\bibinfo {author} {\bibfnamefont {C.}~\bibnamefont
  {{Froese Fischer}}}, \bibinfo {author} {\bibfnamefont {G.}~\bibnamefont
  {Gaigalas}}, \bibinfo {author} {\bibfnamefont {P.}~\bibnamefont
  {J{\"{o}}nsson}},\ and\ \bibinfo {author} {\bibfnamefont {J.}~\bibnamefont
  {Biero{\'{n}}}},\ }\href {https://doi.org/10.1016/j.cpc.2018.10.032}
  {\bibfield  {journal} {\bibinfo  {journal} {Computer Physics Communications}\
  }\textbf {\bibinfo {volume} {237}},\ \bibinfo {pages} {184} (\bibinfo {year}
  {2019})}\BibitemShut {NoStop}%
\bibitem [{\citenamefont {Grant}(2007)}]{Grant2007}%
  \BibitemOpen
  \bibfield  {author} {\bibinfo {author} {\bibfnamefont {I.~P.}\ \bibnamefont
  {Grant}},\ }\href {https://doi.org/10.1007/978-0-387-35069-1} {\emph
  {\bibinfo {title} {{Relativistic Quantum Theory of Atoms and Molecules}}}},\
  edited by\ \bibinfo {editor} {\bibfnamefont {I.~P.}\ \bibnamefont {Grant}},\
  \bibinfo {series} {Springer Series on Atomic, Optical, and Plasma Physics},
  Vol.~\bibinfo {volume} {40}\ (\bibinfo  {publisher} {Springer},\ \bibinfo
  {address} {New York, NY},\ \bibinfo {year} {2007})\BibitemShut {NoStop}%
\bibitem [{\citenamefont {{Froese Fischer}}\ \emph {et~al.}(2016)\citenamefont
  {{Froese Fischer}}, \citenamefont {Godefroid}, \citenamefont {Brage},
  \citenamefont {J{\"{o}}nsson},\ and\ \citenamefont
  {Gaigalas}}]{FroeseFischer2016}%
  \BibitemOpen
  \bibfield  {author} {\bibinfo {author} {\bibfnamefont {C.}~\bibnamefont
  {{Froese Fischer}}}, \bibinfo {author} {\bibfnamefont {M.~R.}\ \bibnamefont
  {Godefroid}}, \bibinfo {author} {\bibfnamefont {T.}~\bibnamefont {Brage}},
  \bibinfo {author} {\bibfnamefont {P.}~\bibnamefont {J{\"{o}}nsson}},\ and\
  \bibinfo {author} {\bibfnamefont {G.}~\bibnamefont {Gaigalas}},\ }\href
  {https://doi.org/10.1088/0953-4075/49/18/182004} {\bibfield  {journal}
  {\bibinfo  {journal} {Journal of Physics B: Atomic, Molecular and Optical
  Physics}\ }\textbf {\bibinfo {volume} {49}},\ \bibinfo {pages} {182004}
  (\bibinfo {year} {2016})}\BibitemShut {NoStop}%
\bibitem [{\citenamefont {Kozio{\l}}\ and\ \citenamefont
  {Rzadkiewicz}(2018)}]{Kozio2018}%
  \BibitemOpen
  \bibfield  {author} {\bibinfo {author} {\bibfnamefont {K.}~\bibnamefont
  {Kozio{\l}}}\ and\ \bibinfo {author} {\bibfnamefont {J.}~\bibnamefont
  {Rzadkiewicz}},\ }\href {https://doi.org/10.1103/PhysRevA.98.062504}
  {\bibfield  {journal} {\bibinfo  {journal} {Physical Review A}\ }\textbf
  {\bibinfo {volume} {98}},\ \bibinfo {pages} {062504} (\bibinfo {year}
  {2018})}\BibitemShut {NoStop}%
\bibitem [{\citenamefont {Grant}(1984)}]{Grant1984}%
  \BibitemOpen
  \bibfield  {author} {\bibinfo {author} {\bibfnamefont {I.~P.}\ \bibnamefont
  {Grant}},\ }\href {https://doi.org/10.1002/qua.560250104} {\bibfield
  {journal} {\bibinfo  {journal} {International Journal of Quantum Chemistry}\
  }\textbf {\bibinfo {volume} {25}},\ \bibinfo {pages} {23} (\bibinfo {year}
  {1984})}\BibitemShut {NoStop}%
\bibitem [{\citenamefont {Si}\ \emph {et~al.}(2018)\citenamefont {Si},
  \citenamefont {Guo}, \citenamefont {Brage}, \citenamefont {Chen},
  \citenamefont {Hutton},\ and\ \citenamefont {{Froese Fischer}}}]{Si2018}%
  \BibitemOpen
  \bibfield  {author} {\bibinfo {author} {\bibfnamefont {R.}~\bibnamefont
  {Si}}, \bibinfo {author} {\bibfnamefont {X.~L.}\ \bibnamefont {Guo}},
  \bibinfo {author} {\bibfnamefont {T.}~\bibnamefont {Brage}}, \bibinfo
  {author} {\bibfnamefont {C.~Y.}\ \bibnamefont {Chen}}, \bibinfo {author}
  {\bibfnamefont {R.}~\bibnamefont {Hutton}},\ and\ \bibinfo {author}
  {\bibfnamefont {C.}~\bibnamefont {{Froese Fischer}}},\ }\href
  {https://doi.org/10.1103/PhysRevA.98.012504} {\bibfield  {journal} {\bibinfo
  {journal} {Physical Review A}\ }\textbf {\bibinfo {volume} {98}},\ \bibinfo
  {pages} {012504} (\bibinfo {year} {2018})}\BibitemShut {NoStop}%
\bibitem [{\citenamefont {Li}\ \emph {et~al.}(2020)\citenamefont {Li},
  \citenamefont {Grumer}, \citenamefont {Brage},\ and\ \citenamefont
  {J{\"{o}}nsson}}]{Li2020}%
  \BibitemOpen
  \bibfield  {author} {\bibinfo {author} {\bibfnamefont {W.}~\bibnamefont
  {Li}}, \bibinfo {author} {\bibfnamefont {J.}~\bibnamefont {Grumer}}, \bibinfo
  {author} {\bibfnamefont {T.}~\bibnamefont {Brage}},\ and\ \bibinfo {author}
  {\bibfnamefont {P.}~\bibnamefont {J{\"{o}}nsson}},\ }\href
  {https://doi.org/10.1016/j.cpc.2020.107211} {\bibfield  {journal} {\bibinfo
  {journal} {Computer Physics Communications}\ }\textbf {\bibinfo {volume}
  {253}},\ \bibinfo {pages} {107211} (\bibinfo {year} {2020})}\BibitemShut
  {NoStop}%
\bibitem [{\citenamefont {Harston}\ and\ \citenamefont
  {Chemin}(1999)}]{Harston1999}%
  \BibitemOpen
  \bibfield  {author} {\bibinfo {author} {\bibfnamefont {M.~R.}\ \bibnamefont
  {Harston}}\ and\ \bibinfo {author} {\bibfnamefont {J.~F.}\ \bibnamefont
  {Chemin}},\ }\href {https://doi.org/10.1103/PhysRevC.59.2462} {\bibfield
  {journal} {\bibinfo  {journal} {Physical Review C}\ }\textbf {\bibinfo
  {volume} {59}},\ \bibinfo {pages} {2462} (\bibinfo {year}
  {1999})}\BibitemShut {NoStop}%
\bibitem [{\citenamefont {Harston}(2001)}]{Harston2001}%
  \BibitemOpen
  \bibfield  {author} {\bibinfo {author} {\bibfnamefont {M.}~\bibnamefont
  {Harston}},\ }\href {https://doi.org/10.1016/S0375-9474(01)00358-X}
  {\bibfield  {journal} {\bibinfo  {journal} {Nuclear Physics A}\ }\textbf
  {\bibinfo {volume} {690}},\ \bibinfo {pages} {447} (\bibinfo {year}
  {2001})}\BibitemShut {NoStop}%
\bibitem [{\citenamefont {Denis-Petit}\ \emph {et~al.}(2017)\citenamefont
  {Denis-Petit}, \citenamefont {Gosselin}, \citenamefont {Hannachi},
  \citenamefont {Tarisien}, \citenamefont {Bonnet}, \citenamefont {Comet},
  \citenamefont {Gobet}, \citenamefont {Versteegen}, \citenamefont {Morel},
  \citenamefont {M{\'{e}}ot},\ and\ \citenamefont {Matea}}]{Denis-Petit2017}%
  \BibitemOpen
  \bibfield  {author} {\bibinfo {author} {\bibfnamefont {D.}~\bibnamefont
  {Denis-Petit}}, \bibinfo {author} {\bibfnamefont {G.}~\bibnamefont
  {Gosselin}}, \bibinfo {author} {\bibfnamefont {F.}~\bibnamefont {Hannachi}},
  \bibinfo {author} {\bibfnamefont {M.}~\bibnamefont {Tarisien}}, \bibinfo
  {author} {\bibfnamefont {T.}~\bibnamefont {Bonnet}}, \bibinfo {author}
  {\bibfnamefont {M.}~\bibnamefont {Comet}}, \bibinfo {author} {\bibfnamefont
  {F.}~\bibnamefont {Gobet}}, \bibinfo {author} {\bibfnamefont
  {M.}~\bibnamefont {Versteegen}}, \bibinfo {author} {\bibfnamefont
  {P.}~\bibnamefont {Morel}}, \bibinfo {author} {\bibfnamefont
  {V.}~\bibnamefont {M{\'{e}}ot}},\ and\ \bibinfo {author} {\bibfnamefont
  {I.}~\bibnamefont {Matea}},\ }\href
  {https://doi.org/10.1103/PhysRevC.96.024604} {\bibfield  {journal} {\bibinfo
  {journal} {Physical Review C}\ }\textbf {\bibinfo {volume} {96}},\ \bibinfo
  {pages} {024604} (\bibinfo {year} {2017})}\BibitemShut {NoStop}%
\bibitem [{\citenamefont {Ring}\ and\ \citenamefont {Schuck}(1980)}]{Ring1980}%
  \BibitemOpen
  \bibfield  {author} {\bibinfo {author} {\bibfnamefont {P.}~\bibnamefont
  {Ring}}\ and\ \bibinfo {author} {\bibfnamefont {P.}~\bibnamefont {Schuck}},\
  }\href@noop {} {\emph {\bibinfo {title} {{The Nuclear Many-Body Problem}}}}\
  (\bibinfo  {publisher} {Springer-Verlag},\ \bibinfo {year}
  {1980})\BibitemShut {NoStop}%
\bibitem [{\citenamefont {Minkov}\ and\ \citenamefont
  {P{\'{a}}lffy}(2019)}]{Minkov2019}%
  \BibitemOpen
  \bibfield  {author} {\bibinfo {author} {\bibfnamefont {N.}~\bibnamefont
  {Minkov}}\ and\ \bibinfo {author} {\bibfnamefont {A.}~\bibnamefont
  {P{\'{a}}lffy}},\ }\href {https://doi.org/10.1103/PhysRevLett.122.162502}
  {\bibfield  {journal} {\bibinfo  {journal} {Physical Review Letters}\
  }\textbf {\bibinfo {volume} {122}},\ \bibinfo {pages} {162502} (\bibinfo
  {year} {2019})}\BibitemShut {NoStop}%
\bibitem [{\citenamefont {Schweiger}\ \emph {et~al.}(2019)\citenamefont
  {Schweiger}, \citenamefont {K{\"{o}}nig}, \citenamefont {{Crespo
  L{\'{o}}pez-Urrutia}}, \citenamefont {Door}, \citenamefont {Dorrer},
  \citenamefont {D{\"{u}}llmann}, \citenamefont {Eliseev}, \citenamefont
  {Filianin}, \citenamefont {Huang}, \citenamefont {Kromer}, \citenamefont
  {Micke}, \citenamefont {M{\"{u}}ller}, \citenamefont {Renisch}, \citenamefont
  {Rischka}, \citenamefont {Sch{\"{u}}ssler},\ and\ \citenamefont
  {Blaum}}]{Schweiger2019}%
  \BibitemOpen
  \bibfield  {author} {\bibinfo {author} {\bibfnamefont {C.}~\bibnamefont
  {Schweiger}}, \bibinfo {author} {\bibfnamefont {C.~M.}\ \bibnamefont
  {K{\"{o}}nig}}, \bibinfo {author} {\bibfnamefont {J.~R.}\ \bibnamefont
  {{Crespo L{\'{o}}pez-Urrutia}}}, \bibinfo {author} {\bibfnamefont
  {M.}~\bibnamefont {Door}}, \bibinfo {author} {\bibfnamefont {H.}~\bibnamefont
  {Dorrer}}, \bibinfo {author} {\bibfnamefont {C.~E.}\ \bibnamefont
  {D{\"{u}}llmann}}, \bibinfo {author} {\bibfnamefont {S.}~\bibnamefont
  {Eliseev}}, \bibinfo {author} {\bibfnamefont {P.}~\bibnamefont {Filianin}},
  \bibinfo {author} {\bibfnamefont {W.}~\bibnamefont {Huang}}, \bibinfo
  {author} {\bibfnamefont {K.}~\bibnamefont {Kromer}}, \bibinfo {author}
  {\bibfnamefont {P.}~\bibnamefont {Micke}}, \bibinfo {author} {\bibfnamefont
  {M.}~\bibnamefont {M{\"{u}}ller}}, \bibinfo {author} {\bibfnamefont
  {D.}~\bibnamefont {Renisch}}, \bibinfo {author} {\bibfnamefont
  {A.}~\bibnamefont {Rischka}}, \bibinfo {author} {\bibfnamefont {R.~X.}\
  \bibnamefont {Sch{\"{u}}ssler}},\ and\ \bibinfo {author} {\bibfnamefont
  {K.}~\bibnamefont {Blaum}},\ }\href {https://doi.org/10.1063/1.5128331}
  {\bibfield  {journal} {\bibinfo  {journal} {Review of Scientific
  Instruments}\ }\textbf {\bibinfo {volume} {90}},\ \bibinfo {pages} {123201}
  (\bibinfo {year} {2019})}\BibitemShut {NoStop}%
\bibitem [{\citenamefont {Gu}(2008)}]{Gu2008}%
  \BibitemOpen
  \bibfield  {author} {\bibinfo {author} {\bibfnamefont {M.~F.}\ \bibnamefont
  {Gu}},\ }\href {https://doi.org/10.1139/P07-197} {\bibfield  {journal}
  {\bibinfo  {journal} {Canadian Journal of Physics}\ }\textbf {\bibinfo
  {volume} {86}},\ \bibinfo {pages} {675} (\bibinfo {year} {2008})}\BibitemShut
  {NoStop}%
\end{thebibliography}


\end{document}

\title{Nuclear Excitation by Near-Resonant Electron Transition in \ce{^{229}Th^{39+}} Ions -- Supplementary Material}
\author{Karol Kozio{\l}}
\email{Karol.Koziol@ncbj.gov.pl}
\author{Jacek Rzadkiewicz} 
\email{Jacek.Rzadkiewicz@ncbj.gov.pl}
\affiliation{Narodowe Centrum Bada\'{n} J\k{a}drowych (NCBJ), Andrzeja So{\l}tana 7, 05-400 Otwock-\'{S}wierk, Poland}

\maketitle

%%%%%

\onecolumngrid

\section{Energies of all \NoCaseChange{\ce{[Kr]{4d}^{10}{4f}^{5}}} levels in \NoCaseChange{\ce{Th^39+}}}

Table~\ref{tab:en_all} lists the energies and $LS$-compositions of all \ce{[Kr]{4d}^{10}{4f}^{5}} levels, calculated at the AS2(4df) level of theory and by using one radial wavefunction for all 198 states (the EOL scheme). Note that the values in Table I of the main manuscript are calculated by using radial wavefunctions optimized separately for the ground state and the first excited state, in order to take into account the orbital relaxation effect. Therefore, the energy of the first excited state from Table~\ref{tab:en_all} (8.126~eV) is not equal to the energy marked as ''AS2(4df)'' from Table I of the main manuscript (8.245~eV). 

The energy uncertainties are approximately \SI{0.2}{eV} for states below \SI{16}{eV}, \SI{0.5}{eV} for states in the range \SIrange{16}{60}{eV}, \SI{1}{eV} for states in the range \SIrange{60}{80}{eV}, and \SI{2}{eV} for states above \SI{80}{eV}. 

\begin{figure}[!t]
\centering
\includegraphics[width=0.75\linewidth]{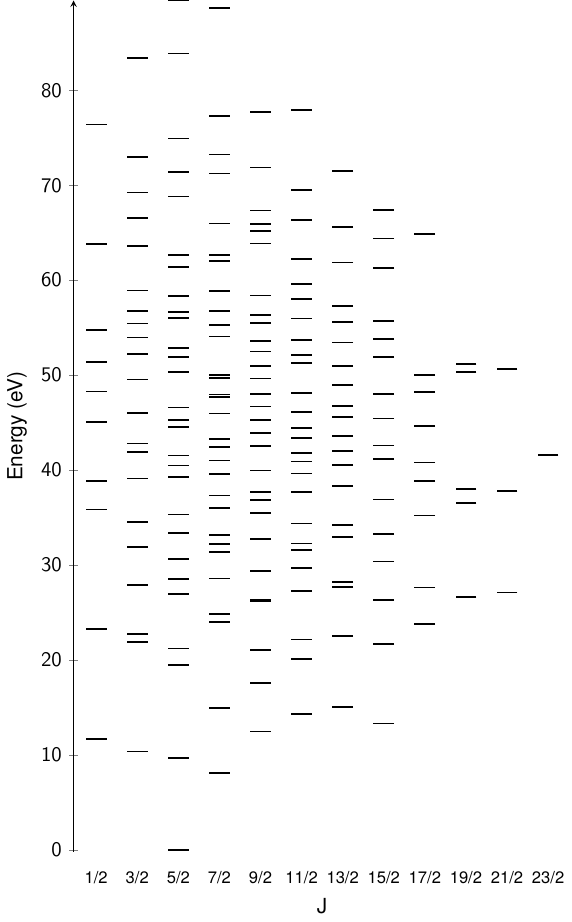}
\caption{Energy level diagram for all states of the \ce{[Kr]{4d}^{10}{4f}^{5}} configuration in \ce{Th^39+}.}
\label{fig:levels}
\end{figure}

A graphical representation of the level arrangement is presented in Figure~\ref{fig:levels}. 

\setlength{\LTcapwidth}{\linewidth}
\setlength{\LTleft}{0pt}
\setlength{\LTright}{0pt} 
\renewcommand{\arraystretch}{1.6}

\begin{longtable*}{@{} r @{\extracolsep{\fill}} l l S[table-format=6.0] S[table-format=4.3,round-mode=places,round-precision=3] @{}}
\caption{\label{tab:en_all}Energies of all \ce{[Kr]{4d}^{10}{4f}^{5}} levels in \ce{Th^39+}. }\\  
\toprule
No. & $J$ & $LS$-composition & {$E$ (\cmm)} & {$E$ (eV)} \\ 

\midrule
\endfirsthead
\caption{Continued.}\\  
\toprule
No. & $J$ & $LS$-composition & {$E$ (\cmm)} & {$E$ (eV)} \\ 

\midrule
\endhead
\bottomrule
\endfoot
1   & 5/2   & 0.27~$4d^{10}(^{1}_{0}S)\,4f^{5}(^{6}_{0}H)~^{6}H$ + 0.18~$4d^{10}(^{1}_{0}S)\,4f^{5}(^{4}_{4}G)~^{4}G$ + 0.15~$4d^{10}(^{1}_{0}S)\,4f^{5}(^{4}_{1}G)~^{4}G$ & 0            & 0.0000   \\ 
2   & 7/2   & 0.53~$4d^{10}(^{1}_{0}S)\,4f^{5}(^{6}_{0}H)~^{6}H$ + 0.17~$4d^{10}(^{1}_{0}S)\,4f^{5}(^{4}_{4}G)~^{4}G$ + 0.13~$4d^{10}(^{1}_{0}S)\,4f^{5}(^{4}_{1}G)~^{4}G$ & 65541        & 8.1260   \\ 
3   & 5/2   & 0.31~$4d^{10}(^{1}_{0}S)\,4f^{5}(^{6}_{0}H)~^{6}H$ + 0.16~$4d^{10}(^{1}_{0}S)\,4f^{5}(^{6}_{0}F)~^{6}F$ + 0.08~$4d^{10}(^{1}_{0}S)\,4f^{5}(^{4}_{3}F)~^{4}F$ & 78003        & 9.6712   \\ 
4   & 3/2   & 0.40~$4d^{10}(^{1}_{0}S)\,4f^{5}(^{6}_{0}F)~^{6}F$ + 0.23~$4d^{10}(^{1}_{0}S)\,4f^{5}(^{4}_{3}F)~^{4}F$ + 0.07~$4d^{10}(^{1}_{0}S)\,4f^{5}(^{4}_{1}F)~^{4}F$ & 83788        & 10.3884  \\ 
5   & 1/2   & 0.52~$4d^{10}(^{1}_{0}S)\,4f^{5}(^{6}_{0}F)~^{6}F$ + 0.12~$4d^{10}(^{1}_{0}S)\,4f^{5}(^{4}_{3}D)~^{4}D$ + 0.10~$4d^{10}(^{1}_{0}S)\,4f^{5}(^{4}_{1}D)~^{4}D$ & 94082        & 11.6647  \\ 
6   & 9/2   & 0.48~$4d^{10}(^{1}_{0}S)\,4f^{5}(^{6}_{0}H)~^{6}H$ + 0.10~$4d^{10}(^{1}_{0}S)\,4f^{5}(^{4}_{3}I)~^{4}I$ + 0.06~$4d^{10}(^{1}_{0}S)\,4f^{5}(^{4}_{4}G)~^{4}G$ & 100531       & 12.4643  \\ 
7   & 15/2  & 0.21~$4d^{10}(^{1}_{0}S)\,4f^{5}(^{4}_{0}M)~^{4}M$ + 0.17~$4d^{10}(^{1}_{0}S)\,4f^{5}(^{2}_{3}L)~^{2}L$ + 0.12~$4d^{10}(^{1}_{0}S)\,4f^{5}(^{4}_{3}I)~^{4}I$ & 107412       & 13.3174  \\ 
8   & 11/2  & 0.28~$4d^{10}(^{1}_{0}S)\,4f^{5}(^{6}_{0}H)~^{6}H$ + 0.14~$4d^{10}(^{1}_{0}S)\,4f^{5}(^{4}_{3}I)~^{4}I$ + 0.12~$4d^{10}(^{1}_{0}S)\,4f^{5}(^{4}_{1}K)~^{4}K$ & 115286       & 14.2937  \\ 
9   & 7/2   & 0.34~$4d^{10}(^{1}_{0}S)\,4f^{5}(^{6}_{0}F)~^{6}F$ + 0.14~$4d^{10}(^{1}_{0}S)\,4f^{5}(^{6}_{0}H)~^{6}H$ + 0.12~$4d^{10}(^{1}_{0}S)\,4f^{5}(^{4}_{3}F)~^{4}F$ & 120776       & 14.9743  \\ 
10  & 13/2  & 0.17~$4d^{10}(^{1}_{0}S)\,4f^{5}(^{4}_{3}I)~^{4}I$ + 0.17~$4d^{10}(^{1}_{0}S)\,4f^{5}(^{4}_{1}L)~^{4}L$ + 0.15~$4d^{10}(^{1}_{0}S)\,4f^{5}(^{6}_{0}H)~^{6}H$ & 121603       & 15.0769  \\ 
11  & 9/2   & 0.23~$4d^{10}(^{1}_{0}S)\,4f^{5}(^{6}_{0}F)~^{6}F$ + 0.14~$4d^{10}(^{1}_{0}S)\,4f^{5}(^{6}_{0}H)~^{6}H$ + 0.09~$4d^{10}(^{1}_{0}S)\,4f^{5}(^{4}_{3}G)~^{4}G$ & 141565       & 17.5518  \\ 
12  & 5/2   & 0.35~$4d^{10}(^{1}_{0}S)\,4f^{5}(^{6}_{0}F)~^{6}F$ + 0.17~$4d^{10}(^{1}_{0}S)\,4f^{5}(^{4}_{2}P)~^{4}P$ + 0.11~$4d^{10}(^{1}_{0}S)\,4f^{5}(^{6}_{0}P)~^{6}P$ & 157053       & 19.4721  \\ 
13  & 11/2  & 0.17~$4d^{10}(^{1}_{0}S)\,4f^{5}(^{6}_{0}F)~^{6}F$ + 0.17~$4d^{10}(^{1}_{0}S)\,4f^{5}(^{6}_{0}H)~^{6}H$ + 0.15~$4d^{10}(^{1}_{0}S)\,4f^{5}(^{4}_{2}G)~^{4}G$ & 162253       & 20.1168  \\ 
14  & 9/2   & 0.25~$4d^{10}(^{1}_{0}S)\,4f^{5}(^{4}_{3}I)~^{4}I$ + 0.15~$4d^{10}(^{1}_{0}S)\,4f^{5}(^{6}_{0}H)~^{6}H$ + 0.12~$4d^{10}(^{1}_{0}S)\,4f^{5}(^{6}_{0}F)~^{6}F$ & 169886       & 21.0631  \\ 
15  & 5/2   & 0.22~$4d^{10}(^{1}_{0}S)\,4f^{5}(^{6}_{0}H)~^{6}H$ + 0.13~$4d^{10}(^{1}_{0}S)\,4f^{5}(^{6}_{0}F)~^{6}F$ + 0.09~$4d^{10}(^{1}_{0}S)\,4f^{5}(^{6}_{0}P)~^{6}P$ & 171184       & 21.2241  \\ 
16  & 15/2  & 0.33~$4d^{10}(^{1}_{0}S)\,4f^{5}(^{4}_{0}M)~^{4}M$ + 0.26~$4d^{10}(^{1}_{0}S)\,4f^{5}(^{6}_{0}H)~^{6}H$ + 0.23~$4d^{10}(^{1}_{0}S)\,4f^{5}(^{4}_{3}I)~^{4}I$ & 174999       & 21.6971  \\ 
17  & 3/2   & 0.33~$4d^{10}(^{1}_{0}S)\,4f^{5}(^{6}_{0}F)~^{6}F$ + 0.32~$4d^{10}(^{1}_{0}S)\,4f^{5}(^{4}_{3}F)~^{4}F$ + 0.09~$4d^{10}(^{1}_{0}S)\,4f^{5}(^{4}_{4}F)~^{4}F$ & 176628       & 21.8991  \\ 
18  & 11/2  & 0.28~$4d^{10}(^{1}_{0}S)\,4f^{5}(^{4}_{1}K)~^{4}K$ + 0.27~$4d^{10}(^{1}_{0}S)\,4f^{5}(^{6}_{0}H)~^{6}H$ + 0.10~$4d^{10}(^{1}_{0}S)\,4f^{5}(^{6}_{0}F)~^{6}F$ & 178909       & 22.1818  \\ 
19  & 13/2  & 0.38~$4d^{10}(^{1}_{0}S)\,4f^{5}(^{6}_{0}H)~^{6}H$ + 0.36~$4d^{10}(^{1}_{0}S)\,4f^{5}(^{4}_{1}L)~^{4}L$ + 0.07~$4d^{10}(^{1}_{0}S)\,4f^{5}(^{4}_{3}I)~^{4}I$ & 181751       & 22.5343  \\ 
20  & 3/2   & 0.30~$4d^{10}(^{1}_{0}S)\,4f^{5}(^{6}_{0}P)~^{6}P$ + 0.28~$4d^{10}(^{1}_{0}S)\,4f^{5}(^{4}_{2}P)~^{4}P$ + 0.07~$4d^{10}(^{1}_{0}S)\,4f^{5}(^{2}_{3}D)~^{2}D$ & 183329       & 22.7299  \\ 
21  & 1/2   & 0.32~$4d^{10}(^{1}_{0}S)\,4f^{5}(^{6}_{0}F)~^{6}F$ + 0.31~$4d^{10}(^{1}_{0}S)\,4f^{5}(^{2}_{4}P)~^{2}P$ + 0.20~$4d^{10}(^{1}_{0}S)\,4f^{5}(^{2}_{3}P)~^{2}P$ & 187634       & 23.2637  \\ 
22  & 17/2  & 0.56~$4d^{10}(^{1}_{0}S)\,4f^{5}(^{4}_{0}M)~^{4}M$ + 0.18~$4d^{10}(^{1}_{0}S)\,4f^{5}(^{4}_{1}L)~^{4}L$ + 0.13~$4d^{10}(^{1}_{0}S)\,4f^{5}(^{2}_{3}L)~^{2}L$ & 192130       & 23.8211  \\ 
23  & 7/2   & 0.23~$4d^{10}(^{1}_{0}S)\,4f^{5}(^{6}_{0}F)~^{6}F$ + 0.18~$4d^{10}(^{1}_{0}S)\,4f^{5}(^{4}_{1}H)~^{4}H$ + 0.17~$4d^{10}(^{1}_{0}S)\,4f^{5}(^{4}_{3}H)~^{4}H$ & 193852       & 24.0346  \\ 
24  & 7/2   & 0.22~$4d^{10}(^{1}_{0}S)\,4f^{5}(^{4}_{2}G)~^{4}G$ + 0.16~$4d^{10}(^{1}_{0}S)\,4f^{5}(^{6}_{0}H)~^{6}H$ + 0.10~$4d^{10}(^{1}_{0}S)\,4f^{5}(^{4}_{4}G)~^{4}G$ & 200513       & 24.8604  \\ 
25  & 9/2   & 0.26~$4d^{10}(^{1}_{0}S)\,4f^{5}(^{6}_{0}F)~^{6}F$ + 0.23~$4d^{10}(^{1}_{0}S)\,4f^{5}(^{4}_{2}I)~^{4}I$ + 0.11~$4d^{10}(^{1}_{0}S)\,4f^{5}(^{2}_{4}H)~^{2}H$ & 211314       & 26.1996  \\ 
26  & 15/2  & 0.51~$4d^{10}(^{1}_{0}S)\,4f^{5}(^{4}_{1}L)~^{4}L$ + 0.14~$4d^{10}(^{1}_{0}S)\,4f^{5}(^{4}_{1}K)~^{4}K$ + 0.10~$4d^{10}(^{1}_{0}S)\,4f^{5}(^{4}_{2}K)~^{4}K$ & 212502       & 26.3468  \\ 
27  & 9/2   & 0.22~$4d^{10}(^{1}_{0}S)\,4f^{5}(^{4}_{1}H)~^{4}H$ + 0.20~$4d^{10}(^{1}_{0}S)\,4f^{5}(^{4}_{2}G)~^{4}G$ + 0.10~$4d^{10}(^{1}_{0}S)\,4f^{5}(^{4}_{3}I)~^{4}I$ & 212855       & 26.3907  \\ 
28  & 19/2  & 0.39~$4d^{10}(^{1}_{0}S)\,4f^{5}(^{4}_{0}M)~^{4}M$ + 0.30~$4d^{10}(^{1}_{0}S)\,4f^{5}(^{2}_{1}N)~^{2}N$ + 0.15~$4d^{10}(^{1}_{0}S)\,4f^{5}(^{4}_{1}L)~^{4}L$ & 214609       & 26.6081  \\ 
29  & 5/2   & 0.17~$4d^{10}(^{1}_{0}S)\,4f^{5}(^{2}_{4}D)~^{2}D$ + 0.13~$4d^{10}(^{1}_{0}S)\,4f^{5}(^{4}_{2}D)~^{4}D$ + 0.11~$4d^{10}(^{1}_{0}S)\,4f^{5}(^{2}_{4}F)~^{2}F$ & 217471       & 26.9629  \\ 
30  & 21/2  & 0.45~$4d^{10}(^{1}_{0}S)\,4f^{5}(^{2}_{1}N)~^{2}N$ + 0.29~$4d^{10}(^{1}_{0}S)\,4f^{5}(^{4}_{0}M)~^{4}M$ + 0.26~$4d^{10}(^{1}_{0}S)\,4f^{5}(^{2}_{0}O)~^{2}O$ & 218770       & 27.1240  \\ 
31  & 11/2  & 0.19~$4d^{10}(^{1}_{0}S)\,4f^{5}(^{4}_{1}H)~^{4}H$ + 0.11~$4d^{10}(^{1}_{0}S)\,4f^{5}(^{4}_{2}I)~^{4}I$ + 0.11~$4d^{10}(^{1}_{0}S)\,4f^{5}(^{6}_{0}F)~^{6}F$ & 219853       & 27.2582  \\ 
32  & 17/2  & 0.41~$4d^{10}(^{1}_{0}S)\,4f^{5}(^{2}_{1}M)~^{2}M$ + 0.24~$4d^{10}(^{1}_{0}S)\,4f^{5}(^{4}_{1}L)~^{4}L$ + 0.14~$4d^{10}(^{1}_{0}S)\,4f^{5}(^{2}_{2}L)~^{2}L$ & 222937       & 27.6406  \\ 
33  & 13/2  & 0.51~$4d^{10}(^{1}_{0}S)\,4f^{5}(^{4}_{1}K)~^{4}K$ + 0.09~$4d^{10}(^{1}_{0}S)\,4f^{5}(^{2}_{3}K)~^{2}K$ + 0.09~$4d^{10}(^{1}_{0}S)\,4f^{5}(^{2}_{4}I)~^{2}I$ & 223227       & 27.6766  \\ 
34  & 3/2   & 0.24~$4d^{10}(^{1}_{0}S)\,4f^{5}(^{4}_{1}P)~^{4}P$ + 0.13~$4d^{10}(^{1}_{0}S)\,4f^{5}(^{2}_{3}D)~^{2}D$ + 0.10~$4d^{10}(^{1}_{0}S)\,4f^{5}(^{4}_{2}D)~^{4}D$ & 225076       & 27.9058  \\ 
35  & 13/2  & 0.25~$4d^{10}(^{1}_{0}S)\,4f^{5}(^{4}_{2}I)~^{4}I$ + 0.14~$4d^{10}(^{1}_{0}S)\,4f^{5}(^{4}_{1}H)~^{4}H$ + 0.14~$4d^{10}(^{1}_{0}S)\,4f^{5}(^{2}_{5}K)~^{2}K$ & 227535       & 28.2107  \\ 
36  & 5/2   & 0.25~$4d^{10}(^{1}_{0}S)\,4f^{5}(^{4}_{3}F)~^{4}F$ + 0.18~$4d^{10}(^{1}_{0}S)\,4f^{5}(^{4}_{3}G)~^{4}G$ + 0.13~$4d^{10}(^{1}_{0}S)\,4f^{5}(^{4}_{4}F)~^{4}F$ & 230234       & 28.5454  \\ 
37  & 7/2   & 0.30~$4d^{10}(^{1}_{0}S)\,4f^{5}(^{6}_{0}P)~^{6}P$ + 0.21~$4d^{10}(^{1}_{0}S)\,4f^{5}(^{4}_{1}D)~^{4}D$ + 0.10~$4d^{10}(^{1}_{0}S)\,4f^{5}(^{4}_{2}F)~^{4}F$ & 230506       & 28.5791  \\ 
38  & 9/2   & 0.22~$4d^{10}(^{1}_{0}S)\,4f^{5}(^{4}_{2}F)~^{4}F$ + 0.13~$4d^{10}(^{1}_{0}S)\,4f^{5}(^{4}_{3}I)~^{4}I$ + 0.10~$4d^{10}(^{1}_{0}S)\,4f^{5}(^{2}_{3}G)~^{2}G$ & 236983       & 29.3822  \\ 
39  & 11/2  & 0.22~$4d^{10}(^{1}_{0}S)\,4f^{5}(^{6}_{0}F)~^{6}F$ + 0.19~$4d^{10}(^{1}_{0}S)\,4f^{5}(^{4}_{2}K)~^{4}K$ + 0.12~$4d^{10}(^{1}_{0}S)\,4f^{5}(^{2}_{4}I)~^{2}I$ & 239217       & 29.6591  \\ 
40  & 15/2  & 0.20~$4d^{10}(^{1}_{0}S)\,4f^{5}(^{4}_{2}I)~^{4}I$ + 0.19~$4d^{10}(^{1}_{0}S)\,4f^{5}(^{2}_{2}K)~^{2}K$ + 0.16~$4d^{10}(^{1}_{0}S)\,4f^{5}(^{6}_{0}H)~^{6}H$ & 245087       & 30.3869  \\ 
41  & 5/2   & 0.31~$4d^{10}(^{1}_{0}S)\,4f^{5}(^{6}_{0}P)~^{6}P$ + 0.20~$4d^{10}(^{1}_{0}S)\,4f^{5}(^{4}_{3}G)~^{4}G$ + 0.11~$4d^{10}(^{1}_{0}S)\,4f^{5}(^{4}_{2}G)~^{4}G$ & 247229       & 30.6525  \\ 
42  & 7/2   & 0.14~$4d^{10}(^{1}_{0}S)\,4f^{5}(^{4}_{3}G)~^{4}G$ + 0.12~$4d^{10}(^{1}_{0}S)\,4f^{5}(^{4}_{3}H)~^{4}H$ + 0.09~$4d^{10}(^{1}_{0}S)\,4f^{5}(^{4}_{1}G)~^{4}G$ & 252805       & 31.3438  \\ 
43  & 11/2  & 0.18~$4d^{10}(^{1}_{0}S)\,4f^{5}(^{2}_{6}H)~^{2}H$ + 0.17~$4d^{10}(^{1}_{0}S)\,4f^{5}(^{4}_{2}I)~^{4}I$ + 0.14~$4d^{10}(^{1}_{0}S)\,4f^{5}(^{2}_{4}H)~^{2}H$ & 254995       & 31.6153  \\ 
44  & 3/2   & 0.19~$4d^{10}(^{1}_{0}S)\,4f^{5}(^{6}_{0}P)~^{6}P$ + 0.15~$4d^{10}(^{1}_{0}S)\,4f^{5}(^{4}_{2}F)~^{4}F$ + 0.14~$4d^{10}(^{1}_{0}S)\,4f^{5}(^{2}_{3}D)~^{2}D$ & 257399       & 31.9133  \\ 
45  & 7/2   & 0.14~$4d^{10}(^{1}_{0}S)\,4f^{5}(^{6}_{0}P)~^{6}P$ + 0.11~$4d^{10}(^{1}_{0}S)\,4f^{5}(^{4}_{2}D)~^{4}D$ + 0.09~$4d^{10}(^{1}_{0}S)\,4f^{5}(^{2}_{5}G)~^{2}G$ & 259854       & 32.2178  \\ 
46  & 11/2  & 0.42~$4d^{10}(^{1}_{0}S)\,4f^{5}(^{4}_{3}I)~^{4}I$ + 0.08~$4d^{10}(^{1}_{0}S)\,4f^{5}(^{2}_{4}I)~^{2}I$ + 0.08~$4d^{10}(^{1}_{0}S)\,4f^{5}(^{4}_{4}G)~^{4}G$ & 260266       & 32.2689  \\ 
47  & 9/2   & 0.24~$4d^{10}(^{1}_{0}S)\,4f^{5}(^{4}_{2}F)~^{4}F$ + 0.13~$4d^{10}(^{1}_{0}S)\,4f^{5}(^{4}_{3}H)~^{4}H$ + 0.10~$4d^{10}(^{1}_{0}S)\,4f^{5}(^{4}_{2}I)~^{4}I$ & 264199       & 32.7565  \\ 
48  & 13/2  & 0.34~$4d^{10}(^{1}_{0}S)\,4f^{5}(^{6}_{0}H)~^{6}H$ + 0.24~$4d^{10}(^{1}_{0}S)\,4f^{5}(^{4}_{1}L)~^{4}L$ + 0.12~$4d^{10}(^{1}_{0}S)\,4f^{5}(^{2}_{4}I)~^{2}I$ & 265732       & 32.9465  \\ 
49  & 7/2   & 0.22~$4d^{10}(^{1}_{0}S)\,4f^{5}(^{4}_{2}G)~^{4}G$ + 0.14~$4d^{10}(^{1}_{0}S)\,4f^{5}(^{4}_{3}D)~^{4}D$ + 0.13~$4d^{10}(^{1}_{0}S)\,4f^{5}(^{6}_{0}F)~^{6}F$ & 267657       & 33.1853  \\ 
50  & 15/2  & 0.25~$4d^{10}(^{1}_{0}S)\,4f^{5}(^{6}_{0}H)~^{6}H$ + 0.16~$4d^{10}(^{1}_{0}S)\,4f^{5}(^{4}_{0}M)~^{4}M$ + 0.13~$4d^{10}(^{1}_{0}S)\,4f^{5}(^{2}_{2}L)~^{2}L$ & 268448       & 33.2833  \\ 
51  & 5/2   & 0.33~$4d^{10}(^{1}_{0}S)\,4f^{5}(^{4}_{2}F)~^{4}F$ + 0.24~$4d^{10}(^{1}_{0}S)\,4f^{5}(^{2}_{3}D)~^{2}D$ + 0.12~$4d^{10}(^{1}_{0}S)\,4f^{5}(^{4}_{1}P)~^{4}P$ & 269210       & 33.3777  \\ 
52  & 13/2  & 0.18~$4d^{10}(^{1}_{0}S)\,4f^{5}(^{4}_{2}K)~^{4}K$ + 0.18~$4d^{10}(^{1}_{0}S)\,4f^{5}(^{2}_{5}K)~^{2}K$ + 0.17~$4d^{10}(^{1}_{0}S)\,4f^{5}(^{2}_{4}I)~^{2}I$ & 276183       & 34.2424  \\ 
53  & 11/2  & 0.09~$4d^{10}(^{1}_{0}S)\,4f^{5}(^{2}_{4}H)~^{2}H$ + 0.09~$4d^{10}(^{1}_{0}S)\,4f^{5}(^{2}_{3}H)~^{2}H$ + 0.09~$4d^{10}(^{1}_{0}S)\,4f^{5}(^{2}_{6}H)~^{2}H$ & 277228       & 34.3719  \\ 
54  & 3/2   & 0.26~$4d^{10}(^{1}_{0}S)\,4f^{5}(^{4}_{3}D)~^{4}D$ + 0.15~$4d^{10}(^{1}_{0}S)\,4f^{5}(^{6}_{0}F)~^{6}F$ + 0.15~$4d^{10}(^{1}_{0}S)\,4f^{5}(^{4}_{1}D)~^{4}D$ & 278577       & 34.5392  \\ 
55  & 17/2  & 0.52~$4d^{10}(^{1}_{0}S)\,4f^{5}(^{4}_{1}K)~^{4}K$ + 0.30~$4d^{10}(^{1}_{0}S)\,4f^{5}(^{4}_{0}M)~^{4}M$ + 0.06~$4d^{10}(^{1}_{0}S)\,4f^{5}(^{4}_{1}L)~^{4}L$ & 284025       & 35.2146  \\ 
56  & 5/2   & 0.23~$4d^{10}(^{1}_{0}S)\,4f^{5}(^{4}_{3}G)~^{4}G$ + 0.20~$4d^{10}(^{1}_{0}S)\,4f^{5}(^{4}_{2}G)~^{4}G$ + 0.17~$4d^{10}(^{1}_{0}S)\,4f^{5}(^{4}_{3}F)~^{4}F$ & 285011       & 35.3369  \\ 
57  & 9/2   & 0.15~$4d^{10}(^{1}_{0}S)\,4f^{5}(^{4}_{1}G)~^{4}G$ + 0.10~$4d^{10}(^{1}_{0}S)\,4f^{5}(^{4}_{4}G)~^{4}G$ + 0.09~$4d^{10}(^{1}_{0}S)\,4f^{5}(^{4}_{3}H)~^{4}H$ & 285961       & 35.4546  \\ 
58  & 1/2   & 0.66~$4d^{10}(^{1}_{0}S)\,4f^{5}(^{4}_{2}P)~^{4}P$ + 0.16~$4d^{10}(^{1}_{0}S)\,4f^{5}(^{4}_{3}D)~^{4}D$ + 0.05~$4d^{10}(^{1}_{0}S)\,4f^{5}(^{4}_{1}D)~^{4}D$ & 289121       & 35.8464  \\ 
59  & 7/2   & 0.17~$4d^{10}(^{1}_{0}S)\,4f^{5}(^{4}_{1}H)~^{4}H$ + 0.12~$4d^{10}(^{1}_{0}S)\,4f^{5}(^{6}_{0}F)~^{6}F$ + 0.11~$4d^{10}(^{1}_{0}S)\,4f^{5}(^{4}_{3}H)~^{4}H$ & 290271       & 35.9890  \\ 
60  & 19/2  & 0.53~$4d^{10}(^{1}_{0}S)\,4f^{5}(^{4}_{0}M)~^{4}M$ + 0.20~$4d^{10}(^{1}_{0}S)\,4f^{5}(^{2}_{1}M)~^{2}M$ + 0.15~$4d^{10}(^{1}_{0}S)\,4f^{5}(^{2}_{1}N)~^{2}N$ & 294456       & 36.5079  \\ 
61  & 9/2   & 0.22~$4d^{10}(^{1}_{0}S)\,4f^{5}(^{4}_{1}H)~^{4}H$ + 0.16~$4d^{10}(^{1}_{0}S)\,4f^{5}(^{2}_{5}G)~^{2}G$ + 0.15~$4d^{10}(^{1}_{0}S)\,4f^{5}(^{4}_{2}G)~^{4}G$ & 297307       & 36.8614  \\ 
62  & 15/2  & 0.51~$4d^{10}(^{1}_{0}S)\,4f^{5}(^{4}_{1}K)~^{4}K$ + 0.24~$4d^{10}(^{1}_{0}S)\,4f^{5}(^{4}_{1}L)~^{4}L$ + 0.07~$4d^{10}(^{1}_{0}S)\,4f^{5}(^{2}_{5}K)~^{2}K$ & 297555       & 36.8921  \\ 
63  & 7/2   & 0.15~$4d^{10}(^{1}_{0}S)\,4f^{5}(^{6}_{0}P)~^{6}P$ + 0.14~$4d^{10}(^{1}_{0}S)\,4f^{5}(^{4}_{4}F)~^{4}F$ + 0.11~$4d^{10}(^{1}_{0}S)\,4f^{5}(^{4}_{2}F)~^{4}F$ & 301128       & 37.3352  \\ 
64  & 9/2   & 0.15~$4d^{10}(^{1}_{0}S)\,4f^{5}(^{4}_{2}I)~^{4}I$ + 0.14~$4d^{10}(^{1}_{0}S)\,4f^{5}(^{4}_{3}F)~^{4}F$ + 0.10~$4d^{10}(^{1}_{0}S)\,4f^{5}(^{6}_{0}F)~^{6}F$ & 303898       & 37.6786  \\ 
65  & 11/2  & 0.24~$4d^{10}(^{1}_{0}S)\,4f^{5}(^{4}_{2}G)~^{4}G$ + 0.16~$4d^{10}(^{1}_{0}S)\,4f^{5}(^{6}_{0}F)~^{6}F$ + 0.13~$4d^{10}(^{1}_{0}S)\,4f^{5}(^{4}_{2}K)~^{4}K$ & 304135       & 37.7079  \\ 
66  & 21/2  & 0.50~$4d^{10}(^{1}_{0}S)\,4f^{5}(^{4}_{0}M)~^{4}M$ + 0.49~$4d^{10}(^{1}_{0}S)\,4f^{5}(^{2}_{0}O)~^{2}O$ & 305062       & 37.8228  \\ 
67  & 19/2  & 0.53~$4d^{10}(^{1}_{0}S)\,4f^{5}(^{4}_{1}L)~^{4}L$ + 0.39~$4d^{10}(^{1}_{0}S)\,4f^{5}(^{2}_{1}N)~^{2}N$ + 0.06~$4d^{10}(^{1}_{0}S)\,4f^{5}(^{2}_{1}M)~^{2}M$ & 306697       & 38.0256  \\ 
68  & 13/2  & 0.22~$4d^{10}(^{1}_{0}S)\,4f^{5}(^{4}_{1}H)~^{4}H$ + 0.15~$4d^{10}(^{1}_{0}S)\,4f^{5}(^{2}_{2}K)~^{2}K$ + 0.12~$4d^{10}(^{1}_{0}S)\,4f^{5}(^{4}_{1}I)~^{4}I$ & 309276       & 38.3454  \\ 
69  & 1/2   & 0.50~$4d^{10}(^{1}_{0}S)\,4f^{5}(^{4}_{1}P)~^{4}P$ + 0.32~$4d^{10}(^{1}_{0}S)\,4f^{5}(^{4}_{2}D)~^{4}D$ + 0.05~$4d^{10}(^{1}_{0}S)\,4f^{5}(^{6}_{0}F)~^{6}F$ & 313235       & 38.8362  \\ 
70  & 17/2  & 0.43~$4d^{10}(^{1}_{0}S)\,4f^{5}(^{2}_{1}M)~^{2}M$ + 0.27~$4d^{10}(^{1}_{0}S)\,4f^{5}(^{4}_{1}L)~^{4}L$ + 0.08~$4d^{10}(^{1}_{0}S)\,4f^{5}(^{4}_{2}K)~^{4}K$ & 313543       & 38.8743  \\ 
71  & 3/2   & 0.24~$4d^{10}(^{1}_{0}S)\,4f^{5}(^{4}_{2}D)~^{4}D$ + 0.11~$4d^{10}(^{1}_{0}S)\,4f^{5}(^{4}_{1}F)~^{4}F$ + 0.10~$4d^{10}(^{1}_{0}S)\,4f^{5}(^{6}_{0}P)~^{6}P$ & 315440       & 39.1096  \\ 
72  & 5/2   & 0.16~$4d^{10}(^{1}_{0}S)\,4f^{5}(^{4}_{3}D)~^{4}D$ + 0.11~$4d^{10}(^{1}_{0}S)\,4f^{5}(^{2}_{3}D)~^{2}D$ + 0.11~$4d^{10}(^{1}_{0}S)\,4f^{5}(^{2}_{4}F)~^{2}F$ & 316602       & 39.2537  \\ 
73  & 7/2   & 0.16~$4d^{10}(^{1}_{0}S)\,4f^{5}(^{4}_{3}H)~^{4}H$ + 0.14~$4d^{10}(^{1}_{0}S)\,4f^{5}(^{4}_{3}D)~^{4}D$ + 0.12~$4d^{10}(^{1}_{0}S)\,4f^{5}(^{4}_{2}F)~^{4}F$ & 319409       & 39.6016  \\ 
74  & 11/2  & 0.31~$4d^{10}(^{1}_{0}S)\,4f^{5}(^{4}_{3}H)~^{4}H$ + 0.12~$4d^{10}(^{1}_{0}S)\,4f^{5}(^{4}_{4}G)~^{4}G$ + 0.12~$4d^{10}(^{1}_{0}S)\,4f^{5}(^{4}_{1}K)~^{4}K$ & 319805       & 39.6508  \\ 
75  & 9/2   & 0.20~$4d^{10}(^{1}_{0}S)\,4f^{5}(^{4}_{3}G)~^{4}G$ + 0.10~$4d^{10}(^{1}_{0}S)\,4f^{5}(^{4}_{2}G)~^{4}G$ + 0.08~$4d^{10}(^{1}_{0}S)\,4f^{5}(^{4}_{3}H)~^{4}H$ & 322151       & 39.9416  \\ 
76  & 5/2   & 0.17~$4d^{10}(^{1}_{0}S)\,4f^{5}(^{4}_{2}F)~^{4}F$ + 0.15~$4d^{10}(^{1}_{0}S)\,4f^{5}(^{6}_{0}P)~^{6}P$ + 0.14~$4d^{10}(^{1}_{0}S)\,4f^{5}(^{4}_{2}P)~^{4}P$ & 326660       & 40.5007  \\ 
77  & 13/2  & 0.21~$4d^{10}(^{1}_{0}S)\,4f^{5}(^{2}_{5}K)~^{2}K$ + 0.20~$4d^{10}(^{1}_{0}S)\,4f^{5}(^{4}_{3}H)~^{4}H$ + 0.17~$4d^{10}(^{1}_{0}S)\,4f^{5}(^{2}_{3}K)~^{2}K$ & 326936       & 40.5348  \\ 
78  & 17/2  & 0.30~$4d^{10}(^{1}_{0}S)\,4f^{5}(^{4}_{2}K)~^{4}K$ + 0.24~$4d^{10}(^{1}_{0}S)\,4f^{5}(^{2}_{3}L)~^{2}L$ + 0.19~$4d^{10}(^{1}_{0}S)\,4f^{5}(^{2}_{2}M)~^{2}M$ & 329178       & 40.8129  \\ 
79  & 11/2  & 0.16~$4d^{10}(^{1}_{0}S)\,4f^{5}(^{4}_{2}I)~^{4}I$ + 0.15~$4d^{10}(^{1}_{0}S)\,4f^{5}(^{2}_{4}H)~^{2}H$ + 0.14~$4d^{10}(^{1}_{0}S)\,4f^{5}(^{4}_{1}H)~^{4}H$ & 329877       & 40.8995  \\ 
80  & 7/2   & 0.24~$4d^{10}(^{1}_{0}S)\,4f^{5}(^{4}_{2}H)~^{4}H$ + 0.13~$4d^{10}(^{1}_{0}S)\,4f^{5}(^{4}_{2}F)~^{4}F$ + 0.10~$4d^{10}(^{1}_{0}S)\,4f^{5}(^{4}_{3}H)~^{4}H$ & 330821       & 41.0166  \\ 
81  & 15/2  & 0.44~$4d^{10}(^{1}_{0}S)\,4f^{5}(^{4}_{2}I)~^{4}I$ + 0.25~$4d^{10}(^{1}_{0}S)\,4f^{5}(^{2}_{2}L)~^{2}L$ + 0.09~$4d^{10}(^{1}_{0}S)\,4f^{5}(^{2}_{1}L)~^{2}L$ & 331886       & 41.1486  \\ 
82  & 5/2   & 0.15~$4d^{10}(^{1}_{0}S)\,4f^{5}(^{4}_{3}D)~^{4}D$ + 0.13~$4d^{10}(^{1}_{0}S)\,4f^{5}(^{4}_{1}P)~^{4}P$ + 0.10~$4d^{10}(^{1}_{0}S)\,4f^{5}(^{4}_{3}G)~^{4}G$ & 334913       & 41.5239  \\ 
83  & 23/2  & 0.99~$4d^{10}(^{1}_{0}S)\,4f^{5}(^{2}_{0}O)~^{2}O$                                 & 335551       & 41.6031  \\ 
84  & 11/2  & 0.20~$4d^{10}(^{1}_{0}S)\,4f^{5}(^{4}_{3}G)~^{4}G$ + 0.09~$4d^{10}(^{1}_{0}S)\,4f^{5}(^{2}_{5}H)~^{2}H$ + 0.07~$4d^{10}(^{1}_{0}S)\,4f^{5}(^{4}_{2}K)~^{4}K$ & 337400       & 41.8322  \\ 
85  & 3/2   & 0.15~$4d^{10}(^{1}_{0}S)\,4f^{5}(^{4}_{2}F)~^{4}F$ + 0.15~$4d^{10}(^{1}_{0}S)\,4f^{5}(^{6}_{0}P)~^{6}P$ + 0.11~$4d^{10}(^{1}_{0}S)\,4f^{5}(^{4}_{1}S)~^{4}S$ & 338191       & 41.9304  \\ 
86  & 13/2  & 0.29~$4d^{10}(^{1}_{0}S)\,4f^{5}(^{4}_{2}I)~^{4}I$ + 0.17~$4d^{10}(^{1}_{0}S)\,4f^{5}(^{4}_{1}H)~^{4}H$ + 0.11~$4d^{10}(^{1}_{0}S)\,4f^{5}(^{2}_{1}I)~^{2}I$ & 338913       & 42.0199  \\ 
87  & 7/2   & 0.33~$4d^{10}(^{1}_{0}S)\,4f^{5}(^{2}_{3}G)~^{2}G$ + 0.08~$4d^{10}(^{1}_{0}S)\,4f^{5}(^{2}_{6}G)~^{2}G$ + 0.08~$4d^{10}(^{1}_{0}S)\,4f^{5}(^{4}_{1}H)~^{4}H$ & 342206       & 42.4281  \\ 
88  & 9/2   & 0.19~$4d^{10}(^{1}_{0}S)\,4f^{5}(^{2}_{3}G)~^{2}G$ + 0.18~$4d^{10}(^{1}_{0}S)\,4f^{5}(^{2}_{3}H)~^{2}H$ + 0.16~$4d^{10}(^{1}_{0}S)\,4f^{5}(^{4}_{2}F)~^{4}F$ & 343071       & 42.5353  \\ 
89  & 15/2  & 0.37~$4d^{10}(^{1}_{0}S)\,4f^{5}(^{4}_{2}K)~^{4}K$ + 0.17~$4d^{10}(^{1}_{0}S)\,4f^{5}(^{2}_{3}L)~^{2}L$ + 0.12~$4d^{10}(^{1}_{0}S)\,4f^{5}(^{4}_{3}I)~^{4}I$ & 343380       & 42.5737  \\ 
90  & 3/2   & 0.22~$4d^{10}(^{1}_{0}S)\,4f^{5}(^{2}_{5}D)~^{2}D$ + 0.13~$4d^{10}(^{1}_{0}S)\,4f^{5}(^{4}_{1}P)~^{4}P$ + 0.10~$4d^{10}(^{1}_{0}S)\,4f^{5}(^{2}_{2}D)~^{2}D$ & 345345       & 42.8174  \\ 
91  & 7/2   & 0.31~$4d^{10}(^{1}_{0}S)\,4f^{5}(^{2}_{3}F)~^{2}F$ + 0.15~$4d^{10}(^{1}_{0}S)\,4f^{5}(^{4}_{2}F)~^{4}F$ + 0.10~$4d^{10}(^{1}_{0}S)\,4f^{5}(^{4}_{3}F)~^{4}F$ & 348864       & 43.2537  \\ 
92  & 11/2  & 0.16~$4d^{10}(^{1}_{0}S)\,4f^{5}(^{4}_{2}H)~^{4}H$ + 0.15~$4d^{10}(^{1}_{0}S)\,4f^{5}(^{2}_{3}H)~^{2}H$ + 0.12~$4d^{10}(^{1}_{0}S)\,4f^{5}(^{2}_{1}I)~^{2}I$ & 349695       & 43.3566  \\ 
93  & 13/2  & 0.18~$4d^{10}(^{1}_{0}S)\,4f^{5}(^{4}_{2}I)~^{4}I$ + 0.17~$4d^{10}(^{1}_{0}S)\,4f^{5}(^{4}_{3}H)~^{4}H$ + 0.13~$4d^{10}(^{1}_{0}S)\,4f^{5}(^{4}_{2}K)~^{4}K$ & 351677       & 43.6024  \\ 
94  & 9/2   & 0.32~$4d^{10}(^{1}_{0}S)\,4f^{5}(^{4}_{2}H)~^{4}H$ + 0.11~$4d^{10}(^{1}_{0}S)\,4f^{5}(^{2}_{2}G)~^{2}G$ + 0.06~$4d^{10}(^{1}_{0}S)\,4f^{5}(^{4}_{4}G)~^{4}G$ & 354141       & 43.9079  \\ 
95  & 11/2  & 0.24~$4d^{10}(^{1}_{0}S)\,4f^{5}(^{2}_{3}H)~^{2}H$ + 0.18~$4d^{10}(^{1}_{0}S)\,4f^{5}(^{2}_{3}I)~^{2}I$ + 0.08~$4d^{10}(^{1}_{0}S)\,4f^{5}(^{4}_{4}G)~^{4}G$ & 358623       & 44.4635  \\ 
96  & 5/2   & 0.13~$4d^{10}(^{1}_{0}S)\,4f^{5}(^{4}_{1}F)~^{4}F$ + 0.13~$4d^{10}(^{1}_{0}S)\,4f^{5}(^{2}_{3}D)~^{2}D$ + 0.10~$4d^{10}(^{1}_{0}S)\,4f^{5}(^{4}_{4}F)~^{4}F$ & 359230       & 44.5389  \\ 
97  & 17/2  & 0.72~$4d^{10}(^{1}_{0}S)\,4f^{5}(^{2}_{2}L)~^{2}L$ + 0.19~$4d^{10}(^{1}_{0}S)\,4f^{5}(^{2}_{3}L)~^{2}L$ + 0.02~$4d^{10}(^{1}_{0}S)\,4f^{5}(^{4}_{1}L)~^{4}L$ & 360334       & 44.6757  \\ 
98  & 1/2   & 0.32~$4d^{10}(^{1}_{0}S)\,4f^{5}(^{4}_{1}D)~^{4}D$ + 0.20~$4d^{10}(^{1}_{0}S)\,4f^{5}(^{4}_{2}D)~^{4}D$ + 0.15~$4d^{10}(^{1}_{0}S)\,4f^{5}(^{4}_{2}P)~^{4}P$ & 363353       & 45.0500  \\ 
99  & 5/2   & 0.15~$4d^{10}(^{1}_{0}S)\,4f^{5}(^{2}_{3}F)~^{2}F$ + 0.12~$4d^{10}(^{1}_{0}S)\,4f^{5}(^{2}_{4}D)~^{2}D$ + 0.12~$4d^{10}(^{1}_{0}S)\,4f^{5}(^{4}_{3}D)~^{4}D$ & 365337       & 45.2960  \\ 
100 & 9/2   & 0.27~$4d^{10}(^{1}_{0}S)\,4f^{5}(^{4}_{1}I)~^{4}I$ + 0.11~$4d^{10}(^{1}_{0}S)\,4f^{5}(^{4}_{3}F)~^{4}F$ + 0.09~$4d^{10}(^{1}_{0}S)\,4f^{5}(^{4}_{4}G)~^{4}G$ & 365412       & 45.3053  \\ 
101 & 15/2  & 0.21~$4d^{10}(^{1}_{0}S)\,4f^{5}(^{4}_{3}I)~^{4}I$ + 0.15~$4d^{10}(^{1}_{0}S)\,4f^{5}(^{6}_{0}H)~^{6}H$ + 0.13~$4d^{10}(^{1}_{0}S)\,4f^{5}(^{4}_{2}K)~^{4}K$ & 366432       & 45.4318  \\ 
102 & 13/2  & 0.20~$4d^{10}(^{1}_{0}S)\,4f^{5}(^{2}_{2}I)~^{2}I$ + 0.12~$4d^{10}(^{1}_{0}S)\,4f^{5}(^{2}_{4}I)~^{2}I$ + 0.10~$4d^{10}(^{1}_{0}S)\,4f^{5}(^{2}_{5}I)~^{2}I$ & 367945       & 45.6193  \\ 
103 & 7/2   & 0.17~$4d^{10}(^{1}_{0}S)\,4f^{5}(^{4}_{1}F)~^{4}F$ + 0.16~$4d^{10}(^{1}_{0}S)\,4f^{5}(^{2}_{7}F)~^{2}F$ + 0.11~$4d^{10}(^{1}_{0}S)\,4f^{5}(^{4}_{3}D)~^{4}D$ & 370765       & 45.9690  \\ 
104 & 3/2   & 0.18~$4d^{10}(^{1}_{0}S)\,4f^{5}(^{2}_{2}P)~^{2}P$ + 0.13~$4d^{10}(^{1}_{0}S)\,4f^{5}(^{4}_{2}F)~^{4}F$ + 0.12~$4d^{10}(^{1}_{0}S)\,4f^{5}(^{4}_{3}D)~^{4}D$ & 371118       & 46.0128  \\ 
105 & 11/2  & 0.31~$4d^{10}(^{1}_{0}S)\,4f^{5}(^{2}_{2}I)~^{2}I$ + 0.12~$4d^{10}(^{1}_{0}S)\,4f^{5}(^{4}_{2}H)~^{4}H$ + 0.11~$4d^{10}(^{1}_{0}S)\,4f^{5}(^{4}_{1}G)~^{4}G$ & 371873       & 46.1064  \\ 
106 & 5/2   & 0.20~$4d^{10}(^{1}_{0}S)\,4f^{5}(^{4}_{1}P)~^{4}P$ + 0.11~$4d^{10}(^{1}_{0}S)\,4f^{5}(^{4}_{1}D)~^{4}D$ + 0.09~$4d^{10}(^{1}_{0}S)\,4f^{5}(^{2}_{4}D)~^{2}D$ & 375770       & 46.5895  \\ 
107 & 9/2   & 0.33~$4d^{10}(^{1}_{0}S)\,4f^{5}(^{2}_{5}H)~^{2}H$ + 0.14~$4d^{10}(^{1}_{0}S)\,4f^{5}(^{4}_{1}H)~^{4}H$ + 0.09~$4d^{10}(^{1}_{0}S)\,4f^{5}(^{4}_{2}H)~^{4}H$ & 376724       & 46.7078  \\ 
108 & 13/2  & 0.23~$4d^{10}(^{1}_{0}S)\,4f^{5}(^{4}_{3}I)~^{4}I$ + 0.16~$4d^{10}(^{1}_{0}S)\,4f^{5}(^{2}_{1}K)~^{2}K$ + 0.12~$4d^{10}(^{1}_{0}S)\,4f^{5}(^{2}_{4}K)~^{2}K$ & 376931       & 46.7334  \\ 
109 & 7/2   & 0.17~$4d^{10}(^{1}_{0}S)\,4f^{5}(^{2}_{4}G)~^{2}G$ + 0.12~$4d^{10}(^{1}_{0}S)\,4f^{5}(^{4}_{2}H)~^{4}H$ + 0.08~$4d^{10}(^{1}_{0}S)\,4f^{5}(^{4}_{3}D)~^{4}D$ & 384710       & 47.6980  \\ 
110 & 7/2   & 0.13~$4d^{10}(^{1}_{0}S)\,4f^{5}(^{4}_{4}F)~^{4}F$ + 0.09~$4d^{10}(^{1}_{0}S)\,4f^{5}(^{4}_{2}F)~^{4}F$ + 0.09~$4d^{10}(^{1}_{0}S)\,4f^{5}(^{4}_{3}H)~^{4}H$ & 386749       & 47.9507  \\ 
111 & 9/2   & 0.17~$4d^{10}(^{1}_{0}S)\,4f^{5}(^{2}_{1}H)~^{2}H$ + 0.12~$4d^{10}(^{1}_{0}S)\,4f^{5}(^{2}_{4}G)~^{2}G$ + 0.09~$4d^{10}(^{1}_{0}S)\,4f^{5}(^{4}_{1}G)~^{4}G$ & 387146       & 48.0000  \\ 
112 & 15/2  & 0.24~$4d^{10}(^{1}_{0}S)\,4f^{5}(^{2}_{5}K)~^{2}K$ + 0.14~$4d^{10}(^{1}_{0}S)\,4f^{5}(^{2}_{3}K)~^{2}K$ + 0.12~$4d^{10}(^{1}_{0}S)\,4f^{5}(^{2}_{4}K)~^{2}K$ & 387150       & 48.0004  \\ 
113 & 11/2  & 0.18~$4d^{10}(^{1}_{0}S)\,4f^{5}(^{4}_{2}G)~^{4}G$ + 0.11~$4d^{10}(^{1}_{0}S)\,4f^{5}(^{2}_{3}I)~^{2}I$ + 0.10~$4d^{10}(^{1}_{0}S)\,4f^{5}(^{2}_{2}I)~^{2}I$ & 388268       & 48.1391  \\ 
114 & 17/2  & 0.28~$4d^{10}(^{1}_{0}S)\,4f^{5}(^{4}_{1}K)~^{4}K$ + 0.25~$4d^{10}(^{1}_{0}S)\,4f^{5}(^{2}_{2}M)~^{2}M$ + 0.13~$4d^{10}(^{1}_{0}S)\,4f^{5}(^{2}_{1}L)~^{2}L$ & 388995       & 48.2293  \\ 
115 & 1/2   & 0.36~$4d^{10}(^{1}_{0}S)\,4f^{5}(^{4}_{1}P)~^{4}P$ + 0.19~$4d^{10}(^{1}_{0}S)\,4f^{5}(^{2}_{2}P)~^{2}P$ + 0.14~$4d^{10}(^{1}_{0}S)\,4f^{5}(^{4}_{2}D)~^{4}D$ & 389298       & 48.2668  \\ 
116 & 13/2  & 0.28~$4d^{10}(^{1}_{0}S)\,4f^{5}(^{4}_{2}H)~^{4}H$ + 0.13~$4d^{10}(^{1}_{0}S)\,4f^{5}(^{4}_{2}K)~^{4}K$ + 0.11~$4d^{10}(^{1}_{0}S)\,4f^{5}(^{2}_{3}I)~^{2}I$ & 394668       & 48.9326  \\ 
117 & 3/2   & 0.16~$4d^{10}(^{1}_{0}S)\,4f^{5}(^{4}_{4}F)~^{4}F$ + 0.13~$4d^{10}(^{1}_{0}S)\,4f^{5}(^{2}_{5}D)~^{2}D$ + 0.10~$4d^{10}(^{1}_{0}S)\,4f^{5}(^{2}_{2}D)~^{2}D$ & 399680       & 49.5540  \\ 
118 & 9/2   & 0.14~$4d^{10}(^{1}_{0}S)\,4f^{5}(^{2}_{6}G)~^{2}G$ + 0.12~$4d^{10}(^{1}_{0}S)\,4f^{5}(^{4}_{3}H)~^{4}H$ + 0.10~$4d^{10}(^{1}_{0}S)\,4f^{5}(^{2}_{2}G)~^{2}G$ & 400497       & 49.6553  \\ 
119 & 7/2   & 0.11~$4d^{10}(^{1}_{0}S)\,4f^{5}(^{4}_{1}H)~^{4}H$ + 0.10~$4d^{10}(^{1}_{0}S)\,4f^{5}(^{4}_{1}F)~^{4}F$ + 0.07~$4d^{10}(^{1}_{0}S)\,4f^{5}(^{2}_{3}F)~^{2}F$ & 400987       & 49.7160  \\ 
120 & 7/2   & 0.12~$4d^{10}(^{1}_{0}S)\,4f^{5}(^{4}_{2}H)~^{4}H$ + 0.09~$4d^{10}(^{1}_{0}S)\,4f^{5}(^{4}_{4}G)~^{4}G$ + 0.08~$4d^{10}(^{1}_{0}S)\,4f^{5}(^{4}_{3}G)~^{4}G$ & 403280       & 50.0003  \\ 
121 & 17/2  & 0.48~$4d^{10}(^{1}_{0}S)\,4f^{5}(^{4}_{2}K)~^{4}K$ + 0.26~$4d^{10}(^{1}_{0}S)\,4f^{5}(^{2}_{2}M)~^{2}M$ + 0.09~$4d^{10}(^{1}_{0}S)\,4f^{5}(^{4}_{1}L)~^{4}L$ & 403661       & 50.0475  \\ 
122 & 19/2  & 0.32~$4d^{10}(^{1}_{0}S)\,4f^{5}(^{2}_{1}M)~^{2}M$ + 0.30~$4d^{10}(^{1}_{0}S)\,4f^{5}(^{4}_{1}L)~^{4}L$ + 0.18~$4d^{10}(^{1}_{0}S)\,4f^{5}(^{2}_{2}M)~^{2}M$ & 406027       & 50.3410  \\ 
123 & 5/2   & 0.16~$4d^{10}(^{1}_{0}S)\,4f^{5}(^{4}_{2}P)~^{4}P$ + 0.12~$4d^{10}(^{1}_{0}S)\,4f^{5}(^{4}_{2}F)~^{4}F$ + 0.11~$4d^{10}(^{1}_{0}S)\,4f^{5}(^{4}_{4}F)~^{4}F$ & 406121       & 50.3525  \\ 
124 & 21/2  & 0.54~$4d^{10}(^{1}_{0}S)\,4f^{5}(^{2}_{1}N)~^{2}N$ + 0.24~$4d^{10}(^{1}_{0}S)\,4f^{5}(^{2}_{0}O)~^{2}O$ + 0.21~$4d^{10}(^{1}_{0}S)\,4f^{5}(^{4}_{0}M)~^{4}M$ & 408548       & 50.6536  \\ 
125 & 13/2  & 0.29~$4d^{10}(^{1}_{0}S)\,4f^{5}(^{2}_{2}K)~^{2}K$ + 0.24~$4d^{10}(^{1}_{0}S)\,4f^{5}(^{2}_{2}I)~^{2}I$ + 0.17~$4d^{10}(^{1}_{0}S)\,4f^{5}(^{2}_{3}I)~^{2}I$ & 411044       & 50.9630  \\ 
126 & 9/2   & 0.13~$4d^{10}(^{1}_{0}S)\,4f^{5}(^{2}_{6}G)~^{2}G$ + 0.13~$4d^{10}(^{1}_{0}S)\,4f^{5}(^{4}_{3}G)~^{4}G$ + 0.09~$4d^{10}(^{1}_{0}S)\,4f^{5}(^{4}_{1}G)~^{4}G$ & 411064       & 50.9655  \\ 
127 & 19/2  & 0.61~$4d^{10}(^{1}_{0}S)\,4f^{5}(^{2}_{2}M)~^{2}M$ + 0.36~$4d^{10}(^{1}_{0}S)\,4f^{5}(^{2}_{1}M)~^{2}M$ & 412462       & 51.1388  \\ 
128 & 11/2  & 0.23~$4d^{10}(^{1}_{0}S)\,4f^{5}(^{4}_{1}H)~^{4}H$ + 0.12~$4d^{10}(^{1}_{0}S)\,4f^{5}(^{2}_{3}H)~^{2}H$ + 0.07~$4d^{10}(^{1}_{0}S)\,4f^{5}(^{2}_{4}H)~^{2}H$ & 413336       & 51.2471  \\ 
129 & 1/2   & 0.30~$4d^{10}(^{1}_{0}S)\,4f^{5}(^{2}_{2}P)~^{2}P$ + 0.18~$4d^{10}(^{1}_{0}S)\,4f^{5}(^{4}_{3}D)~^{4}D$ + 0.15~$4d^{10}(^{1}_{0}S)\,4f^{5}(^{2}_{4}P)~^{2}P$ & 414336       & 51.3711  \\ 
130 & 5/2   & 0.17~$4d^{10}(^{1}_{0}S)\,4f^{5}(^{4}_{1}D)~^{4}D$ + 0.16~$4d^{10}(^{1}_{0}S)\,4f^{5}(^{2}_{4}F)~^{2}F$ + 0.14~$4d^{10}(^{1}_{0}S)\,4f^{5}(^{2}_{1}D)~^{2}D$ & 418426       & 51.8782  \\ 
131 & 15/2  & 0.34~$4d^{10}(^{1}_{0}S)\,4f^{5}(^{2}_{3}K)~^{2}K$ + 0.14~$4d^{10}(^{1}_{0}S)\,4f^{5}(^{2}_{4}K)~^{2}K$ + 0.14~$4d^{10}(^{1}_{0}S)\,4f^{5}(^{4}_{2}K)~^{4}K$ & 418804       & 51.9251  \\ 
132 & 11/2  & 0.22~$4d^{10}(^{1}_{0}S)\,4f^{5}(^{2}_{5}H)~^{2}H$ + 0.15~$4d^{10}(^{1}_{0}S)\,4f^{5}(^{2}_{5}I)~^{2}I$ + 0.11~$4d^{10}(^{1}_{0}S)\,4f^{5}(^{4}_{2}H)~^{4}H$ & 420383       & 52.1209  \\ 
133 & 3/2   & 0.29~$4d^{10}(^{1}_{0}S)\,4f^{5}(^{4}_{1}D)~^{4}D$ + 0.29~$4d^{10}(^{1}_{0}S)\,4f^{5}(^{4}_{1}F)~^{4}F$ + 0.15~$4d^{10}(^{1}_{0}S)\,4f^{5}(^{4}_{2}D)~^{4}D$ & 421095       & 52.2091  \\ 
134 & 9/2   & 0.28~$4d^{10}(^{1}_{0}S)\,4f^{5}(^{4}_{4}F)~^{4}F$ + 0.18~$4d^{10}(^{1}_{0}S)\,4f^{5}(^{4}_{2}H)~^{4}H$ + 0.10~$4d^{10}(^{1}_{0}S)\,4f^{5}(^{4}_{1}F)~^{4}F$ & 423220       & 52.4726  \\ 
135 & 5/2   & 0.13~$4d^{10}(^{1}_{0}S)\,4f^{5}(^{2}_{1}F)~^{2}F$ + 0.13~$4d^{10}(^{1}_{0}S)\,4f^{5}(^{2}_{7}F)~^{2}F$ + 0.12~$4d^{10}(^{1}_{0}S)\,4f^{5}(^{4}_{1}F)~^{4}F$ & 426066       & 52.8255  \\ 
136 & 13/2  & 0.19~$4d^{10}(^{1}_{0}S)\,4f^{5}(^{2}_{3}I)~^{2}I$ + 0.14~$4d^{10}(^{1}_{0}S)\,4f^{5}(^{2}_{2}I)~^{2}I$ + 0.11~$4d^{10}(^{1}_{0}S)\,4f^{5}(^{2}_{1}I)~^{2}I$ & 431042       & 53.4424  \\ 
137 & 9/2   & 0.18~$4d^{10}(^{1}_{0}S)\,4f^{5}(^{2}_{4}H)~^{2}H$ + 0.17~$4d^{10}(^{1}_{0}S)\,4f^{5}(^{4}_{1}F)~^{4}F$ + 0.10~$4d^{10}(^{1}_{0}S)\,4f^{5}(^{2}_{7}H)~^{2}H$ & 431992       & 53.5602  \\ 
138 & 11/2  & 0.25~$4d^{10}(^{1}_{0}S)\,4f^{5}(^{2}_{4}I)~^{2}I$ + 0.20~$4d^{10}(^{1}_{0}S)\,4f^{5}(^{4}_{2}H)~^{4}H$ + 0.18~$4d^{10}(^{1}_{0}S)\,4f^{5}(^{4}_{3}G)~^{4}G$ & 433367       & 53.7307  \\ 
139 & 15/2  & 0.30~$4d^{10}(^{1}_{0}S)\,4f^{5}(^{2}_{2}K)~^{2}K$ + 0.17~$4d^{10}(^{1}_{0}S)\,4f^{5}(^{2}_{1}L)~^{2}L$ + 0.14~$4d^{10}(^{1}_{0}S)\,4f^{5}(^{2}_{2}L)~^{2}L$ & 433924       & 53.7998  \\ 
140 & 3/2   & 0.19~$4d^{10}(^{1}_{0}S)\,4f^{5}(^{4}_{2}P)~^{4}P$ + 0.18~$4d^{10}(^{1}_{0}S)\,4f^{5}(^{4}_{2}F)~^{4}F$ + 0.09~$4d^{10}(^{1}_{0}S)\,4f^{5}(^{2}_{4}P)~^{2}P$ & 435170       & 53.9542  \\ 
141 & 7/2   & 0.22~$4d^{10}(^{1}_{0}S)\,4f^{5}(^{2}_{3}G)~^{2}G$ + 0.20~$4d^{10}(^{1}_{0}S)\,4f^{5}(^{2}_{5}F)~^{2}F$ + 0.10~$4d^{10}(^{1}_{0}S)\,4f^{5}(^{2}_{2}F)~^{2}F$ & 436081       & 54.0672  \\ 
142 & 1/2   & 0.28~$4d^{10}(^{1}_{0}S)\,4f^{5}(^{2}_{2}P)~^{2}P$ + 0.26~$4d^{10}(^{1}_{0}S)\,4f^{5}(^{4}_{3}D)~^{4}D$ + 0.25~$4d^{10}(^{1}_{0}S)\,4f^{5}(^{4}_{1}D)~^{4}D$ & 441616       & 54.7534  \\ 
143 & 7/2   & 0.30~$4d^{10}(^{1}_{0}S)\,4f^{5}(^{2}_{4}F)~^{2}F$ + 0.11~$4d^{10}(^{1}_{0}S)\,4f^{5}(^{2}_{5}F)~^{2}F$ + 0.11~$4d^{10}(^{1}_{0}S)\,4f^{5}(^{4}_{3}D)~^{4}D$ & 445673       & 55.2563  \\ 
144 & 3/2   & 0.21~$4d^{10}(^{1}_{0}S)\,4f^{5}(^{4}_{1}P)~^{4}P$ + 0.15~$4d^{10}(^{1}_{0}S)\,4f^{5}(^{4}_{1}S)~^{4}S$ + 0.13~$4d^{10}(^{1}_{0}S)\,4f^{5}(^{4}_{4}F)~^{4}F$ & 447007       & 55.4219  \\ 
145 & 9/2   & 0.17~$4d^{10}(^{1}_{0}S)\,4f^{5}(^{2}_{3}G)~^{2}G$ + 0.16~$4d^{10}(^{1}_{0}S)\,4f^{5}(^{2}_{4}H)~^{2}H$ + 0.11~$4d^{10}(^{1}_{0}S)\,4f^{5}(^{4}_{3}F)~^{4}F$ & 447321       & 55.4607  \\ 
146 & 13/2  & 0.23~$4d^{10}(^{1}_{0}S)\,4f^{5}(^{2}_{4}I)~^{2}I$ + 0.17~$4d^{10}(^{1}_{0}S)\,4f^{5}(^{2}_{2}I)~^{2}I$ + 0.14~$4d^{10}(^{1}_{0}S)\,4f^{5}(^{2}_{5}K)~^{2}K$ & 448301       & 55.5823  \\ 
147 & 15/2  & 0.24~$4d^{10}(^{1}_{0}S)\,4f^{5}(^{2}_{3}L)~^{2}L$ + 0.21~$4d^{10}(^{1}_{0}S)\,4f^{5}(^{4}_{3}I)~^{4}I$ + 0.20~$4d^{10}(^{1}_{0}S)\,4f^{5}(^{4}_{1}I)~^{4}I$ & 449525       & 55.7340  \\ 
148 & 11/2  & 0.29~$4d^{10}(^{1}_{0}S)\,4f^{5}(^{4}_{1}I)~^{4}I$ + 0.16~$4d^{10}(^{1}_{0}S)\,4f^{5}(^{2}_{6}H)~^{2}H$ + 0.13~$4d^{10}(^{1}_{0}S)\,4f^{5}(^{4}_{1}G)~^{4}G$ & 451255       & 55.9485  \\ 
149 & 5/2   & 0.32~$4d^{10}(^{1}_{0}S)\,4f^{5}(^{2}_{3}F)~^{2}F$ + 0.17~$4d^{10}(^{1}_{0}S)\,4f^{5}(^{2}_{7}F)~^{2}F$ + 0.14~$4d^{10}(^{1}_{0}S)\,4f^{5}(^{2}_{3}D)~^{2}D$ & 451765       & 56.0117  \\ 
150 & 9/2   & 0.14~$4d^{10}(^{1}_{0}S)\,4f^{5}(^{4}_{3}F)~^{4}F$ + 0.11~$4d^{10}(^{1}_{0}S)\,4f^{5}(^{2}_{3}H)~^{2}H$ + 0.10~$4d^{10}(^{1}_{0}S)\,4f^{5}(^{2}_{3}G)~^{2}G$ & 454196       & 56.3131  \\ 
151 & 5/2   & 0.19~$4d^{10}(^{1}_{0}S)\,4f^{5}(^{4}_{2}D)~^{4}D$ + 0.18~$4d^{10}(^{1}_{0}S)\,4f^{5}(^{4}_{1}G)~^{4}G$ + 0.14~$4d^{10}(^{1}_{0}S)\,4f^{5}(^{4}_{1}P)~^{4}P$ & 456724       & 56.6266  \\ 
152 & 3/2   & 0.20~$4d^{10}(^{1}_{0}S)\,4f^{5}(^{2}_{3}P)~^{2}P$ + 0.12~$4d^{10}(^{1}_{0}S)\,4f^{5}(^{4}_{1}S)~^{4}S$ + 0.11~$4d^{10}(^{1}_{0}S)\,4f^{5}(^{2}_{1}D)~^{2}D$ & 457495       & 56.7222  \\ 
153 & 7/2   & 0.17~$4d^{10}(^{1}_{0}S)\,4f^{5}(^{2}_{5}G)~^{2}G$ + 0.13~$4d^{10}(^{1}_{0}S)\,4f^{5}(^{2}_{4}G)~^{2}G$ + 0.11~$4d^{10}(^{1}_{0}S)\,4f^{5}(^{2}_{6}G)~^{2}G$ & 457513       & 56.7244  \\ 
154 & 13/2  & 0.38~$4d^{10}(^{1}_{0}S)\,4f^{5}(^{2}_{5}I)~^{2}I$ + 0.10~$4d^{10}(^{1}_{0}S)\,4f^{5}(^{4}_{3}H)~^{4}H$ + 0.09~$4d^{10}(^{1}_{0}S)\,4f^{5}(^{4}_{2}K)~^{4}K$ & 461925       & 57.2714  \\ 
155 & 11/2  & 0.27~$4d^{10}(^{1}_{0}S)\,4f^{5}(^{2}_{1}I)~^{2}I$ + 0.14~$4d^{10}(^{1}_{0}S)\,4f^{5}(^{4}_{1}G)~^{4}G$ + 0.14~$4d^{10}(^{1}_{0}S)\,4f^{5}(^{4}_{3}G)~^{4}G$ & 468166       & 58.0452  \\ 
156 & 5/2   & 0.21~$4d^{10}(^{1}_{0}S)\,4f^{5}(^{4}_{1}F)~^{4}F$ + 0.18~$4d^{10}(^{1}_{0}S)\,4f^{5}(^{2}_{3}F)~^{2}F$ + 0.10~$4d^{10}(^{1}_{0}S)\,4f^{5}(^{2}_{7}F)~^{2}F$ & 470554       & 58.3413  \\ 
157 & 9/2   & 0.18~$4d^{10}(^{1}_{0}S)\,4f^{5}(^{2}_{7}H)~^{2}H$ + 0.15~$4d^{10}(^{1}_{0}S)\,4f^{5}(^{2}_{3}H)~^{2}H$ + 0.12~$4d^{10}(^{1}_{0}S)\,4f^{5}(^{2}_{6}H)~^{2}H$ & 471029       & 58.4002  \\ 
158 & 7/2   & 0.24~$4d^{10}(^{1}_{0}S)\,4f^{5}(^{2}_{4}G)~^{2}G$ + 0.14~$4d^{10}(^{1}_{0}S)\,4f^{5}(^{2}_{6}G)~^{2}G$ + 0.10~$4d^{10}(^{1}_{0}S)\,4f^{5}(^{2}_{4}F)~^{2}F$ & 474761       & 58.8629  \\ 
159 & 3/2   & 0.25~$4d^{10}(^{1}_{0}S)\,4f^{5}(^{2}_{4}D)~^{2}D$ + 0.14~$4d^{10}(^{1}_{0}S)\,4f^{5}(^{2}_{2}P)~^{2}P$ + 0.11~$4d^{10}(^{1}_{0}S)\,4f^{5}(^{4}_{1}S)~^{4}S$ & 475131       & 58.9087  \\ 
160 & 11/2  & 0.23~$4d^{10}(^{1}_{0}S)\,4f^{5}(^{2}_{5}H)~^{2}H$ + 0.12~$4d^{10}(^{1}_{0}S)\,4f^{5}(^{2}_{6}H)~^{2}H$ + 0.10~$4d^{10}(^{1}_{0}S)\,4f^{5}(^{2}_{5}I)~^{2}I$ & 480938       & 59.6287  \\ 
161 & 15/2  & 0.25~$4d^{10}(^{1}_{0}S)\,4f^{5}(^{2}_{1}L)~^{2}L$ + 0.14~$4d^{10}(^{1}_{0}S)\,4f^{5}(^{2}_{4}K)~^{2}K$ + 0.14~$4d^{10}(^{1}_{0}S)\,4f^{5}(^{4}_{1}I)~^{4}I$ & 494281       & 61.2830  \\ 
162 & 5/2   & 0.19~$4d^{10}(^{1}_{0}S)\,4f^{5}(^{2}_{5}F)~^{2}F$ + 0.12~$4d^{10}(^{1}_{0}S)\,4f^{5}(^{4}_{4}G)~^{4}G$ + 0.11~$4d^{10}(^{1}_{0}S)\,4f^{5}(^{4}_{1}D)~^{4}D$ & 495243       & 61.4023  \\ 
163 & 13/2  & 0.35~$4d^{10}(^{1}_{0}S)\,4f^{5}(^{4}_{1}I)~^{4}I$ + 0.12~$4d^{10}(^{1}_{0}S)\,4f^{5}(^{2}_{5}I)~^{2}I$ + 0.08~$4d^{10}(^{1}_{0}S)\,4f^{5}(^{2}_{1}K)~^{2}K$ & 498809       & 61.8444  \\ 
164 & 7/2   & 0.18~$4d^{10}(^{1}_{0}S)\,4f^{5}(^{4}_{1}G)~^{4}G$ + 0.15~$4d^{10}(^{1}_{0}S)\,4f^{5}(^{4}_{1}F)~^{4}F$ + 0.13~$4d^{10}(^{1}_{0}S)\,4f^{5}(^{2}_{6}G)~^{2}G$ & 500479       & 62.0515  \\ 
165 & 11/2  & 0.37~$4d^{10}(^{1}_{0}S)\,4f^{5}(^{2}_{1}H)~^{2}H$ + 0.14~$4d^{10}(^{1}_{0}S)\,4f^{5}(^{2}_{1}I)~^{2}I$ + 0.10~$4d^{10}(^{1}_{0}S)\,4f^{5}(^{2}_{7}H)~^{2}H$ & 502118       & 62.2547  \\ 
166 & 7/2   & 0.27~$4d^{10}(^{1}_{0}S)\,4f^{5}(^{4}_{1}D)~^{4}D$ + 0.26~$4d^{10}(^{1}_{0}S)\,4f^{5}(^{2}_{2}G)~^{2}G$ + 0.07~$4d^{10}(^{1}_{0}S)\,4f^{5}(^{4}_{2}H)~^{4}H$ & 505232       & 62.6408  \\ 
167 & 5/2   & 0.25~$4d^{10}(^{1}_{0}S)\,4f^{5}(^{2}_{6}F)~^{2}F$ + 0.15~$4d^{10}(^{1}_{0}S)\,4f^{5}(^{4}_{1}G)~^{4}G$ + 0.12~$4d^{10}(^{1}_{0}S)\,4f^{5}(^{2}_{2}D)~^{2}D$ & 505252       & 62.6432  \\ 
168 & 3/2   & 0.18~$4d^{10}(^{1}_{0}S)\,4f^{5}(^{4}_{1}D)~^{4}D$ + 0.18~$4d^{10}(^{1}_{0}S)\,4f^{5}(^{2}_{4}P)~^{2}P$ + 0.12~$4d^{10}(^{1}_{0}S)\,4f^{5}(^{2}_{3}P)~^{2}P$ & 513160       & 63.6238  \\ 
169 & 1/2   & 0.34~$4d^{10}(^{1}_{0}S)\,4f^{5}(^{2}_{4}P)~^{2}P$ + 0.23~$4d^{10}(^{1}_{0}S)\,4f^{5}(^{2}_{3}P)~^{2}P$ + 0.17~$4d^{10}(^{1}_{0}S)\,4f^{5}(^{2}_{1}P)~^{2}P$ & 514872       & 63.8360  \\ 
170 & 9/2   & 0.19~$4d^{10}(^{1}_{0}S)\,4f^{5}(^{4}_{1}F)~^{4}F$ + 0.15~$4d^{10}(^{1}_{0}S)\,4f^{5}(^{2}_{5}G)~^{2}G$ + 0.15~$4d^{10}(^{1}_{0}S)\,4f^{5}(^{2}_{2}H)~^{2}H$ & 515040       & 63.8569  \\ 
171 & 15/2  & 0.48~$4d^{10}(^{1}_{0}S)\,4f^{5}(^{2}_{1}K)~^{2}K$ + 0.17~$4d^{10}(^{1}_{0}S)\,4f^{5}(^{4}_{1}I)~^{4}I$ + 0.14~$4d^{10}(^{1}_{0}S)\,4f^{5}(^{2}_{4}K)~^{2}K$ & 519238       & 64.3773  \\ 
172 & 17/2  & 0.58~$4d^{10}(^{1}_{0}S)\,4f^{5}(^{2}_{1}L)~^{2}L$ + 0.21~$4d^{10}(^{1}_{0}S)\,4f^{5}(^{2}_{2}M)~^{2}M$ + 0.10~$4d^{10}(^{1}_{0}S)\,4f^{5}(^{4}_{2}K)~^{4}K$ & 522902       & 64.8315  \\ 
173 & 9/2   & 0.17~$4d^{10}(^{1}_{0}S)\,4f^{5}(^{2}_{2}G)~^{2}G$ + 0.10~$4d^{10}(^{1}_{0}S)\,4f^{5}(^{4}_{1}F)~^{4}F$ + 0.09~$4d^{10}(^{1}_{0}S)\,4f^{5}(^{2}_{1}G)~^{2}G$ & 525777       & 65.1881  \\ 
174 & 13/2  & 0.44~$4d^{10}(^{1}_{0}S)\,4f^{5}(^{2}_{1}I)~^{2}I$ + 0.16~$4d^{10}(^{1}_{0}S)\,4f^{5}(^{2}_{2}K)~^{2}K$ + 0.12~$4d^{10}(^{1}_{0}S)\,4f^{5}(^{4}_{2}H)~^{4}H$ & 529134       & 65.6043  \\ 
175 & 9/2   & 0.21~$4d^{10}(^{1}_{0}S)\,4f^{5}(^{2}_{7}H)~^{2}H$ + 0.17~$4d^{10}(^{1}_{0}S)\,4f^{5}(^{2}_{1}H)~^{2}H$ + 0.11~$4d^{10}(^{1}_{0}S)\,4f^{5}(^{2}_{2}G)~^{2}G$ & 531591       & 65.9088  \\ 
176 & 7/2   & 0.15~$4d^{10}(^{1}_{0}S)\,4f^{5}(^{4}_{4}G)~^{4}G$ + 0.13~$4d^{10}(^{1}_{0}S)\,4f^{5}(^{4}_{1}F)~^{4}F$ + 0.11~$4d^{10}(^{1}_{0}S)\,4f^{5}(^{2}_{2}F)~^{2}F$ & 532063       & 65.9674  \\ 
177 & 11/2  & 0.16~$4d^{10}(^{1}_{0}S)\,4f^{5}(^{2}_{1}I)~^{2}I$ + 0.13~$4d^{10}(^{1}_{0}S)\,4f^{5}(^{4}_{1}G)~^{4}G$ + 0.11~$4d^{10}(^{1}_{0}S)\,4f^{5}(^{2}_{6}H)~^{2}H$ & 535205       & 66.3569  \\ 
178 & 3/2   & 0.21~$4d^{10}(^{1}_{0}S)\,4f^{5}(^{2}_{2}P)~^{2}P$ + 0.20~$4d^{10}(^{1}_{0}S)\,4f^{5}(^{2}_{4}D)~^{2}D$ + 0.11~$4d^{10}(^{1}_{0}S)\,4f^{5}(^{2}_{3}D)~^{2}D$ & 536820       & 66.5572  \\ 
179 & 9/2   & 0.24~$4d^{10}(^{1}_{0}S)\,4f^{5}(^{4}_{1}G)~^{4}G$ + 0.12~$4d^{10}(^{1}_{0}S)\,4f^{5}(^{2}_{1}G)~^{2}G$ + 0.09~$4d^{10}(^{1}_{0}S)\,4f^{5}(^{2}_{2}H)~^{2}H$ & 543184       & 67.3463  \\ 
180 & 15/2  & 0.28~$4d^{10}(^{1}_{0}S)\,4f^{5}(^{2}_{5}K)~^{2}K$ + 0.18~$4d^{10}(^{1}_{0}S)\,4f^{5}(^{2}_{1}K)~^{2}K$ + 0.14~$4d^{10}(^{1}_{0}S)\,4f^{5}(^{4}_{1}I)~^{4}I$ & 543477       & 67.3826  \\ 
181 & 5/2   & 0.12~$4d^{10}(^{1}_{0}S)\,4f^{5}(^{2}_{4}D)~^{2}D$ + 0.12~$4d^{10}(^{1}_{0}S)\,4f^{5}(^{2}_{7}F)~^{2}F$ + 0.10~$4d^{10}(^{1}_{0}S)\,4f^{5}(^{2}_{4}F)~^{2}F$ & 555103       & 68.8240  \\ 
182 & 3/2   & 0.39~$4d^{10}(^{1}_{0}S)\,4f^{5}(^{2}_{1}D)~^{2}D$ + 0.12~$4d^{10}(^{1}_{0}S)\,4f^{5}(^{4}_{1}F)~^{4}F$ + 0.10~$4d^{10}(^{1}_{0}S)\,4f^{5}(^{2}_{4}P)~^{2}P$ & 558270       & 69.2167  \\ 
183 & 11/2  & 0.17~$4d^{10}(^{1}_{0}S)\,4f^{5}(^{2}_{4}I)~^{2}I$ + 0.17~$4d^{10}(^{1}_{0}S)\,4f^{5}(^{2}_{7}H)~^{2}H$ + 0.13~$4d^{10}(^{1}_{0}S)\,4f^{5}(^{2}_{1}H)~^{2}H$ & 560638       & 69.5103  \\ 
184 & 7/2   & 0.21~$4d^{10}(^{1}_{0}S)\,4f^{5}(^{2}_{1}G)~^{2}G$ + 0.16~$4d^{10}(^{1}_{0}S)\,4f^{5}(^{2}_{6}F)~^{2}F$ + 0.13~$4d^{10}(^{1}_{0}S)\,4f^{5}(^{2}_{2}G)~^{2}G$ & 574583       & 71.2392  \\ 
185 & 5/2   & 0.45~$4d^{10}(^{1}_{0}S)\,4f^{5}(^{2}_{2}F)~^{2}F$ + 0.17~$4d^{10}(^{1}_{0}S)\,4f^{5}(^{2}_{6}F)~^{2}F$ + 0.08~$4d^{10}(^{1}_{0}S)\,4f^{5}(^{4}_{4}F)~^{4}F$ & 575751       & 71.3841  \\ 
186 & 13/2  & 0.37~$4d^{10}(^{1}_{0}S)\,4f^{5}(^{2}_{4}K)~^{2}K$ + 0.26~$4d^{10}(^{1}_{0}S)\,4f^{5}(^{2}_{1}K)~^{2}K$ + 0.20~$4d^{10}(^{1}_{0}S)\,4f^{5}(^{2}_{5}I)~^{2}I$ & 576453       & 71.4711  \\ 
187 & 9/2   & 0.31~$4d^{10}(^{1}_{0}S)\,4f^{5}(^{2}_{1}G)~^{2}G$ + 0.22~$4d^{10}(^{1}_{0}S)\,4f^{5}(^{2}_{4}G)~^{2}G$ + 0.21~$4d^{10}(^{1}_{0}S)\,4f^{5}(^{2}_{1}H)~^{2}H$ & 579634       & 71.8655  \\ 
188 & 3/2   & 0.46~$4d^{10}(^{1}_{0}S)\,4f^{5}(^{2}_{1}P)~^{2}P$ + 0.15~$4d^{10}(^{1}_{0}S)\,4f^{5}(^{2}_{2}D)~^{2}D$ + 0.13~$4d^{10}(^{1}_{0}S)\,4f^{5}(^{2}_{5}D)~^{2}D$ & 588320       & 72.9423  \\ 
189 & 7/2   & 0.38~$4d^{10}(^{1}_{0}S)\,4f^{5}(^{2}_{2}F)~^{2}F$ + 0.19~$4d^{10}(^{1}_{0}S)\,4f^{5}(^{2}_{6}G)~^{2}G$ + 0.13~$4d^{10}(^{1}_{0}S)\,4f^{5}(^{2}_{1}F)~^{2}F$ & 590723       & 73.2403  \\ 
190 & 5/2   & 0.24~$4d^{10}(^{1}_{0}S)\,4f^{5}(^{2}_{5}D)~^{2}D$ + 0.17~$4d^{10}(^{1}_{0}S)\,4f^{5}(^{2}_{7}F)~^{2}F$ + 0.06~$4d^{10}(^{1}_{0}S)\,4f^{5}(^{2}_{5}F)~^{2}F$ & 604231       & 74.9151  \\ 
191 & 1/2   & 0.56~$4d^{10}(^{1}_{0}S)\,4f^{5}(^{2}_{1}P)~^{2}P$ + 0.23~$4d^{10}(^{1}_{0}S)\,4f^{5}(^{2}_{3}P)~^{2}P$ + 0.14~$4d^{10}(^{1}_{0}S)\,4f^{5}(^{2}_{2}P)~^{2}P$ & 616076       & 76.3837  \\ 
192 & 7/2   & 0.27~$4d^{10}(^{1}_{0}S)\,4f^{5}(^{2}_{1}G)~^{2}G$ + 0.17~$4d^{10}(^{1}_{0}S)\,4f^{5}(^{2}_{6}F)~^{2}F$ + 0.12~$4d^{10}(^{1}_{0}S)\,4f^{5}(^{2}_{5}F)~^{2}F$ & 623479       & 77.3015  \\ 
193 & 9/2   & 0.24~$4d^{10}(^{1}_{0}S)\,4f^{5}(^{2}_{2}H)~^{2}H$ + 0.20~$4d^{10}(^{1}_{0}S)\,4f^{5}(^{2}_{6}G)~^{2}G$ + 0.09~$4d^{10}(^{1}_{0}S)\,4f^{5}(^{2}_{1}G)~^{2}G$ & 626993       & 77.7372  \\ 
194 & 11/2  & 0.54~$4d^{10}(^{1}_{0}S)\,4f^{5}(^{2}_{2}H)~^{2}H$ + 0.25~$4d^{10}(^{1}_{0}S)\,4f^{5}(^{2}_{7}H)~^{2}H$ + 0.06~$4d^{10}(^{1}_{0}S)\,4f^{5}(^{2}_{1}H)~^{2}H$ & 628421       & 77.9142  \\ 
195 & 3/2   & 0.40~$4d^{10}(^{1}_{0}S)\,4f^{5}(^{2}_{2}D)~^{2}D$ + 0.18~$4d^{10}(^{1}_{0}S)\,4f^{5}(^{2}_{5}D)~^{2}D$ + 0.13~$4d^{10}(^{1}_{0}S)\,4f^{5}(^{2}_{4}P)~^{2}P$ & 672448       & 83.3729  \\ 
196 & 5/2   & 0.25~$4d^{10}(^{1}_{0}S)\,4f^{5}(^{2}_{2}D)~^{2}D$ + 0.23~$4d^{10}(^{1}_{0}S)\,4f^{5}(^{2}_{1}D)~^{2}D$ + 0.18~$4d^{10}(^{1}_{0}S)\,4f^{5}(^{2}_{1}F)~^{2}F$ & 676505       & 83.8760  \\ 
197 & 7/2   & 0.59~$4d^{10}(^{1}_{0}S)\,4f^{5}(^{2}_{1}F)~^{2}F$ + 0.21~$4d^{10}(^{1}_{0}S)\,4f^{5}(^{2}_{6}F)~^{2}F$ + 0.04~$4d^{10}(^{1}_{0}S)\,4f^{5}(^{2}_{2}F)~^{2}F$ & 715351       & 88.6922  \\ 
198 & 5/2   & 0.33~$4d^{10}(^{1}_{0}S)\,4f^{5}(^{2}_{1}F)~^{2}F$ + 0.27~$4d^{10}(^{1}_{0}S)\,4f^{5}(^{2}_{2}D)~^{2}D$ + 0.14~$4d^{10}(^{1}_{0}S)\,4f^{5}(^{2}_{5}D)~^{2}D$ & 721528       & 89.4580  \\
\end{longtable*}

\clearpage

\section{Collisional-Radiative Model Simulations}

In our Collisional-Radiative Model (CRM) simulations, conducted using the \fac code \cite{Gu2008}, we employed the so-called 'one-ion model' \cite{Gu2003}. In this model, only radiative transition rates and cross sections for collisional excitations are considered in calculating the level populations for a given ion. The electron configurations considered in the CRM simulations are listed in Table~\ref{tab:crm-conf}. The model includes radiative transitions of types E1, E2, M1, and M2 between levels of the 'gr' group, as well as radiative transitions of type E1 between levels of the 'exc' and 'gr' groups. Collisional excitations between levels within the 'gr' group and between levels of the 'gr' and 'exc' groups are also included. 
Since the ionization energy for the \ce{Th^38+} ion is \SI{1680}{eV} \cite{NIST_ASD}, we assumed an electron energy of $T_e = \SI{1700}{eV}$ (monoenergetic energy distribution) in the CRM simulations. The assumed electron density of $d_e = \SI{e12}{\per\cm\cubed}$ is a typical value for EBIT devices.

\begin{table}[!htb]
\caption{\label{tab:crm-conf}Configurations included in the CRM simulations.}
\begin{tabular*}{\linewidth}{@{} l @{\extracolsep{\fill}} l r @{}}
\toprule
Group & Electron configurations ([Kr] core) & Number of levels \\
\midrule
gr & \ce{{4d}^{10} {4f}^{5}} & 198\\
exc & \ce{{4d}^{9} {4f}^{6}}, \ce{{4d}^{10} {4f}^{4} {5(s,p,d,f)}^{1}}, \ce{{4d}^{9} {4f}^{5} {5(s,p,d,f)}^{1}} & 61600\\
\bottomrule
\end{tabular*}
\end{table}

%\bibliographystyle{apsrev4-2}
%\bibliography{references}
